\documentclass[a4paper,12pt,justification=centering]{article}
\usepackage{tikz, pgfplots}
\usepackage{tikz-cd} 
\usepackage{latexsym}
\usepackage{rotating}
\usepackage{amsmath,amssymb,amsxtra,amscd,amsthm}
\usepackage[mathscr]{eucal}
\usepackage{graphicx}
\usepackage[justification=centering]{caption}
\usepackage{subfig}
\usepackage{datetime}
\usepackage{pdfsync}
\usepackage{thumbpdf}
\usepackage{color}
\ifx\pdfoutput\undefined

\usepackage{amsmath}
\usepackage{amssymb}
\usepackage{pgfplots}
\usepackage{graphicx} 
\usepackage{caption}
\usepackage{ mathrsfs }
\usepackage{subcaption}
\usepackage{color}
\usepackage[nottoc,notlot,notlof]{tocbibind}

\usepackage{verbatim}
\usepackage{tensor}
\usepackage{xcolor}
\usepackage{titlesec}

\graphicspath {{images/}}
\usepackage{wrapfig}

\usepackage
[dvips,
 verbose=false,
 bookmarks=true,
 colorlinks=true,
 linkcolor=webred,
 filecolor=webbrown,
 citecolor=webgreen,
 pagecolor=webblue,
 urlcolor=webblue,
 pdftitle={},
 pdfauthor={Murad Alim},
 pdfsubject={},
 pdfkeywords={},
 bookmarksopen=false,
 pdfpagemode=None,
 pdfview=FitH,
 pdfstartview=FitH,
 extension=pdf]{hyperref}
\else
\usepackage
[pdftex,
 verbose=false,
 bookmarks=true,
 colorlinks=true,
 linkcolor=webred,
 filecolor=webbrown,
 citecolor=webgreen,
 pagecolor=webblue,
 urlcolor=webblue,
 pdftitle={},
 pdfauthor={Murad Alim},
 pdfsubject={},
 pdfkeywords={},
 bookmarksopen=false,
 pdfpagemode=None,
 pdfview=FitH,
 pdfstartview=FitH,
 extension=pdf]{hyperref}
\fi
\definecolor{webred}{rgb}{.8,0,0}
\definecolor{webbrown}{rgb}{.6,0,0}
\definecolor{webgreen}{rgb}{0,0.5,0}
\definecolor{webdkgreen}{rgb}{0,0.3,0}
\definecolor{webblue}{rgb}{0,0,0.5}

\numberwithin{equation}{section}

\providecommand{\href}[2]{#2}

\allowdisplaybreaks
\usepackage{bbm}
\usepackage{rotating}
\newcommand{\be}{\begin{eqnarray}}
\newcommand{\beq}{\begin{eqnarray}}
\newcommand{\ee}{\end{eqnarray}}

\newcommand{\Z}{\mathbb{Z}}
\newcommand{\C}{\mathbb{C}}

\renewcommand{\b}{a}
\newcommand{\bD}{b}
\newcommand{\A}{A}
\newcommand{\AD}{B}
\newcommand{\holVol}{\omega}

\def\e#1\e{\begin{equation}#1\end{equation}}
\def\ea#1\ea{\begin{align}#1\end{align}}

\theoremstyle{plain}

\newtheorem*{thm*}{Theorem}

\theoremstyle{definition}

\usepackage{enumitem}

\begin{document}

\setlength{\parindent}{0cm}
\setlength{\baselineskip}{1.5em}
\title{Special geometry, quasi-modularity and attractor flow for BPS structures}

\author{Murad Alim$^1$\footnote{\tt{murad.alim@uni-hamburg.de}}, Florian Beck$^1$\footnote{\tt {flrn.beck@gmail.com}}, Anna Biggs$^2$\footnote{\tt {abiggs@princeton.edu}} and Daniel Bryan$^1$\footnote{\tt {daniel.bryan@uni-hamburg.de}}
\\
\small $^1$ Department of Mathematics, University of Hamburg, Bundesstr. 55, 20146, Hamburg, Germany\\
\small $^2$ Jadwin Hall, Princeton University, Princeton, NJ 08540, USA}

\date{}
\maketitle

\abstract{We study mathematical structures on the moduli spaces of BPS structures of $\mathcal{N}=2$ theories. Guided by the realization of BPS structures within type IIB string theory on non-compact Calabi--Yau threefolds, we develop a notion of BPS variation of Hodge structure which gives rise to special K\"ahler geometry as well as to Picard-Fuchs equations governing the central charges of the BPS structure. We focus our study on cases with complex one dimensional moduli spaces and charge lattices of rank two including Argyres--Douglas $A_2$ as well as Seiberg--Witten $SU(2)$ theories. In these cases the moduli spaces are identified with modular curves and we determine the expressions of the central charges in terms of quasi-modular forms of the corresponding duality groups. We furthermore determine the curves of marginal stability and study the attractor flow in these examples, showing that it provides another way of determining the complete BPS spectrum in these cases.}

\clearpage


\tableofcontents


\section{Introduction}

The study of BPS states within supersymmetric field and string theories has been a fruitful source of insights and interactions between mathematics and physics. This is due to the fact that BPS spectra carry rich mathematical structure and can be realized in different mathematical incarnations, see e.~g.~\cite{Aspinwallbook}. The spectrum of BPS states is locally constant in the moduli space of the underlying theory but jumping phenomena can occur at walls of marginal stability, where BPS states can decay or form bound states. These phenomena were studied by Cecotti and Vafa in the context of two-dimensional Landau-Ginzburg theories \cite{Cecotti:1992rm} and were crucial in obtaining the exact low energy effective action in the work of Seiberg and Witten \cite{SW}. See the reviews \cite{cecotti2010trieste,moore2010pitp,Pioline:2013wta}.

The BPS states of Seiberg--Witten theory were realized in type IIB string theory compactified on a non-compact Calabi--Yau threefold in~\cite{KLMVW}. The Seiberg--Witten curve becomes the degeneration locus of the underlying threefold and the Seiberg--Witten differential is obtained from the holomorphic three-form of the threefold geometry. The BPS states correspond in this context to D3 branes wrapping special Lagrangian submanifolds of the threefold. On the curve these special Lagrangians correspond to geodesics of a quadratic differential. An analogous study of BPS states in type IIB string theory for ADE type Argyres--Douglas theories was realized by Shapere and Vafa in~\cite{Shapere:1999xr}, where the existence of geodesics of the quadratic differential was analyzed locally near branching points of the underlying curve and global consistency conditions for the corresponding flows led to solutions of the BPS problem. The Argyres-Douglas theories refer to four-dimensional $\mathcal{N}=2$ superconformal field theories which were found at special slices in higher dimensional moduli spaces of higher rank gauge theories \cite{Argyres:1995jj} as well as within the moduli spaces of  $SU(2)$ theories coupled to matter \cite{Argyres:1995xn}. 

In the context of string theory, BPS states correspond to special objects which exist in the theory. Most notably in type IIA (IIB) string theories, the BPS states originate from supersymmetric D-branes supported on even (odd) dimensional submanifolds of the spatial part of 10 dimensional space-time. Type IIA  and IIB string theories can be considered on ten-dimensional space-times which are of the product form $\mathbb{R}^{1,3} \times X$, where $X$ denotes a Calabi--Yau (CY) threefold. These compactifications give, generically,  effective four dimensional $\mathcal{N}=2$ theories. Mirror symmetry in this setup refers to the fact that there exist mirror families of CY threefolds, such that the effective four dimensional theories originating from type IIA on CY $X$ or type IIB on CY $Y$ are indistinguishable. This process identifies objects supported on even dimensional submanifolds of $X$ with objects supported on odd dimensional submanifolds of $Y$. Mathematically, this can be formulated precisely using the homological mirror symmetry conjecture \cite{KontsevichHMS}, see also \cite{Aspinwallbook} and references therein. 

The stability of the physical BPS states is governed by the $\mathcal{N}=2$ central charge, which associates to every charged state in the theory a complex number which varies over the moduli space of the family. A formulation of the central charge as the relevant notion of stability of objects in the CY compactification setting was given by Douglas, Fiol and R\"omelsberger in~\cite{Douglas_2005} in terms of the notion of $\pi$-stability, see also ~\cite{DouglasStability} for a mathematical account. This stability condition inspired the mathematical notion of stability in triangulated categories put forward by Bridgeland in~\cite{Bridgeland}, which gives the notion of spaces of stability conditions. While $\pi$-stability is naturally associated to a problem of variation of Hodge structure of the underlying family of CY threefolds, Bridgeland's stability condition doesn't require such a constraint. 

The existence of BPS states in the context of supergravity was studied by Denef \cite{Denef:2000nb}, mapping the problem of determining the existence of BPS states within regions in moduli space to the problem of determining solutions to the attractor flow equations. The attractor mechanism was first described in the context of extremal black hole solutions in the context of $\mathcal{N}=2$ supergravity \cite{Ferrara_1995}. The attractor flow equations were then described in detail in \cite{Strominger_1996, Ferrara_1996} and were formulated in terms of a time derivative in \cite{Ferrara:1997tw} such that the BPS mass is minimized in the limit of infinite time.  BPS existence conditions were given in terms of the value of the central charge at the end point of the attractor flow lines \cite{Moore:1998pn}.  The reviews ~\cite{Ferrara:2008hwa,Pioline:2008zz} describe in detail the literature on attractor flow in the context of black hole solutions in supergravity, for example the radial flow of scalars with initial values at infinity to a fixed point at the horizon, the connection to the Bekenstein Hawking entropy, and the OSV conjecture. A wall-crossing formula was developed in this context by Denef and Moore \cite{Denef:2007vg} giving a quantitative handle on the problem of determining the decay and recombination of BPS solutions. This was then further developed by \cite{Alexandrov:2018iao} where the BPS index is explicitly determined from the attractor indices.

The method of using split attractor flows to determine the existence of BPS states was used in \cite{Denef:2000nb}, for example including type IIB string theory on the mirror quintic and Seiberg-Witten theory in the low energy $\mathcal{N}=2$ supergravity limit of type II string theory and only weakly coupled to gravity as described in \cite{Denef:1998sv}. The interpretation of the existence conditions was further developed by finding the explicit black hole and empty hole solutions to the supergravity equations at the end point of the flow. Importantly, the existence conditions for Seiberg-Witten theory were determined in the context of monodromies around the singular points and the known spectrum of BPS states reproduced by splitting the attractor flow of composite states at the wall of marginal stability and by allowing the flow to pass through the appropriate branch cuts.   

In \cite{Denef:2001xn} the attractor flow method is extended to a detailed analysis of BPS type IIA D-branes on the quintic. Here, the attractor equations were written in terms of gradient flow of the central charges. The equations were solved approximately by minimizing the central charge using iterative methods and then plotting the attractor flow lines. The existence conditions on the endpoint of the flow were then considered as before. More recently in \cite{Bousseau:2022snm} this attractor flow analysis was applied on the type IIA side to local $\mathbbm{P}^2$, which has its BPS spectrum represented on a 3 node quiver.

The goal of our work is to further develop the mathematical structures which encode the existence and stability data of BPS structures on the moduli space of a given theory itself. Guided by the realization of simple BPS structures within compactifications of string theory on Calabi--Yau threefolds we put forward a notion of BPS variation of Hodge structure (BPS-VHS) \footnote{A more mathematical account of this structure will appear in \cite{BPS-VHS}} which gives rise to a special K\"ahler geometry structure on the moduli space associated to a given BPS structure. In the context of Calabi--Yau threefolds this is just the usual special K\"ahler geometry \cite{FreedSK}, our goal is however to formulate the ingredients of this additional geometric structure constraining the space of stability in terms of the language of variation of Hodge structure. The BPS-VHS allows us to put forward Picard-Fuchs differential equations for the central charges associated to a BPS structure, we use techniques due to Griffiths to derive the corresponding Picard-Fuchs equations explicitly for the examples which we study, complementing known results in the literature. We focus our study on simple examples with a complex one-dimensional moduli space and a rank two lattice of charges, in these cases the moduli space can be further identified with a modular curve and the central charges have expressions in terms of quasi-modular forms which we provide. We revisit the attractor flow equations in the context of our examples and re-derive the results of \cite{Denef:2000nb} for Seiberg--Witten theory by explicitly producing plots of iterative solutions to the attractor flow equations, finding all possible flows between covers of the moduli space, and using the existence conditions on these flows to determine the spectrum. We then extend this method to two parameterizations of the Argyres--Douglas $A_{2}$ theory to determine the spectrum in a novel example. The $A_{1}$ model is also briefly discussed.

The structure of this paper is as follows. In Sec.~\ref{sec:BPSstructures} we introduce the notion of BPS structures following ~\cite{GMN1,Bridgeland}. We proceed in Sec.~\ref{sec:BPSVHS} by introducing the notion of BPS variation of Hodge structure and show how this leads to Picard-Fuchs differential equations governing the central charge of the BPS structure. We introduce the geometric realizations of the BPS structures which we study in detail in this work originating from Calabi--Yau threefolds. We proceed in Sec.~\ref{sec:modularity} by reviewing quasi-modular forms for subgrougps of $SL(2,\mathbb{Z})$ as well as the Fricke involution acting both on the moduli space as well as on the quasi-modular forms. In Sec.~\ref{section:attractorflowproject2examplesdiagrams} we study the attractor flow for our examples showing that in these cases this provides a complete answer to the problem of determining the BPS spectrum in all chambers of the moduli space.

\subsubsection*{\it{Relation to other work}}
The BPS content of Seiberg--Witten theory was already part of the original paper \cite{SW}. The modular structure as well as the Picard-Fuchs equations governing the central charges of SW theory were studied in \cite{Klemm:1994qs,Klemm:1995wp} as well as in the context of topological string theory in \cite{Huang:2006si,Aganagic:2006wq}. The realization of SW theory in the context of type IIB string theory was put forward in \cite{KLMVW} where it was also realized that BPS states of the SW physical theory correspond to special Lagrangians of the non-compact CY threefold and to geodesics of a quadratic differential on the Seiberg-Witten curve which appeared as a degeneration locus of the CY threefold, see also \cite{Lerche:1996xu} and references therein. This method of studying BPS spectra was applied in \cite{Shapere:1999xr} to study the BPS spectrum of Argyres-Douglas $A_2$ theory which features prominently in our work, this reference also contains two different realizations of the curve associated to the theory which we study in detail in our work. The study of the geometric realization of BPS states as geodesics of quadratic differentials also a cornerstone of the works of Gaiotto, Moore and Neitzke (GMN) \cite{GMN1,GMN2} as well as of Bridgeland and Smith \cite{BridgelandSmith}. The attractor flow as a method of studying the BPS spectrum of theories decoupled from gravity and hence from the Black Hole context was considered by Denef in \cite{Denef:1998sv}. The BPS spectrum of SW theory using quiver representation theory was also studied by Denef in \cite{Denef:2002ru}, BPS quivers were also subject of \cite{Douglas:2000ah,Douglas:2000qw} where the notion of $\pi-$stability was put forward. Quiver representation theory as a tool to study BPS spectra of $\mathcal{N}=2$ theories was furthermore used in \cite{Cecotti:2011rv,Alim:2011ae,Alim:2011kw} which also include the study of the $A_2$ quiver, the latter was also studied in \cite{BridgelandA2} in the context of stability conditions. The curve of marginal stability for $SU(2)$ SW theory was studied in \cite{Fayyazuddin:1995pk,Argyres:1995gd,Matone:1995jr}, parts of the curve of marginal stability for the Argyres-Douglas theory in one geometric realization of our work appears in \cite{Shapere:1999xr}, for another realization the curve appears in \cite{Neitzke:2013tca}. Recently, curves of marginal stability of $SU(3)$ SW theory were also studied in \cite{DDN}. 

We remark that the wall-crossing and BPS structures have a related but different incarnation in terms of the study of the exact WKB analysis within quantum mechanics which deals with the asymptotic nature the WKB method using Borel resummation. The WKB setup which corresponds to the $A_2$ theory studied in our work is the one of the cubic oscillator, studied in detail by Delabaere, Dillinger and Pham in \cite{DDP}, the relation to the BPS context of quadratic differentials was given in \cite{IwakiNakanishi}. This exact WKB setup is in turn related to topological string theory in the Nekrasov-Shatashvili limit. This was studied for the cubic oscillator in \cite{Codesido:2017dns} and for Seiberg--Witten theory in \cite{Grassi:2019coc}. In \cite{Codesido:2017dns}, the exact WKB of the cubic oscillator was studied using quasi modular forms by first mapping the cubic curve to a SW form and using the quasi-modular structure of the latter giving rise to a different quasi-modular structure from the ones in our work which are based on two different realizations of the cubic curve and the Picard-Fuchs equations which we derive for these. The WKB analysis of the cubic oscillator further admits connections to Painlev\'e equations, this is subject of various works including \cite{Bonelli:2016qwg,Grassi:2018spf,Iwaki:2019zeq}. The walls of marginal stability which we analyze in our work appear as Stokes lines in the exact WKB context as well as in the direct analysis of Borel resummation in connection with Painlev\'e transcendents, see \cite{Aniceto:2018bis} and references therein as well as \cite{vanSpaendonck:2022kit} for a recent treatment.

\section{BPS structures and wall-crossing} \label{sec:BPSstructures}
In this section we introduce the relevant notions characterizing the set of BPS states of a given physical theory, their geometric realizations, as well as the data determining their wall-crossing structures.

\subsection{BPS structures}\label{ss:bps}
We follow ~\cite{GMN1,Bridgeland} in introducing the necessary data for studying the BPS problem. 

Let $\mathcal{B}$ be a complex manifold, the \emph{moduli space}, of dimension $r$. 
We will denote any local coordinate on $\mathcal{B}$ by $(u_1, \dots, u_r)$. 
In physical realizations of $\mathcal{N} = 2$ gauge theory, $\mathcal{B}$ corresponds to the Coulomb branch and $r$ to the rank of the gauge group. 
We assume that $\mathcal{B}$ carries a local system $\Gamma$ with fiber $\Gamma_u \cong \mathbb{Z}^{2r}$ at each $u\in \mathcal{B}$.
In physical realizations, $\Gamma_u$ is the lattice of electric and magnetic charges of the corresponding theory. 
(Local) sections of $\Gamma$ are often informally denoted by $\gamma \in \Gamma$. 

We further assume that $\Gamma$ carries a symplectic pairing\footnote{Here $\underline{\mathbb{Z}}$ is the constant sheaf with fiber $\mathbb{Z}$.}
\begin{equation}\label{pairing}
\langle -, - \rangle \colon \Gamma \times \Gamma \to \underline{\mathbb{Z}}.
\end{equation}
Locally on $\mathcal{B}$ we can split $\Gamma = \Gamma^e \oplus \Gamma^m$ into Lagrangian sub-lattices.
These are the \emph{electric} and \emph{magnetic charge lattices} respectively. 
Concretely, we may find bases $\{ \alpha_1,\dots,\alpha_r\}$ for $\Gamma^m$ and $\{ \beta^1,\dots,\beta^r\}$ for $\Gamma^e$ such that
\begin{equation}
\langle \alpha_I , \beta^J \rangle =\delta_I^J \,, \quad  \langle \alpha_I , \alpha_J \rangle = 0 = \langle \beta^I , \beta^J \rangle
\end{equation}
for $I, J = 1, \dots, r$.
Such a basis of $\Gamma$ is called an electric-magnetic basis. 
Then every $\gamma \in \Gamma$ is expressed as
\begin{equation}
\gamma=\sum_{I=1}^r p^I \alpha_I + q_I \beta^I\,,
\end{equation}
with $(q,p)$ denoting the electric and magnetic charges respectively. 
The pairing between $\gamma^i \in \Gamma$ ($i=1,2$) is the Dirac pairing of the corresponding charges $(q^i, p^i)$:
\begin{equation}
\langle \gamma^i, \gamma^j\rangle = \sum_a (p^i)^a (q^j)_a - (p^j)^a (q^i)_a\,, \quad i, j\in \{1,2\}\,.
\end{equation}
We also assume that $\Gamma$ comes with a holomorphic map 
\begin{equation}
Z \colon \mathcal{B} \to \mathrm{Hom}(\Gamma, \mathbb{C})\,,
\end{equation}
subject to the following two conditions\footnote{Technically, here the brackets stand for the dual pairing on $\mathrm{Hom}(\Gamma, \mathbb{C})$.}
\begin{equation}\label{eq:special}
\begin{gathered}
\langle dZ \wedge dZ \rangle = 0, \\
\langle dZ \wedge d\bar{Z} \rangle \mbox{ is positive 2-form.}
\end{gathered}
\end{equation}

The last condition means that for any non-zero tangent vector $V\in T^{1,0}_u \mathcal{B}$, $u\in \mathcal{B}$, we have
\begin{equation}\label{eq:special2}
-i\, \langle dZ \wedge d\bar{Z}\rangle (V, \bar{V}) > 0.
\end{equation}
The holomorphic function
$
Z_\gamma(u) := Z(u) \cdot \gamma
$
is the \emph{central charge of $\gamma$} since in physical realizations it corresponds to the central charge of the $\mathcal{N}=2$ supersymmetry. 
Finally, the \emph{mass function} is a map 
\begin{equation}
M \colon \mathcal{B} \to \mathrm{Map}(\Gamma, \mathbb{R})\,,
\end{equation}
such that for each $\gamma$ and $u\in \mathcal{B}$ the BPS bound is satisfied
\begin{equation}\label{eq:bpsbound}
M_\gamma(u):= M(u)\cdot \gamma \geq |Z_\gamma (u)|\,.
\end{equation}
The BPS states are states in the theory labeled by their charge $\gamma \in \Gamma$ such that \eqref{eq:bpsbound} is saturated. 
We will denote by $\mathcal{S}_u$ the BPS spectrum at $u \in \mathcal{B}$, i.e. all BPS states in $\Gamma_u$.

More generally, we consider triples $((\hat\Gamma, \langle -, - \rangle), \hat{Z})$ where 
\begin{itemize}
    \item $\hat\Gamma$ is a finite rank lattice over $\mathcal{B}$ with a skew-symmetric pairing $\langle -, - \rangle$.
    Its radical, often called the \emph{flavour lattice}, 
    is denoted by $\Gamma_{fl}$ so that we obtain the extension
    \begin{equation}
    \begin{tikzcd}
    0 \ar[r] & \Gamma_{fl} \ar[r] & \hat\Gamma \ar[r] & \Gamma \ar[r] & 0. 
    \end{tikzcd}
    \end{equation}
    \item $\hat Z$ is a holomorphic map $\hat{Z} \colon \mathcal{B} \to \mathrm{Hom}(\hat\Gamma, \C)$ such that $\hat{Z}_{\gamma}(-)$ is locally constant for each $\gamma \in \Gamma_{fl}$.
    Thus $d\hat{Z}$ gives a well-defined $1$-form on $\Gamma$ which we denote by $dZ$.
\end{itemize}
Then the symplectic lattice $(\Gamma, \langle -, - \rangle)$ together with $dZ$ is required to satisfy the above properties. 

\subsection{Induced special geometry and attractor flow}\label{ss:specialgeom}
The previous structures define an affine special K\"ahler geometry on $\mathcal{B}$ which we briefly recall (for example, see \cite{NeitzkeHK}, \cite{FreedSK}).
We first consider the case where $\Gamma_{fl} = 0$.
The second condition in \eqref{eq:special} implies that for an electro-magnetic frame $(\alpha_I, \beta^J)$ of $\Gamma$ the central charges
\begin{align}
\b^J := Z_{\beta^J}, \quad \bD_I:= Z_{\alpha_I}
\end{align} 
form a pair of local coordinates on $\mathcal{B}$.
Let $(\A^I,\AD_J)$ be the dual basis of $(\alpha_I, \beta^J)$ so that 
\begin{align*}
    Z&= Z_{\alpha_I}\, \A^I + Z_{\beta^J}\, \AD_J = \bD_I\, \A^I + \b^J\, \AD_J, \\
    dZ& = d\bD_I \, A^I + d\b^J \, \AD_J. 
\end{align*}
Using $\langle \A^I, \AD_J \rangle = \delta_{I}^J$ and the first condition in \eqref{eq:special}, we have
\begin{align}
0 &= \langle dZ\left( \tfrac{\partial}{\partial \b^I}\right), dZ\left(\tfrac{\partial}{\partial \b^J}\right) \rangle  
\\
&= \tfrac{\partial \bD_J}{\partial \b^I} - \tfrac{\partial \bD_I}{\partial \b^J}\,. 
\end{align}
Hence there is a local prepotential $\mathcal{F}(\b^J)$ such that 
\begin{equation}
\bD_I = \tfrac{\partial \mathcal{F}}{\partial \b^I}.
\end{equation}
In particular, $(\b^J)$ and $(\bD_I)$ is a pair of local \emph{special} coordinates on $\mathcal{B}$.
Moreover, $\omega:=\langle dZ \wedge d\bar{Z}\rangle$ is the K\"ahler form of this special K\"ahler structure. 
In the previous coordinates, it is given by 
\begin{equation}
\omega = \omega_{I\bar{J}}\, d\b^I \wedge d\bar{\b}^J = 2i \, \mathrm{Im}\left(\tfrac{\partial^2 \mathcal{F}}{\partial \b^I \partial \b^J}\right) \, d\b^I \wedge d\bar{\b}^J. 
\end{equation}
In particular, $\mathcal{K}:=2\,\mathrm{Im}\left(\bD_I \bar{\b}^I\right)$ is a K\"ahler potential, i.e. $\omega = i\partial \bar{\partial} \mathcal{K}$.
We denote by $g$ the corresponding K\"ahler metric.
Its coefficients with respect to the local coordinates $\b^I$ are 
\begin{equation}
    g_{I \bar{J}} = -i \, \omega_{I \bar{J}}.
\end{equation}
For each $\gamma \in \Gamma$ we define its \emph{attractor flow} as the flow of the vector field
\begin{equation}\label{eq:attractorflow}
    - \mathrm{grad}^g\, |Z_{\gamma}|,
\end{equation}
i.e. the gradient flow of $|Z_{\gamma}|$ with respect to $g$.

If $\Gamma_{fl} \neq 0$, then then special geometry is determined by $\Gamma = \hat{\Gamma}/\Gamma_{fl}$ and $dZ$ in the same way as before. 
For the attractor flows, we work with $\hat{Z}$.

\subsection{One-dimensional case}\label{One-dimensional case}
In our examples, we focus on the case $\dim_{\mathbb{C}}\mathcal{B} = 1$, in particular $\Gamma$ is of rank $2$ (but $\hat{\Gamma}$ might have higher rank). 
We next give explicit formulas for the attractor flow in this case. 

Let $(\alpha, \beta)$ be an electro-magnetic frame of $\Gamma$ and $(\b, \bD)$ the corresponding pair of special coordinates. 
Then $\mathcal{F} = 2\, \mathrm{Im} (\bD \bar{\b})$ is a K\"ahler potential so that we write 
\begin{equation}
\omega = \omega_{\b \bar{\b}}\, d\b \wedge d \bar{\b}
\end{equation}
and similarly for the K\"ahler metric $g$.
A (local) flow line $u(\tau)$ for the attractor flow \eqref{eq:attractorflow} satisfies the local equation
\begin{align}\label{eq:attractor2}
    \dot{u}(\tau) &= - g^{\b \bar{\b}}\, \bar{\partial}_{\bar{\b}} |Z_\gamma| \circ u(\tau) \\
    &= - \frac{i}{ \omega_{\b \bar{\b}}} \, \bar{\partial}_{\bar{\b}} |Z_\gamma| \circ u(\tau).
\end{align}
The reparameterization $\tau \to -(\omega_{\b \bar{\bD}}/i) \tau$ of $u(\tau)$ solves the gradient flow of $|Z_{\gamma}|$ for the flat metric. 
In particular, the flow lines do not qualitatively differ between these two metrics. 
This is useful for plotting the flow lines in our examples. 
For that purpose we write $|Z_\gamma(u)|$ as a function on $\mathbb{R}^{2}$ as $|Z_\gamma(x,y)|$.
Hence the standard gradient gives 
\begin{align}
\nabla |Z_\gamma(x,y)| = \frac{\partial |Z_\gamma(x,y)|}{\partial x} \hat{e}_{x}+\frac{\partial |Z_\gamma(x,y)|}{\partial y} \hat{e}_{y}\,.
\end{align} 
Therefore the required differential equation reads
\begin{align}
\frac{dy}{dx} =  \frac{\left(\frac{\partial |Z_\gamma(x,y)|}{\partial y}\right)}{\Big(\frac{\partial |Z_\gamma(x,y)|}{\partial x}\Big)}\,.
\end{align}
When the expression for $|Z_\gamma(x,y)|$ is written on in terms of $x$ and $y$ this equation should be integrated to find the attractor flow lines. Where this is not possible we compute the gradient flow iteratively and plot the attractor flow lines, e.g. with \textsc{Mathematica}, from this. 

If $\Gamma \neq \hat{\Gamma}$ the same formulas hold with $Z$ replaced by $\hat{Z}$.

\subsection{Wall crossing} \label{sec:wall}
A BPS particle of charge $\gamma$ can only decay into two BPS particles (at some $u\in \mathcal{B}$) if both its mass and charge can split into the masses and charges of its decay constituents, i.e. the following has to be satisfied
\begin{eqnarray}
&\gamma = \gamma^1+ \gamma^2\,, \\
&M_\gamma(u) = M_{\gamma^1}(u) + M_{\gamma^2}(u)\,.
\end{eqnarray}

Since $\hat Z(u)$ is a homomorphism from $\hat\Gamma$ to $\mathbbm{C}$ we have:
\begin{equation*}
\hat{Z}_\gamma(u)= \hat{Z}_{\gamma^1}(u) + \hat{Z}_{\gamma^2}(u).
\end{equation*}
For BPS states we have furthermore that $M_\gamma(u) = |\hat{Z}_\gamma(u)|$, but in general,
\begin{equation*}
|\hat{Z}_\gamma(u)| \le |\hat{Z}_{\gamma^1}(u)| + |\hat{Z}_{\gamma^2}(u)|,
\end{equation*}
hence a decay can only happen when the phases of the central charges of the particles are aligned. This only happens at co-dimension one loci in the moduli space called the \emph{walls of marginal stability}.
We denote by\footnote{Of course, we implicitly require $\hat{Z}_{\gamma^2}(u) \neq 0$ here.}
\begin{equation}
    \mathcal{W}_{\gamma^1,\gamma^2} := \left\{ u \in \mathcal{B} \,|\, \hat{Z}_{\gamma^1}(u)/  \hat{Z}_{\gamma^2}(u)\in \mathbb{R}\right\}
\end{equation}
the wall of marginal stability for a decay into constituent BPS particles of charges $\gamma^1$ and $\gamma^2$.

\subsection{Existence conditions}
\label{sec:existenceconditions}

One can determine which BPS states exist in the spectrum in each part of the moduli space and the corresponding wall crossing phenomena. Existence conditions of BPS states have been studied using several methods including spectral networks, quiver representation theory or wall crossing formulae, see \cite{Aspinwallbook,neitzkelectures,Alim:2011kw}. The method of attractor flow has been developed and used by Denef \cite{Denef:2000nb} and Denef, Green and Raugas \cite{Denef:2001xn} to determine BPS spectra for the quintic and re-derive it for Seiberg--Witten theory.
We briefly recall this method for $\Gamma_{fl} = 0$. 
Note that this case includes the situation for $\hat{Z}_\gamma = 0$ for all $\gamma \in \Gamma_{fl}$ (which does occur in examples). 
The general case works similarly. 

Once the basis BPS states, in our case $\alpha$ and $\beta$, are fixed one can derive the full spectrum by considering all possible linear combinations of the basis states of the form $n \alpha+m \beta$ and determining which states exist and which can be excluded. To distinguish the existing BPS spectrum from the excluded states existence conditions used by Moore and Denef can be applied. These are based on the solutions of the attractor flow equations for each individual central charge  $Z_{\gamma}(u)$ where $\gamma = n \alpha +m \beta$. 

The attractor flow can then be worked out for the BPS central charges in the theory. The existence conditions state that that if a solution to the supergravity equations, in our case the gradient flow lines, exists the BPS state exists. Whether the solution exists depends on the end point of the flow. There are 3 possibilities:   \cite{Denef:2000nb,Denef:2001xn}.  

\begin{enumerate}

	\item  If the flow terminates at a point and $|Z_{\gamma}(u)| > 0$ it forms black hole. Here $\tau \rightarrow \infty$ and the metric becomes near horizon around a massive state.
	
	\item  If the flow terminates at a regular point and $|Z_{\gamma}(u)| =0$ no solution exists. This is because the periods can vanish only at a singular point as the cycles pinch here.

	\item  If the flow terminates at a singular point and $|Z_{\gamma}(u)| =0$ the solution forms an empty hole.
	
\end{enumerate}

As non compact theories are decoupled from gravity the existing states in our examples correspond to the empty hole case. For such a solution the potential vanishes at the radius within that of vanishing central charge whereas the moduli stays constant \cite{Denef:2000nb,Denef:2000ar}.

\section{BPS Variation of Hodge structure}\label{sec:BPSVHS}
In this section we introduce the notion of BPS variation of Hodge structure (BPS-VHS) which will be treated in more mathematical detail in \cite{BPS-VHS}. The VHS setup allows up to put forward Picard-Fuchs equations for the examples which are discussed in our paper.

\subsection{BPS-VHS and Picard--Fuchs equations}
In our examples, the central charges are of a special form. 
Assume first that the symplectic lattice $(\Gamma, \langle -, - \rangle)$ over $\mathcal{B}$ admits a holomorphic section $\lambda \in H^0(\mathcal{B}, \Gamma^\vee\otimes \mathcal{O}_{\mathcal{B}})$ such that 
\begin{equation}
    T\mathcal{B} \to \Gamma^\vee\otimes \mathcal{O}_{\mathcal{B}},\quad V \mapsto \nabla_V \lambda
\end{equation}
is an isomorphism onto a Lagrangian subbundle $F^\lambda \subset \Gamma^\vee\otimes \mathcal{O}_{\mathcal{B}}$.
Here $\nabla$ is the canonical flat connection on $\Gamma^\vee\otimes \mathcal{O}_{\mathcal{B}}$. 
Then the tuple $((\Gamma, \langle -, - \rangle), \lambda)$ is called a \emph{BPS-variation of Hodge structures} (BPS-VHS). 
In the forthcoming \cite{BPS-VHS} it is shown that 
\begin{equation}
    Z(u)(-):= \lambda_u(-)
\end{equation}
satisfies \eqref{eq:special}. 
Moreover, $(\Gamma^\vee, F^\lambda)$ is a variation of Hodge structure of weight $1$, in particular 
\begin{equation}
    \Gamma_u^\vee\otimes \mathbb{C} = F^\lambda_u \oplus \bar{F}^\lambda_u, \quad u\in \mathcal{B}.
\end{equation}
In this case, we can explicitly determine the special geometry (introduced in Subsection \ref{ss:specialgeom}) as follows. 
For simplicity, we assume $\dim_{\mathbb{C}} \mathcal{B} = 1$ which is the relevant case in our examples. 
Let $(\alpha, \beta)$ be a local electro-magnetic frame of $\Gamma$ so that the dual special coordinates are given by 
\begin{equation}
    a = Z_\beta =  \lambda(\beta), \quad b = Z_\alpha = \lambda(\alpha). 
\end{equation}
Since $\Gamma$ is of rank $2$, there are (at worst) meromorphic functions $c_1, c_2$ such that 
\begin{equation}
    \nabla_V^2 \lambda + c_1\, \nabla_V \lambda + c_2\, \lambda = 0.
\end{equation}
Here $V = \partial_u$ is the local frame of $T\mathcal{B}$ induced by a local coordinate $u$. 
Since $\alpha, \beta$ and the symplectic pairing are flat with respect to $\nabla$, we deduce that $a$ and $b$ satisfy the second order differential equation 
\begin{equation}
    \partial_u^2 f + c_1\, \partial_u f + c_2 \, f = 0.
\end{equation}
We call it the \emph{Picard--Fuchs equation of the BPS-VHS}, in analogy to Picard--Fuchs equations of usual variations of Hodge structures. 
Note that $\partial_u a$ and $\partial_u b$ are periods of the VHS $(\Gamma^\vee, F^\lambda)$.

As before, we also consider a generalization with degenerate pairings. 
In this context, a BPS-VHS is a triple $((\hat{\Gamma}, \langle -,- \rangle), \lambda)$ consisting of
\begin{itemize}
    \item a finite rank lattice $\hat{\Gamma}$ with a (degenerate) skew-symmetric pairing $\langle -,- \rangle$
    \item a holomorphic section $\lambda \in H^0(\mathcal{B}, \hat\Gamma^\vee \otimes \mathcal{O}_{\mathcal{B}})$ such that\footnote{We informally denote the image of $\lambda$ in $\Gamma\otimes \mathcal{O}_\mathcal{B}$ by the same symbol.} 
    \begin{equation}
        T\mathcal{B} \to \hat{\Gamma}^\vee\otimes \mathcal{O}_{\mathcal{B}}, \quad V \mapsto \nabla_V \lambda,
    \end{equation}
    is an isomorphism onto a Lagrangian subbundle $F^\lambda$ of $\hat{\Gamma}_{fl}^\perp\otimes \mathcal{O}_{\mathcal{B}}$. 
    Note that $\hat\Gamma_{fl}^\perp \cong \Gamma^\vee$ (for $\Gamma=\hat{\Gamma}/\hat{\Gamma}_{fl}$) is a symplectic lattice. 
\end{itemize}
We further define 
\begin{equation}
   \hat{Z}\colon \mathcal{B} \to \mathrm{Hom}(\hat\Gamma, \mathbb{C}), \quad u \mapsto \hat{Z}(u)(-) =  \lambda_u(-)
\end{equation}
as before.
By identifying $\Gamma^\vee = \hat{\Gamma}_{fl}^\perp$ for $\Gamma = \hat{\Gamma}/\hat\Gamma_{fl}$, we get the relation to the non-degenerate case.
In particular, we obtain a special geometry on $\mathcal{B}$ as well as Picard--Fuchs equations as above.

Using Griffiths' pole order reduction, we explicitly determine the Picard--Fuchs equations of the BPS-VHS in our examples and hence their special geometry.   
Before doing so, we motivate BPS-VHS geometrically and give a large class of concrete examples.

\subsection{Geometric realization of BPS structures}
We will consider the following geometric realization of the previous structures in type IIB string theory on $\mathbbm{R}^{1,3}\times X$ for certain non-compact Calabi--Yau threefolds $X$.
These fiber over $\mathbb{CP}^1$ and are constructed as follows: 
consider a polynomial 
\begin{equation}\label{eq:poly}
    f(z) = \prod_{i=1}^3 (z - a_i), \quad a_i \neq a_j \mbox{ for } i \neq j,
\end{equation}
of degree $3$ with pairwise distinct roots.
For simplicity, we assume $a_i \neq 0$ for $i = 1,2,3$. 
Then we obtain the affine and smooth threefold 
\begin{equation}\label{eq:X0}
    X^0:= \{ (y,z,v_1, v_2) \in \mathbb{C}^4 ~|~ z^4(v_1^2 + v_2^2) + y^2 = f(z) \}. 
\end{equation}
Now let $z':= 1/z$ for $z\in \mathbb{C}^*$ and set 
\begin{equation}
g(z'):= \prod_{i=1}^3 (1 - a_i z').
\end{equation}
Then we define 
\begin{equation}\label{eq:Xinfty}
    X^\infty:= \{ (y', z', v_1', v_2') \in \mathbb{C}^4 ~|~ v_1'^2 + v_2'^2 + y'^2 = z'g(z') \}.
\end{equation}
Now we may glue $X^0$ and $X^\infty$ via 
\begin{equation}\label{eq:gluing}
    (y, z, v_1, v_2) \mapsto (y' = y/z^2, z' = 1/z, v_1' = v_1, v_2' = v_2).
\end{equation}
The result is a threefold $X$ with a natural map $\pi \colon X \to \mathbb{CP}^1$ (induced by the projection to $z$ and $z'$). 
Its fiber over $z\in \mathbb{CP}^1$ is given as follows:
\begin{itemize}
    \item $z=0$: a disjoint union of two copies of $\mathbb{C}^2$;
    \item $z=a_i$: a degenerate quadric isomorphic to $v_1^2 + v_2^2 + y^2 = 0$;
    \item $z \neq a_i$ or $0$: a smooth quadric isomorphic to $v_1^2 + v_2^2 + y^2 = w$ for some $w \neq 0$.
\end{itemize}
The gluing data show that there is a natural inclusion $X \hookrightarrow \mathrm{tot}(\mathcal{V})$ for the rank three vector bundle
\begin{equation}
    \mathcal{V}:= \mathcal{O}(2) \oplus \mathcal{O} \oplus \mathcal{O}
\end{equation}
over $\mathbb{CP}^1$.
Here each summand corresponds to $y$, $v_1$, $v_2$ and $y'$, $v_1'$, $v_2'$ respectively.
Moreover, we consider the equalities in \eqref{eq:X0} and \eqref{eq:Xinfty} as equalities in $\mathrm{tot}(\mathcal{O}(4))$.

We next show that $X$ is a Calabi--Yau threefold, i.e. that it has trivial canonical class. 
First, let $p \colon \mathrm{tot}(\mathcal{V}) \to \mathbb{CP}^1$ be the projection. 
Then the canonical class of $\mathrm{tot}(\mathcal{V})$ is given by
\begin{equation}
    K_{\mathrm{tot}(\mathcal{V})} = p^*\det \mathcal{V}^* \otimes p^*K_{\mathbb{CP}^1} \cong p^*\mathcal{O}(-4). 
\end{equation}
Since $X \hookrightarrow \mathrm{tot}(\mathcal{V})$, the adjunction formula gives
\begin{align}
    K_X &= (K_{\mathrm{tot}(\mathcal{V})} \otimes \pi^*\mathcal{O}(4))_{|X} \\
    &= (\pi^*\mathcal{O}(-4) \otimes \pi^*\mathcal{O}(4))_{|X}
    \\
    &= \mathcal{O}_X.
\end{align}

More concretely, define $W^0(y,z,v_1,v_2):=z^4(v_1^2+v_2^2) + y^2 - f(z)$ so that $X^0$ is the vanishing locus of $W^0$.
Then 
\begin{equation}
    \holVol^0:= \tfrac{1}{2\pi i} \mathrm{Res}_{X^0}\left( \tfrac{1}{W} ~dy\wedge dz \wedge dv_1 \wedge dv_2 \right)
\end{equation}
is a holomorphic volume form on $X^0$.
For the open subset on $X^0$ where $\partial_y W^0 = 2 y \neq 0$ (so that $(z,v_1,v_2)$ are coordinates), we have 
\begin{eqnarray}
    \holVol^0 = \tfrac{1}{2\pi i} \tfrac{1}{2y} ~dz \wedge dv_1 \wedge dv_2. 
\end{eqnarray}
Analogously, we define $W^\infty(y', z', v_1', v_2') = v_1'^2 + v_2'^2 + y'^2 - z'g(z')$ and the holomorphic volume form $\omega^\infty$. 
On the locus of $X^\infty$ where $\partial_{y'} W^\infty = 2y' \neq 0$ this form is given by 
\begin{align}
    \holVol^\infty & = \tfrac{1}{2\pi i} \tfrac{1}{2y'} ~ dz' \wedge dv'_1 \wedge dv_2' \\
    & = -\tfrac{1}{2 \pi i} \tfrac{z^2}{2y} \tfrac{1}{z^2} ~ dz \wedge dv_1 \wedge dv_2 \\
    & = - \holVol^0.
\end{align}
Here we used the gluing \eqref{eq:gluing} in the second equation. 
Hence $\holVol^0$ and $-\holVol^\infty$ glue to a holomorphic volume form $\holVol$ on $X$. 

The compactificaton of type IIB string theory on $X$ gives rise to a $4d$ theory with $\mathcal{N}=2$ supersymmetry. The BPS states correspond to $D3$ branes wrapping special Lagrangian submanifolds of $X$. 
The theory and its objects give rise to a structure as above:
As moduli space $\mathcal{B}$ we take a subspace of the moduli of complex structures on $X$. 
More precisely, we take the space of polynomials \eqref{eq:poly} (an open subset of $\mathbb{C}^3$). 

Let $X_u$, $u\in \mathcal{B}$, be the corresponding deformation of $X$ and denote by $\holVol_u$ the holomorphic volume form on $X_u$ as constructed above. 
The charge lattice is specified by the integral middle dimensional compactly supported cohomology (or equivalently homology) 
\begin{equation}\label{GammaIso}
\hat\Gamma_u= H^3_c(X_u,\mathbbm{Z}) \cong H_3(X_u, \mathbbm{Z})
\end{equation}
with pairing 
\begin{equation}
\langle \gamma,\gamma' \rangle= \int_{X_u} \gamma \wedge \gamma' \,.
\end{equation}
We will see shortly that this pairing is degenerate. 
Then $((\hat\Gamma, \langle -, -\rangle), \omega)$ is a BPS-VHS over $\mathcal{B}$ where we consider $\omega$ as a section of $\hat{\Gamma}\otimes \mathcal{O}$ via integration. 
The central charge is then
\begin{equation}
\hat{Z}_\gamma(u)=\int_{X_u}  \gamma \wedge \holVol_u = \int_{\gamma} \holVol_u\, ,
\end{equation} 
for $\gamma\in \hat{\Gamma}_u$ and the holomorphic volume form $\holVol_u$ of the CY3 $X_u$. 
The mass function is given by 
\begin{equation}
M_u(\gamma) = \int_{\gamma} |\holVol_u|.
\end{equation}
The BPS inequality is then simply 
\begin{equation}
\int_{\gamma} |\holVol_u| \ge  |\int_{\gamma} \holVol_u|\,.
\end{equation}
BPS states correspond to $\gamma \in H^{3}_c(X_u,\mathbb{Z})$ whose Poincar\'e dual in $H_3(X_u, \mathbb{Z})$ is represented by a special Lagrangian $L_{\gamma}$, i.e. $\holVol_u|_{L_{\gamma}} =e^{i\theta} |\holVol_u |$ for some $\theta \in [0, 2\pi)$.

\subsection{Reduction to curves}
We next recall how the previous structures are encoded in a curve (also compare \cite[\S 3]{Marinowarwick}, \cite{Smith-Fukaya}). 
Let $X$ be a non-compact smooth Calabi--Yau threefold as constructed above. 
The loci $v_1=v_2=0$ and $v_1'=v_2'=0$ in $X^0$ and $X^\infty$ respectively glue to an elliptic curve $\Sigma$. 
It is the compactification of the affine curve 
\begin{equation}
    \{ (y,z)\in \mathbb{C}^2 ~|~ y^2 = f(z) \}. 
\end{equation}
It comes with the natural embedding $\Sigma \hookrightarrow X$ and the double branched covering $\sigma \colon \Sigma \to \mathbb{CP}^1$. 
The latter is just the restriction of $\pi \colon X \to \mathbb{CP}^1$. 

Using the description of the fibers of $\pi$ together with Lemma 3.13 and 3.14 of \cite{Smith-Fukaya}, we see that there is an isomorphism
\begin{equation}\label{IsoHomology}
    \psi\colon \hat\Gamma = H_3(X, \mathbb{Z}) \cong H_1(\Sigma^\circ, \Z)^-
\end{equation}
which preserves the natural pairings. 
Here $\Sigma^\circ = \Sigma - \sigma^{-1}(0)$ and the superindex $-$ stands for the anti-invariants with respect to the involution on $\Sigma^\circ$ induced by $y \mapsto -y$.
Since $f(0) \neq 0$ by assumption, we see that $\Sigma^\circ$ is an elliptic curve with two punctures so that $\hat\Gamma \cong \mathbb{Z}^4$.
Now let $\delta_1, \delta_2$ be cycles around each puncture and $\A$, $\AD$ be cycles which give free generators of $H_1(\Sigma, \Z)$ under $\iota_*\colon \hat \Gamma \rightarrow H_1(\Sigma, \Z)$ for the inclusion $\iota \colon \Sigma^\circ \hookrightarrow \Sigma$. 
Then $\delta_1 - \delta_2$, $\A$, $\AD$ are free generators of $\hat\Gamma$. 
Moreover, $\delta_1 - \delta_2$ is a generator for the radical $\Gamma_{fl}$ of the intersection form $\langle -, - \rangle$ on $\hat\Gamma$ so that 
\begin{equation}
    \Gamma = \hat\Gamma / \Gamma_{fl} \cong H_1(\Sigma, \mathbb{Z}). 
\end{equation}
To complete the reduction to the curve $\Sigma^\circ$, we need a more explicit form of the isomorphism $\psi$ (see \eqref{IsoHomology}).
Let $\gamma_A$, $\gamma_B$ branch cuts in $\mathbb{CP}^1$ giving the cycles $\A$, $\AD$ on $\Sigma$.
Moreover, let $\delta$ be a small loop around $0 \in \mathbb{CP}^1$ (encircling no zeros of $f$). 
Then it can be shown (see \cite[\S 3]{Smith-Fukaya}) that there are Lagrangian spheres $L_{\A}$ and $L_{\AD}$ in $X$ which fiber of $\gamma_{\A}$ and $\Gamma_{\AD}$ respectively. 
The fibers away from the zeros of $f$ are vanishing spheres in the corresponding fiber of $\pi\colon X \to \mathbb{CP}^1$ which shrink at the zeros of $f$.
Similarly, there is a Lagrangian cyclinder $L_\delta$ in $X$ fibering over $\delta$ in two-spheres.
Then we isomorphism in \eqref{IsoHomology} is given by
\begin{equation}
\psi\colon (L_\A, L_{\AD}, L_\delta) \mapsto (\A, \AD, \delta_1 - \delta_2)
\end{equation}
(in homology).
Using these explicit generators, a computation shows that
\begin{equation}
    \int_{L} \holVol = \int_{\psi(L)} \lambda\quad \forall L \in H_3(X, \Z)
\end{equation}
for the meromorphic differential $\lambda:= y\, dz$ on $\Sigma^\circ$ (after a suitable normalization of $\omega$).
It follows that the BPS-VHS associated to the above non-compact CY3s over $\mathcal{B}$ is isomorphic to the BPS-VHS (fiberwise) defined by
\begin{itemize}
    \item $(\hat\Gamma_u =  H_1(\Sigma_u^\circ, \Z)^-, \langle -, - \rangle)$ is $H_1(\Sigma^\circ_u, \Z)$ with its intersection product for $u\in \mathcal{B}$; 
    \item $\lambda_u = y \, dz = \sqrt{f_u(z)} \, dz$ is the natural meromorphic differential on the double cover $\sigma_u \colon \Sigma^\circ_u \to \mathbb{CP}^1-\{0\}$ considered as a section of $\hat{\Gamma}^\vee\otimes \mathcal{O}$ via integration.
\end{itemize}
Finally, we record the central charge for this BPS-VHS, namely 
\begin{equation}
\hat{Z}_\gamma(u) = \int_{\gamma} \lambda_u.
\end{equation}
If $\lambda_u$ is holomorphic at the punctures $\sigma_u^{-1}(0)$, then $\hat{Z}_{\gamma} = 0$ for all $\gamma \in \Gamma_{fl}$.
In particular, $\hat{Z}$ descends to $\Gamma = \hat\Gamma / \Gamma_{fl}$ in this case.

\subsubsection*{Examples} \label{subsec:examplecurves}
In this work we will consider the following examples of BPS structures with complex one-dimensional moduli spaces $\mathcal{B}$.
These are associated to certain physical theories. 
Their geometric realization, as outlined above, are given by the following curves. 
We give the compactified curves $\Sigma$, the open curve $\Sigma^\circ$ are then constructed similarly as before. 

\begin{enumerate}
    \item Argyres--Douglas $A_1$ theory, realized geometrically by the curve 
    \begin{equation}\label{AD1curve}
\Sigma_{A_1} := \{y^2= z^2 - 4 u \in \mathbbm{C}^2 \}\,, \quad u\in \mathcal{B}=\mathbbm{C}^*
\end{equation}

\item Argyres--Douglas $A_2$ realized by:
    \begin{equation}
\Sigma_{A_2}^{I} := \{y^2= 4  z^3 -3 \Lambda^2\,z + u \in \mathbbm{C}^2 \}\,,  \quad u\in \mathcal{B}=\mathbbm{P}^1\setminus \{\pm \Lambda^3,\infty\}
\end{equation}
   
    \item Argyres--Douglas $A_2$ realized by:
   \begin{equation}
\Sigma_{A_2}^{II} := \{y^2= (z-\Lambda^2)(z+\Lambda^2)(z-u) \in \mathbbm{C}^2 \}\,, \quad u\in \mathcal{B}=\mathbbm{P}^1\setminus \{\pm \Lambda^2,\infty\}
\end{equation}

    \item Seiberg--Witten $SU(2)$ realized by:
    \begin{equation}
\Sigma_{SW} := \{y^2= \frac{\Lambda^2}{z^3} +\frac{2u}{z^2} + \frac{\Lambda^2}{z}   \in \mathbbm{C}\times \mathbb{C}^* \}\,, \quad u\in \mathcal{B}=\mathbbm{P}^1\setminus \{\pm \Lambda^2,\infty\}
\end{equation}
\end{enumerate}

As in the previous subsection, we consider in all cases the meromorphic differential
$$\lambda= y\, dz\,.$$


\subsection{Picard--Fuchs equations}

To determine the special geometry of BPS structures in the geometric realizations previously discussed, we will need to derive Picard--Fuchs equations governing the periods of the meromorphic differentials $\lambda$ obtained from the threefolds. We will therefore review the techniques due to Griffiths for finding relations in cohomology \cite{Griffiths}, we will follow the adaptation in \cite{LercheSmit}, see also \cite{CoxKatz} for more details. We will furthermore determine the relation between the meromorphic differentials which govern the BPS-VHS and the holomorphic differentials which lead to the ordinary VHS.

We review Griffiths' pole order reduction technique for determining the Picard-Fuchs equations of the holomorphic differentials on the curves. We therefore start by realizing the curves $\Sigma$ as hypersurfaces defined by the vanishing of a homogeneous polynomial $W$ of degree 3 in $\mathbb{CP}^2$. We will denote the homogeneous coordinates on $\mathbb{CP}^2$ by $x^A,A=1,\dots,3$, we will furthermore consider $W$ to be a function of the moduli $u$.  Since the first Chern class of $\Sigma$ in this realization vanishes, there is a globally defined, holomorphic one-form $\omega_0$ on $\Sigma$. This form can be represented by:

\begin{equation}
\omega_0 = \int_{\gamma} \frac{1}{W} \Xi\,, \quad \Xi= \sum_{A=1}^{3} (-1)^A x^A dx^1\wedge \dots \wedge \widehat{dx^A} \wedge dx^{3}\,, 
\end{equation}
where $\gamma$ is a small, one-dimensional curve winding around the hypersurface $\Sigma$. More generally, the integral
\begin{equation}
    \omega_{\alpha} = \int_{\gamma} \frac{p_{\alpha}}{W^{k+1}} \, \Xi \,,
\end{equation}
where $p_{\alpha}(x^A)$ is a homogeneous polynomial of degree $3k$, represents a rational differential one-form. $\omega_{\alpha}$ represents a non-trivial cohomology element in $H^{1}(\Sigma,\mathbb{C})$ if and only if $p_{\alpha}$ is a non-trivial element of the local ring $\mathcal{R}$ of $W$. We can thus find a basis for the cohomology $H^1$ by taking $\omega_0$, previously defined as well as $\omega_{1}:=\omega_{\alpha}$ with $p_{\alpha}$ homogeneous of degree 3.

For the derivation of Picard--Fuchs equations it is useful to consider the following integration by parts which results in a pole order reduction for forms. We therefore consider the following functions:

\begin{equation}
    \phi =\int_{\gamma} \frac{1}{W^l} \left( \sum_{B<C} (-1)^{B+C} \left(x^B Y_C (x^A)-x^C Y_B (x^A)\right)   dx^1 \wedge \dots \wedge \widehat{dx^B} \wedge \dots \wedge \widehat{dx^C} \wedge \dots \wedge dx^3\right)\,,
\end{equation}
where $Y_B(x^A)$ are homogeneous of degree $3l-2$. We obtain:

\begin{equation} \label{intbyparts}
    d\phi =\int_{\gamma} \frac{1}{W^{l+1}}\left( l \sum_{A=1}^{3} Y_A \frac{\partial W}{\partial x^a} - W \sum_{A=1}^{3} \frac{\partial Y_A}{\partial x^A} \right) \, \Xi \,.
\end{equation}
Since $d\phi$ is an exact form, its expression can be used for integration by parts. To derive the Picard--Fuchs differential equation one considers the following equation in cohomology:
\begin{equation}
    \nabla_{\frac{\partial }{\partial u}} \omega_{\alpha} = \left(\frac{\partial_u p_{\alpha}}{W^{k+1}} - (k+1) \frac{p_{\alpha \partial_u W}}{W^{k+2}} \right) \Xi \,,
\end{equation}
where $\nabla$ is the Gauss-Manin connection and then use partial integration until all elements can be expressed in terms of the local ring of $W$. In the following we apply this method for deriving the Picard-Fuchs equations to our examples

\subsubsection{Argyres--Douglas $A_2$, first realization}\label{AD2firstPF}

We proceed by studying the realization of Argyres--Douglas theory given by the following affine plane curve
\begin{equation}
\Sigma_{A_2}^I := \{y^2= 4  z^3 -3 \Lambda^2\,z + u \in \mathbbm{C}^2 \}\,,  \quad u\in \mathcal{B}=\mathbbm{P}^1\setminus \{\pm \Lambda^3,\infty\}\,.
\end{equation}

To obtain the corresponding projective curve we consider a homogeneous equation for $\Sigma_{I}$ in the projective plane $\mathbbm{P}^2$, we introduce a homogenizing coordinate $x$:
\begin{equation}
y^2 x= 4z^3 - 3 \Lambda^2 x^2 z + u x^3\,,
\end{equation}
i.e. the curve is given by the vanishing locus of 
$$ W = -y^2 x+ 4z^3 - 3 \Lambda^2 x^2 z + u x^3 $$
in $\mathbb{P}^2$ with homogeneous coordinates $x,y,z$.
 A holomorphic form $\omega_0$ on $\Sigma_I$ is given via the residue construction:
$$\omega_0= \textrm{Res} \frac{1}{W} \Xi $$
where 
$$\Xi=- x \,dy \,dz + y \,dx \,dz -z \,dx \,dy.$$
We set
$$ \omega_1:= \nabla_{\frac{\partial}{\partial u}} \omega_0= \textrm{Res}  -\frac{x^3}{W^2} \Xi$$
and the goal is now to determine the linear dependence of $\nabla_{\frac{\partial}{\partial u}} \omega_1$ in terms of $\omega_0$ and $\omega_1$. We use the pole order reduction method to achieve this.

We have
\begin{equation}
\nabla_{\frac{\partial}{\partial u}} \omega_1 = \textrm{Res} \frac{2x^6}{W^3} \xi\,.
\end{equation}
We now want to express  $2 x^6$ as a linear combination:
$$ 2 x^6 = p_1 \partial_x W + p_2  \partial_y W+ p_3 \partial_z W\,,$$
with $p_i\,, i=1,2,3$ homogeneous polynomials of degree 4 in $x,y,z$. We can then use the integration by parts \eqref{intbyparts}, in order to reduce the order of the pole, using:
$$ p_1 \partial_x W + p_2  \partial_y W+ p_3 \partial_z W = \frac{W}{2} \left( \partial_x p_1 +\partial_y p_2 +\partial_z p_3\right)\,,$$

We find the following:
 $$ p_1= \frac{2 x^3 (2 \Lambda^2 z-u x)}{3
   \left(\Lambda^6-u^2\right)}\,, \quad p_2=\frac{x^2 y (u x-2 \Lambda^2 z)}{3
   \left(\Lambda^6-u^2\right)}\,, \quad p_3= \frac{2 \Lambda^4 x^4}{3 \left(\Lambda^6-u^2\right)}\,,
 $$
 and obtain after integration by parts:
\begin{equation}
 (\Lambda^6-u^2)   \nabla_{\frac{\partial}{\partial u}} \omega_1= \frac{7}{6} u \, \omega_1 - \textrm{Res}\frac{10}{6} \frac{\Lambda^2 z x^2}{W^2} \Xi \,,
\end{equation}
for the second term on the r.~h.~s.~we repeat the pole order reduction and find:

$$ \Lambda^2 zx^2 = \frac{u}{2} x^3 + \left(a_1 \partial_x W + a_2 \partial_y W + a_3 \partial_z W\right) \,,$$
with $a_1= -\frac{x}{6}\,,$ $a_2=\frac{y}{12}\,,$ $a_3=0\,.$ After integrating by parts, we obtain

$$ \frac{\Lambda^2 zx^2}{W^2} = \frac{u}{2} \frac{x^3}{W^2} +\frac{1}{W} (\partial_x a_1 + \partial_y a_2 + \partial_z a_3)= \frac{u}{2} \frac{x^3}{W^2} -\frac{1}{12}\frac{1}{W} \,.$$
Putting everything together, we obtain the following relation in cohomology
\begin{equation}
    (\Lambda^6-u^2) (\nabla_{\frac{\partial}{\partial u}})^2 \omega_0 - 2u \nabla_{\frac{\partial}{\partial u}} \omega_0 -\frac{5}{36} \omega_0 =0 \,,
\end{equation}
which becomes the following Picard--Fuchs operator:
\begin{equation}
    \mathcal{L}_h= (\Lambda^6-u^2) \partial^2_u - 2u \partial_u -\frac{5}{36}\,,
\end{equation}
annihilating the periods of $\omega_0$.

To obtain the Picard-Fuchs operator for the periods of the meromorphic differential:
$$ \lambda = y dz\,,$$
we note the relation in cohomology:
\begin{equation}\label{A2relholmero}
\nabla_{\frac{\partial}{\partial u}} \lambda = \omega_0\,,
\end{equation}
and hence a third order operator which annihilates the periods of $\lambda$ is given by:
\begin{equation}
    \mathcal{L}'_m= \left((\Lambda^6-u^2) \partial^2_u - 2u \partial_u -\frac{5}{36}\right)\circ  \partial_u\,,
\end{equation}
this operator can be furthermore factorized as:
$$ \partial_u \circ \mathcal{L}_m = \partial_u \circ \left((\Lambda^6-u^2) \partial^2_u -\frac{5}{36}\right)\,, $$
the latter allows us furthermore to deduce the following relation in cohomology:
\begin{equation}\label{relmeroholA2}
\lambda = \frac{36}{5} (\Lambda^6-u^2) \nabla_{\frac{\partial}{\partial u}} \omega_0\,.
\end{equation}

\subsubsection{Argyres--Douglas $A_2$, second realization}

We proceed the discussion of the examples by the Argyres-Douglas $A_2$ theory, realized by the affine plane curve:
  \begin{equation}
\Sigma_{A_2}^{II} := \{y^2= (z-\Lambda^2)(z+\Lambda^2)(z-u) \in \mathbbm{C}^2 \}\,, \quad u\in \mathcal{B}=\mathbbm{P}^1\setminus \{\pm \Lambda^2,\infty\}
\end{equation} 
We consider the corresponding projective curve by considering the curve as the vanishing locus of:
 
$$ W = -y^2 x+ (z^2- \Lambda^4 x^2)(z-u\, x)\,, $$
in $\mathbb{P}^2$. The holomorphic form $\omega_0$ is given by:
$$\omega_0= \textrm{Res} \frac{1}{W} \Xi \,,$$
and we set:
$$ \omega_1:= \nabla_{\frac{\partial}{\partial u}} \omega_0 = \textrm{Res}\frac{(xz^2-x^3\Lambda^4)}{W^2} \Xi \,,$$
obtaining:
\begin{equation}
\nabla_{\frac{\partial}{\partial u}} \omega_1 = \frac{2 \left(\Lambda^4 x^3-x z^2\right)^2}{W^3}
\end{equation}
We express:
$$ 2 \left(\Lambda^4 x^3-x z^2\right)^2 = p_1 \partial_x W + p_2  \partial_y W+ p_3 \partial_z W\,,$$
with $p_i\,, i=1,2,3$ homogeneous polynomials of degree 4 in $x,y,z$. We can then use the integration by parts \eqref{intbyparts}, in order to reduce the order of the pole, using:
$$ p_1 \partial_x W + p_2  \partial_y W+ p_3 \partial_z W = \frac{W}{2} \left( \partial_x p_1 +\partial_y p_2 +\partial_z p_3\right)\,,$$

We find the following:
 $$ p_1= -\frac{4 x^3 z}{3}\,, \quad p_2=\frac{2}{3} x^2 y z\,, \quad p_3= \frac{2 x^2 z^2}{3}-2 \Lambda^4 x^4\,,
 $$
 and obtain after integration by parts:
\begin{equation}
    \nabla_{\frac{\partial}{\partial u}} \omega_1= \textrm{Res}-\frac{x^2 z}{W^2}  \Xi\,.
\end{equation}

We repeat the pole order reduction and find:

$$ -x^2 z = -\frac{2u}{(u^2-\Lambda^4)} (xz^2-\Lambda^4 x^3) + \left(a_1 \partial_x W + a_2 \partial_y W + a_3 \partial_z W\right) \,,$$
with $a_1= \frac{x}{2 \left(\Lambda^4-u^2\right)}\,,$ $a_2=-\frac{y}{4 \left(\Lambda^4-u^2\right)}\,,$ $a_3=\frac{u x}{2 \left(u^2-\Lambda^4\right)}\,.$ After integrating by parts, we obtain

$$ \frac{ -x^2 z}{W^2} = -\frac{2u}{(u^2-\Lambda^4)} \frac{xz^2-\Lambda^4 x^3)}{W^2} +\frac{1}{W} (\partial_x a_1 + \partial_y a_2 + \partial_z a_3)=  -\frac{2u}{(u^2-\Lambda^4)} \frac{xz^2-\Lambda^4 x^3)}{W^2} -\frac{1}{4(u^2-\Lambda^4)}\frac{1}{W} \,.$$
Putting everything together, we obtain the following relation in cohomology
\begin{equation}
    (u^2-\Lambda^4) (\nabla_{\frac{\partial}{\partial u}})^2 \omega_0 +2u \nabla_{\frac{\partial}{\partial u}} \omega_0 +\frac{1}{4} \omega_0 =0 \,,
\end{equation}
which becomes the following Picard--Fuchs operator:
\begin{equation}\label{holPF}
    \mathcal{L}_h= (\Lambda^4-u^2) \partial^2_u - 2u \partial_u -\frac{1}{4}\,,
\end{equation}
annihilating the periods of $\omega_0$.

We now want to obtain the Picard-Fuchs operator for the periods of the meromorphic differential:
$$ \lambda = y dz\,.$$

We note that the simple relation obtained in the first realization \ref{A2relholmero} does not hold in this case. To obtain a differential operator annihilating the periods of the meromorphic differential we proceed by rederiving a Picard-Fuchs operator, adapting techniques reviewed e.~g.~ in \cite{Carlsonbook}. We work with the affine plane curve:
$$y^2= (z^2-\Lambda^4)(z-u)\,,$$
we have
\begin{eqnarray}
\nabla_{\frac{\partial}{\partial u}} \lambda = -\frac{1}{2}  (z^2-\Lambda^4)^{1/2} (z-u)^{-1/2} dz\,, \\
(\nabla_{\frac{\partial}{\partial u}})^2  \lambda = -\frac{1}{4} (z^2-\Lambda^4)^{1/2} (z-u)^{-3/2} dz \,,
\end{eqnarray}
which implies the relations:
\begin{equation}\label{cohrelations}
(z-u) \nabla_{\frac{\partial}{\partial u}}  \lambda =-\frac{1}{2} \lambda \,,\quad (z-u)(\nabla_{\frac{\partial}{\partial u}} )^2 \lambda = \frac{1}{2} \nabla_{\frac{\partial}{\partial u}}  \lambda\,,
\end{equation}

We now consider $$f= (z^2-\Lambda^4)^{3/2} (z-u)^{-1/2}\,,$$
which gives:
\begin{equation}
\begin{aligned}
df&= 3(z^2-\Lambda^4)^{1/2} (z-u)^{-1/2} dz -\frac{1}{2}(z-u)^{-3/2} (z^2-\Lambda^4)^{3/2} dz \,\\
&= -6 z \nabla_u \lambda+  2 (z^2-\Lambda^4) (\nabla_u)^2 \lambda \, \\
&=(-6 (z-u) -6u )\nabla_u \lambda+  2 ((z-u)^2+2(z-u)+u^2-\Lambda^4) (\nabla_u)^2 \lambda \, \\
&=(u^2-\Lambda^4) (\nabla_{\frac{\partial}{\partial u}})^2  \lambda - 2u \nabla_{\frac{\partial}{\partial u}} \lambda +\frac{5}{4} \lambda \,.
\end{aligned}
\end{equation}
Since $df$ is trivial in cohomology, this gives the desired relation between the meromorphic differential and its derivatives using the Gauss-Manin connection. For the periods of $\lambda$ this gives the following Picard--Fuchs equation:

\begin{equation}\label{meroPF}
    \mathcal{L}_m= \left((u^2-\Lambda^4) \partial^2_u - 2u \partial_u +\frac{5}{4}\right)\,.
\end{equation}

Comparing \ref{holPF} and \ref{meroPF}  we can express the meromorphic differential $\lambda$ in terms of $\omega_0,\nabla_{\frac{\partial}{\partial u}} \omega_0$ and we can also express $\omega_0$ in terms of $\lambda$ and $\nabla_{\frac{\partial}{\partial u}} \lambda$. We find

\begin{align}
    \omega_0 &= \frac{2 \left(u^2+3\right) \nabla_{\frac{\partial}{\partial u}} \lambda-5 u \,\lambda}{4 \left(u^2-1\right)}\, ,  \\
    \lambda &= \frac{4}{15} \left(u^2-1\right) \left(2 \left(u^2+3\right) \nabla_{\frac{\partial}{\partial u}}\omega_0+u\, \omega_0\right)\, , 
\end{align}

which gives in particular

\begin{equation}\label{lambdainomega0AD2nd}
    \nabla_{\frac{\partial}{\partial u}} \lambda =  \frac{2}{3} \left(u^2-1\right) \left(2 u \nabla_{\frac{\partial}{\partial u}} \omega_0+\omega_0\right)\,.
\end{equation}

\subsubsection{Seiberg--Witten $SU(2)$}\label{SWPF}
In the following we revisit the derivation of the Picard-Fuchs equation for the Seiberg-Witten $SU(2)$ theory which is well known, see e.~g.~\cite{Lerche:1996xu} and references therein. The slight difference in our derivation is that it starts from the representation of the Seiberg-Witten curve which is more commonly used in the study of BPS structures, following the works of GMN \cite{GMN1,GMN2}, see \cite{Tachikawareview} for a review of this presentation of the curve. In the following we repeat the previous steps for the derivation of the Picard-Fuchs equation of the $A_2$ theory. We thus introduce a homogenizing coordinate $x$ and write the curve as:
\begin{equation}
y^2 z= \Lambda^2 x^3 + 2 u z x^2 + \Lambda^2 z^2 x
\end{equation}
i.e. the curve is given by the vanishing locus of 
$$ W = - y^2 z +\Lambda^2 x^3 + 2 u z x^2 + \Lambda^2 z^2 x$$
in $\mathbb{P}^2$ with homogeneous coordinates $x,y,z$.
 A holomorphic form $\omega_0$ on $\Sigma_{SW}$ is given via the residue construction:
$$\omega_0= \textrm{Res} \frac{1}{W} \Xi $$
where 
$$\Xi=- x \,dy \,dz + y \,dx \,dz -z \,dx \,dy.$$
We set
$$ \omega_1:= \nabla_{\frac{\partial}{\partial u}} \omega_0= \textrm{Res} -\frac{2zx^2}{W^2} \Xi$$
and the goal is now to determine the linear dependencies of $\nabla_{\frac{\partial}{\partial u}} \omega_1$ in terms of $\omega_0$ and $\omega_1$. We use the pole order reduction method to this end.

We have
\begin{equation}
\nabla_{\frac{\partial}{\partial u}} \omega_1 = \frac{8x^4 z^2}{W^3}
\end{equation}
We now want to express  $8x^4 z^2$ as a linear combination:
$$ 8x^4 z^2 = p_1 \partial_x W + p_2  \partial_y W+ p_3 \partial_z W\,,$$
with $p_i\,, i=1,2,3$ homogeneous polynomials of degree 4 in $x,y,z$. We find the following:
 $$ p_1= \frac{8 u x^3 z}{3 \left(u^2-\Lambda^4\right)}\,, \quad p_2=-\frac{2 x^2 y (3 \Lambda^2 x+u z)}{3
   \left(\Lambda^4-u^2\right)}\,, \quad p_3= \frac{4 x^2 z (3 \Lambda^2 x+u z)}{3
   \left(\Lambda^4-u^2\right)}\,.
 $$
Now we compute:
$$ \frac{1}{2} (\Lambda^4-u^2) \left( \partial_x p_1 + \partial_y p_2 + \partial_z p_3\right)= \Lambda^2 x^3-3u z x^2 \,, $$
and hence after integration by parts we have:
\begin{equation}
 (\Lambda^4-u^2)   \nabla_{\frac{\partial}{\partial u}} \omega_1= \frac{1}{W^2} \left(\Lambda^2 x^3-3u z x^2 \right)\,,
\end{equation}
we can furthermore express the R.~H.~S as:
$$ \Lambda^2 x^3-3u z x^2= 2u W^2\, \omega_1 + \left(a_1 \partial_x W + a_2 \partial_y W + a_3 \partial_z W\right) \,,$$
with $a_1= \frac{x}{3}\,,$ $a_2=\frac{y}{12}\,,$ $a_3=-\frac{z}{6}\,.$ After integrating by parts and putting everything together we obtain the Picard--Fuchs operator:
\begin{equation}
    \mathcal{L}_h= (\Lambda^4-u^2) \partial^2_u - 2u \partial_u -\frac{1}{4}\,.
\end{equation}

To obtain the Picard-Fuchs operator for the periods of the meromorphic differential:
$$ \lambda = y dz\,,$$
we note the relation:
$$ \nabla_{\partial_u} \lambda = \omega_0\,,$$
and hence a third order operator which annihilates the periods of $\lambda$ is given by:
\begin{equation}
    \mathcal{L}'_m= \left((\Lambda^4-u^2) \partial^2_u - 2u \partial_u -\frac{1}{4}\right)\circ  \partial_u\,,
\end{equation}
this operator can be furthermore factorized as:
$$ \partial_u \circ \mathcal{L}_m = \partial_u \circ \left((\Lambda^4-u^2) \partial^2_u -\frac{1}{4}\right)\,, $$
the latter allows us furthermore to deduce the following relation in cohomology:
$$ \lambda = 4 (\Lambda^4-u^2) \nabla_{\frac{\partial}{\partial u}} \omega_0\,.$$

\section{Quasi-modularity of BPS structures} \label{sec:modularity}
In this section we discuss the quasi-modular structure of the central charges of the BPS structures which we consider in this paper. Expressions in terms of quasi-modular forms for the periods of SW theory have been obtained before, see e.~g.~\cite{Klemm:1994qs,Klemm:1995wp,Lerche:1996xu,Huang:2006si}, for $A_2$ Argyres-Douglas theory the modular structure of the periods was considered in \cite{Codesido:2017dns} by mapping the curve to a SW form, the expressions thus obtained are for a different duality group than the ones considered in our work.

\subsection{Quasi modular form}\label{subsec:quasimodular}

We start by summarizing some basic concepts about modular forms and quasi modular forms, following the exposition of \cite{ASYZ}, and we refer to \cite{diamond2005modular,Zagier} and the references therein for more details on the basic theory.

\subsubsection{Modular groups and modular curves}
The generators for the group $SL(2,\mathbb{Z})$ are given by:
\begin{equation}
  \label{eq:monodromies}
T =
  \begin{pmatrix}
    1 & 1\\ 0 & 1
  \end{pmatrix}\,,\quad
  S=
  \begin{pmatrix}
    0& -1 \\ 1 & 0\\
  \end{pmatrix}\,,\quad S^{2}=-I\,,\quad (ST)^{3}=-I\,.
\end{equation}
We consider the genus zero congruence subgroups called Hecke subgroups of $\Gamma(1)=PSL(2,\mathbb{Z})=SL(2,\mathbb{Z})
/\{\pm I\}$ given by:
\begin{equation}
\Gamma_{0}(N)=\left\{
\left.
\begin{pmatrix}
a & b  \\
c & d
\end{pmatrix}
\right\vert\, c\equiv 0\,~ \mathrm{mod} \,~ N\right\}< \Gamma(1)
\end{equation}
with $N=2,3,4$. A further subgroup which will be important in the sequel is the unique normal subgroup in $\Gamma(1)$ of index 2 which is often
denoted $\Gamma_0(1)^*$. We write $N=1^*$ when
listing it together with the groups $\Gamma_0(N)$.

The group $SL(2,\mathbb{Z})$ acts on the upper half plane $\mathcal{H}
= \{ \tau \in \mathbb{C} |\, \mathrm{Im} \tau > 0 \}$ by fractional linear
transformations:
$$\tau \mapsto \gamma\tau=\frac{a\tau+b}{c\tau+d}\quad \mbox{for} \quad
\gamma=\begin{pmatrix} a&b\\c&d\end{pmatrix} \in
SL(2,\mathbb{Z})\,.$$ The quotient space $Y_{0}(N)=
\Gamma_{0}(N)\backslash \mathcal{H}$ is a non-compact orbifold with
certain punctures corresponding to the cusps and orbifold points
corresponding to the elliptic points of the group $\Gamma_{0}(N)$.
By filling the punctures, one then gets a compact orbifold
$X_{0}(N)=\overline{Y_{0}(N)}=\Gamma_{0}(N)\backslash
\mathcal{H}^{*}$ where $\mathcal{H}^* = \mathcal{H} \cup \{i\infty\}
\cup \mathbb{Q}$. The orbifold $X_0(N)$ can be equipped with the
structure of a Riemann surface. The space $X_{0}(N)$ is called a modular curve and is the moduli space
of pairs $(E,C)$, where $E$ is an elliptic curve and $C$ is a cyclic
subgroup of order $N$ of the torsion subgroup
$E_{N} \cong \mathbb{Z}_{N}^{2}$. It classifies each cyclic $N$-isogeny $\phi: E\rightarrow E/C$ up to isomorphism, see for example ~\cite{diamond2005modular,Husemoller} for more details. In the following, we will denote by $\Gamma$ a general subgroup of finite index in $\Gamma(1)$.

\begin{figure}[h!]
  \centering
  \subfloat[ $\Gamma_0(1)^*$]
  {\includegraphics[width=0.3\textwidth]{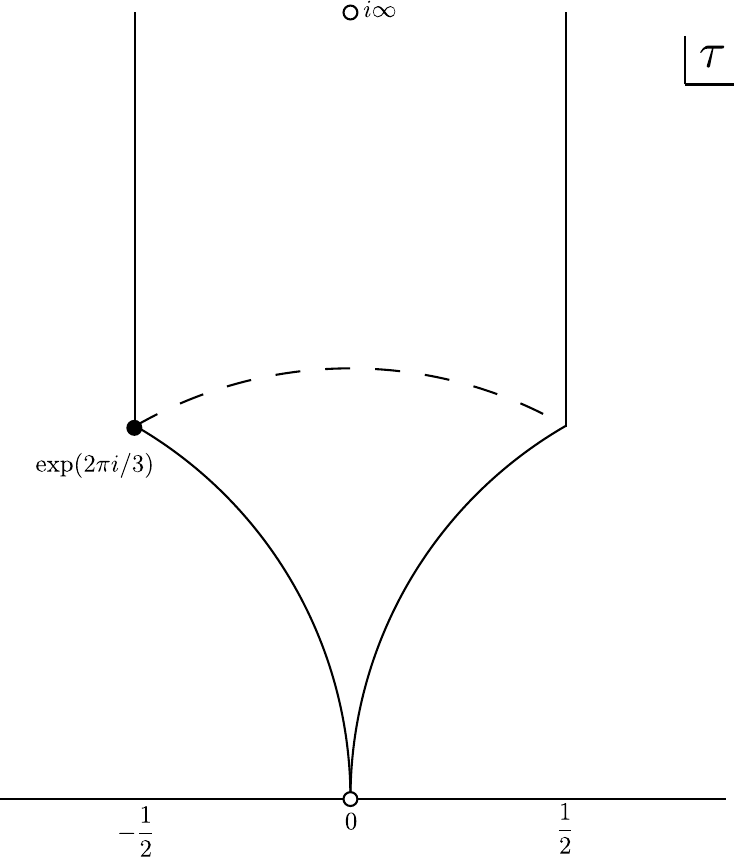}}
  \subfloat[$\Gamma_0(2)$]{\includegraphics[width=0.3\textwidth]{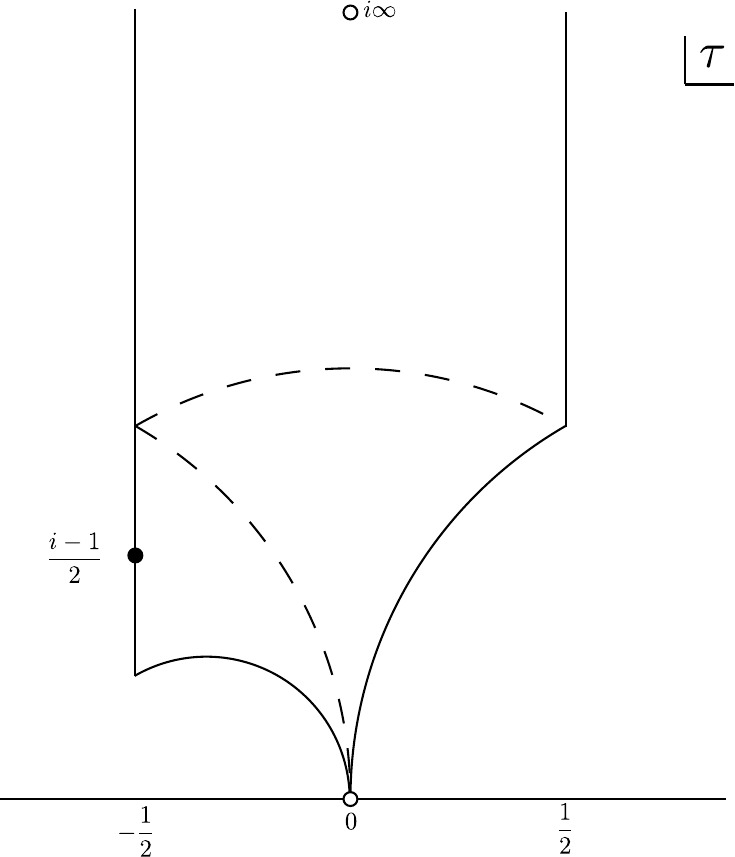}}
  \hfill

  \subfloat[$\Gamma_0(3)$]{\includegraphics[width=0.3\textwidth]{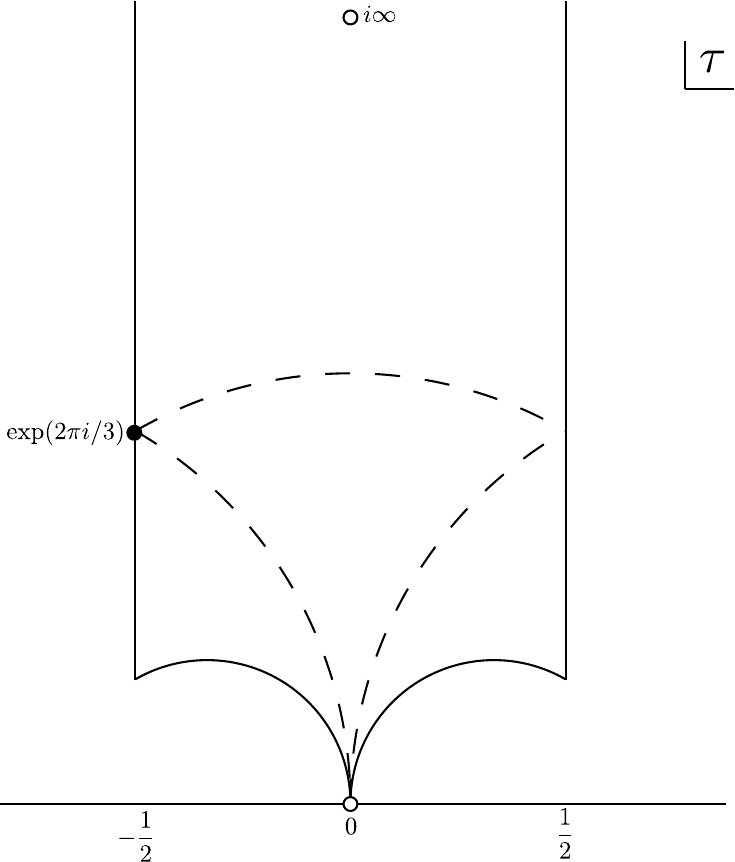}}
  \subfloat[$\Gamma_0(4)$]{\includegraphics[width=0.3\textwidth]{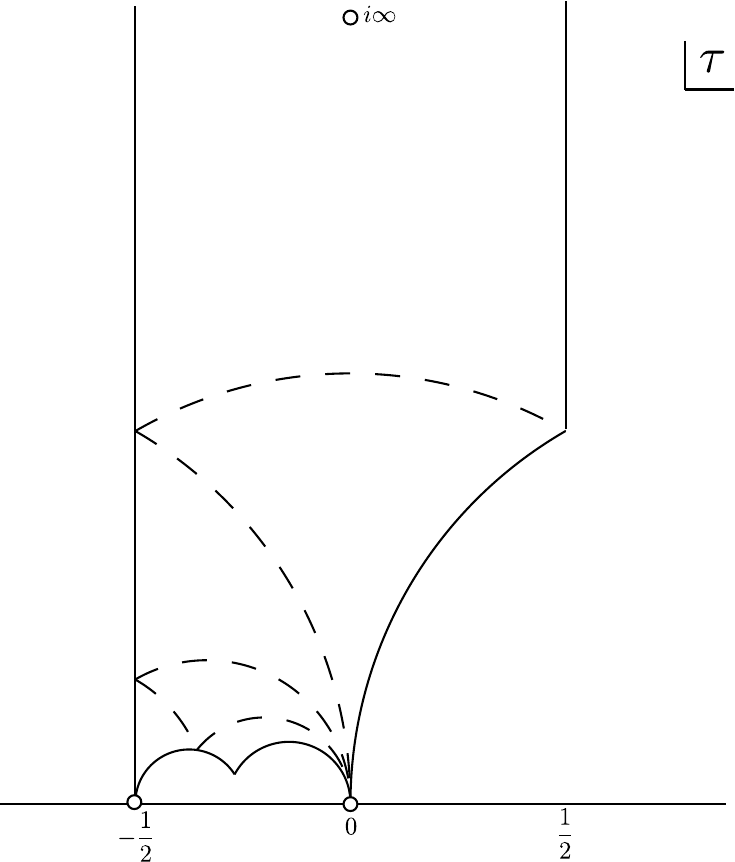}}
\caption{Fundamental domains for $\Gamma_0(1)^*$,
$\Gamma_0(N)$, $N=2,3,4$. \\The empty and full circles stand for cusps and elliptic points respectively.} \label{funddomain}
\end{figure}

The fundamental domains for these groups are depicted in Figure~\ref{funddomain}.

\subsubsection{Modular functions}
\renewcommand{\labelenumi}{(\roman{enumi})}
A (meromorphic) modular function with respect to the a subgroup $\Gamma$ of finite index in $\Gamma(1)$ is a meromorphic function $f : X_{\Gamma}\rightarrow \mathbb{P}^{1}$.
Consider the restriction of $f$ to $Y_{\Gamma}=\Gamma\backslash \mathcal{H}$.
Since the restriction is meromorphic, we know $f$ can be lifted to
a function $f$ on $\mathcal{H}$. Then one gets a function $f: \mathcal{H}\rightarrow \mathbb{P}^{1}$
such that
\begin{enumerate}
\item $\,f(\gamma \tau)=f(\tau), \quad\forall \gamma\in \Gamma\,$.
\item $\,f$ is meromorphic on $\mathcal{H}$.
\item $\,f$ is ``meromorphic at the cusps" in the sense that the function
\begin{equation}
f|_{\gamma}: \tau\mapsto f(\gamma\tau)
\end{equation}
is meromorphic at $\tau=i\infty$ for any $\gamma\in \Gamma(1)$.
\end{enumerate}
The third condition requires more explanation. For any cusp class
$[\sigma] \in \mathcal{H}^*/\Gamma$\footnote{We use the notation
$[\tau]$ to denote the equivalence class of $\tau\in
\mathcal{H}^{*}$ under the group action of $\Gamma$ on
$\mathcal{H}^{*}$.} with respect to the modular group $\Gamma$, one
chooses a representative $\sigma\in \mathbb{Q} \cup \{i\infty\}$.
Then it is easy to see that one can find an element $\gamma\in
\Gamma(1)$ so that $\gamma: i\infty\mapsto \sigma$. Then this
condition means that the function defined by $\tau\mapsto f\circ
\gamma\,(\tau)$ is meromorphic near $\tau=i\infty$ and that the
function $f$ is declared to be ``meromorphic at the cusp $\sigma$"
if this condition is satisfied.

Therefore, equivalently, a (meromorphic) modular function
with respect to the modular group is a meromorphic function $f : \mathcal{H}\rightarrow \mathbb{P}^{1}$ satisfying the above properties on modularity, meromorphicity, and growth condition at the cusps.

\subsubsection{Modular forms}
Similarly, we can define a (meromorphic) modular form of weight $k$ with respect to the group $\Gamma$ to be a (meromorphic) function $f : \mathcal{H}\rightarrow \mathbb{P}^{1}$ satisfying the following conditions:
\begin{enumerate}
\item $\,f(\gamma \tau)=j_{\gamma}(\tau)^{k}f(\tau), \quad \forall \gamma\in \Gamma\,$, where $j$ is called the automorphy factor defined by

$$j: \Gamma \times \mathcal{H}\rightarrow
\mathbb{C},\quad \left(\gamma=\left(
\begin{array}{cc}
a & b  \\
c & d
\end{array}
\right),\tau\right)\mapsto j_{\gamma}(\tau):=(c\tau+d)\,.$$
\item $\,f$ is meromorphic on $\mathcal{H}$.
\item $\,f$ is ``meromorphic at the cusps" in the sense that the function
\begin{equation}
\label{slash}
f|_{\gamma}: \tau\mapsto j_{\gamma}(\tau)^{-k} f(\gamma\tau)
\end{equation}
is meromorphic at $\tau=i\infty$ for any $\gamma\in \Gamma(1)$.
\end{enumerate}

\subsubsection{Quasi modular forms}\label{dfnofquasiform}
A (meromorphic) quasi modular form of weight $k$ with respect to the group $\Gamma$ is a (meromorphic) function $f : \mathcal{H}\rightarrow \mathbb{P}^{1}$ satisfying the following conditions:
\begin{enumerate}
\item $\,$ There exist meromorphic functions $f_{i}, i=0,1,2,3,\dots, k-1$ such that
\begin{equation}
f( \gamma\tau)=j_{\gamma}(\tau)^{k}f(\tau)+\sum_{i=0}^{k-1} c^{k-i}\,j_{\gamma}(\tau)^i  f_{i}(\tau)\,,\quad \forall  \gamma=\left(
\begin{array}{cc}
a & b  \\
c & d
\end{array}
\right)\in \Gamma\,.
\end{equation}
\item $\,f$ is meromorphic on $\mathcal{H}$.
\item $\,f$ is ``meromorphic at the cusps" in the sense that the function
\begin{equation}
f|_{\gamma}: \tau\mapsto j_{\gamma}(\tau)^{-k} f(\gamma\tau)
\end{equation}
is meromorphic at $\tau=i\infty$ for any $\gamma\in \Gamma(1)$.
\end{enumerate}

We proceed by introducing the modular forms which are used in our paper, starting with the Jacobi theta functions with characteristics $(a,b)$ defined by:
\begin{equation}
\vartheta \left[\!\!\! \begin{array}{c}a\\ b\end{array}\!\!\!\right](z,\tau)=\sum_{n\in \mathbbm{Z}}  q^{{\frac{1}{2}}(n+a)^2} e^{2\pi i (n+a)(z+b)} \,.
\end{equation}

for special $(a,b)$ these are denoted by:

\begin{align}
\theta_1(z,\tau)&=\vartheta \left[\!\!\! \begin{array}{c}1/2\\ 1/2\end{array}\!\!\!\right](u,\tau)=\sum_{n\in \mathbbm{Z}+{\frac{1}{2}}}  (-1)^{n}q^{{\frac{1}{2}}n^2}e^{2\pi i n z}\,,\\
\theta_2(z,\tau)&=\vartheta \left[\!\!\! \begin{array}{c}1/2\\ 0\end{array}\!\!\!\right](u,\tau)=\sum_{n\in \mathbbm{Z}+{\frac{1}{2}}}  q^{{\frac{1}{2}}n^2}e^{2\pi i n z}\,,\\
\theta_3(z,\tau)&=\vartheta \left[\!\!\! \begin{array}{c}\,\,\,0\,\,\,\\ \,\,\,0\,\,\,\end{array}\!\!\!\right](u,\tau)=\sum_{n\in \mathbbm{Z}}  q^{{\frac{1}{2}}n^2}e^{2\pi i n z}\,,\\
\theta_4(z,\tau)&=\vartheta \left[\!\!\! \begin{array}{c}0\\ 1/2\end{array}\!\!\!\right](u,\tau)=\sum_{n\in \mathbbm{Z}} (-1)^n q^{{\frac{1}{2}}n^2}e^{2\pi in z}\,.
\end{align}

We further define the following $\theta$--constants:
\begin{equation}
\theta_{2}(\tau)=\theta_2(0,\tau),\quad \theta_{3}(\tau)=\theta_3(0,\tau),\quad \theta_{4}(\tau)=\theta_2(0,\tau)\,.
\end{equation}
The $\eta$--function is defined by
\begin{equation}
\eta(\tau)=q^{\frac{1}{24}}\prod_{n=1}^\infty(1-q^n)\,.
\end{equation}
It transforms according to
\begin{equation}\label{etatrafo}
\eta(\tau+1)=e^{\frac{i\pi}{12}}\eta(\tau),\qquad \eta\left(-\frac{1}{\tau}\right)=\sqrt{\frac{\tau}{i}}\,\eta(\tau)\,.
\end{equation}
The Eisenstein series are defined by
\begin{equation}\label{eisensteinseries}
E_k(\tau)=1-\frac{2k}{B_k}\sum_{n=1}^\infty\frac{n^{k-1}q^n}{1-q^n},
\end{equation}
where $B_k$ denotes the $k$-th Bernoulli number. $E_k$ is a modular form of weight $k$ for $k>2$ and even. The discriminant form and the $j$
invariant are given by
\begin{align}
   \Delta(\tau) &= \frac{1}{1728}\left({E_4}(\tau)^3-{E_6}(\tau)^2\right) = \eta(\tau)^{24},\\
    j(\tau)& = 1728\frac{E_{4}(\tau)^{3}}{ E_{4}(\tau)^3-{E_6}(\tau)^2}\,.
\end{align}

For the subgroups $\Gamma_0(N)$ we introduce three modular forms $A,B,C$ of weight 1, which are given by:
\begin{equation}
    \hspace{2.5em} \begin{array}{c|ccc}\renewcommand{\arraystretch}{0.5}
        N&A&B&C\\[.2ex]
1^{*}&E_{4}(\tau)^{\frac{1}{4}}&(\frac{E_{4}(\tau)^{\frac{3}{2}}+E_{6}(\tau)}{2})^{\frac{1}{6}}&(\frac{E_{4}(\tau)^{\frac{3}{2}}-E_{6}(\tau)}{2})^{\frac{1}{6}}\\[.5ex]
2&\frac{(2^{6}\eta(2\tau)^{24}+\eta(\tau)^{24} )^{\frac{1}{4}}}{
\eta(\tau)^2\eta(2\tau)^2}&\frac{\eta(\tau)^{4}}{\eta(2\tau)^{2}}&2^{\frac{3}{
2}}\frac{\eta(2\tau)^4}{\eta(\tau)^2}\\[1ex]
3&\frac{(3^{3}\eta(3\tau)^{12}+\eta(\tau)^{12} )^{\frac{1}{3}}}{\eta(\tau)\eta(3\tau)}&\frac{\eta(\tau)^{3}}{\eta(3\tau)}&3\frac{\eta(3\tau)^3}{\eta(\tau)}\\[1ex]
4&\frac{(2^{4}\eta(4\tau)^{8}+\eta(\tau)^{8} )^{\frac{1}{2}}}{\eta(2\tau)^2}=
\frac{\eta(2\tau)^{10}}{\eta(\tau)^{4}\eta(4\tau)^{4}}&\frac{\eta(\tau)^{4}}{ \eta(2\tau)^2}&2^2\frac{\eta(4\tau)^4}{ \eta(2\tau)^2}
\end{array}
\end{equation}

These satisfy by definition
\begin{equation}
A^{r}=B^{r}+C^{r}\,.
\end{equation}

with the following values of $r$:
\begin{equation*}
 \begin{array}{c|ccccc}
N&1^{*}&2&3&4&\\
r&6&4&3&2
\end{array}
\end{equation*}
We introduce the analog of the Eisenstein series $E_{2}$ as a quasi-modular form as follows:
\begin{eqnarray}
E=\partial_\tau \log
B^{r}C^{r}\,.
\end{eqnarray}

\subsubsection{Quasi modular forms and Picard-Fuchs equations}\label{quasiPF}
In the following we want to recall the association of the quasi-modular forms for the $\Gamma_0(N)$ subgroups to hypergeometric differential equations which will appear later as the Picard-Fuchs equations governing the central charges of our BPS structures, we will follow \cite{ASYZ} in the exposition which can be consulted for further references.

The relevant data giving the ring of the quasi
modular forms as well as the modular parameter $\tau$ are captured
by the periods $\omega_0$ and $\omega_1$ of the corresponding
families of elliptic curves. The periods satisfy the following Picard-Fuchs
differential equation:
\begin{equation} \label{deflc}
\mathcal{L}_c\,\omega_{i}= (\theta_{\alpha}^{2}-\alpha (\theta_{\alpha}+1/r)(\theta_{\alpha}+1-1/r))\, \omega_i=0\,,\quad i=0\,,1\,.
\end{equation}
The parameter $\alpha$ is the Hauptmodul, and $\theta_{\alpha}=\alpha{\partial \over \partial \alpha}$.
The solutions of this equation are given in terms of the hypergeometric functions:
\begin{equation}\label{hypersols}
\begin{aligned}
    \omega_0(\alpha) &= ~_{2}F_{1}\left(1/r,1-1/r,1;\alpha\right)\,,\\
    \omega_1(\alpha) &=\frac{i}{\sqrt{N}} ~_{2}F_{1}\left(1/r,1-1/r,1;1-\alpha\right)\,.
\end{aligned}
\end{equation}
The numbers $r$ are give by the following:
\begin{equation*}
 \begin{array}{c|ccccc}
N&1^{*}&2&3&4&\\
r&6&4&3&2
\end{array}
\end{equation*}
They are related to the index $\mu$ of $\Gamma_0(N)$ by $r=\frac{12}{\mu}$.
The modular parameter $\tau$ is then given by:
\begin{equation}
\tau=\frac{\omega_1}{\omega_0}\,.
\end{equation}
For the cases $N=1^{*},2,3,4$, the relevant modular forms are given by
\begin{equation}\label{dfnofABC}
A=\omega_0\,, \quad B= (1-\alpha)^{1\over r}\,A\,,  \quad C=\alpha^{1\over r}\,A\,.
\end{equation}
Then by definition one has
\begin{equation}
A^{r}=B^{r}+C^{r}\,.
\end{equation}
By analytic continuation, it is easy to show
that as multi-valued functions on the modular curve $X_{0}(N)$ as an
orbifold, these modular forms (for a multiplier system) have divisors given by

\begin{equation}\label{asympofABC}
\textrm{Div} A={1\over r} (\alpha=\infty)\,,\quad
\textrm{Div} B={1\over r} (\alpha=1)\,,\quad
\textrm{Div} C={1\over r} (\alpha=0)\,.
\end{equation}

The differential ring structure of the quasi modular forms for $N=1^{*},2,3,4$ is given by the relations:
 
\begin{equation}\label{quasiring}
\begin{aligned}
\partial_\tau A&={1\over 2r}A(E+{C^{r}-B^{r}\over A^{r-2}})\,,\\
\partial_\tau B&={1\over 2r}B(E-A^{2})\,,\\
\partial_\tau C&={1\over 2r}C(E+A^{2})\,,\\
\partial_\tau E&={1\over
2r}(E^{2}-A^{4})\,.
  \end{aligned}
\end{equation}


\subsubsection{Fricke involution}
For each of the modular curves $X_{0}(N)$ with $N=1^{*},2,3,4$,  as
a covering of the $j$--plane $\Gamma(1)\backslash\mathcal{H}^{*}$,
there are three branch points. There
are two distinguished cusps given by $[i\infty]=[1/N]$ and $[0]=
[1/1]$. The third branch point is a cubic elliptic point, quad\-ratic
elliptic point, cubic elliptic point and a cusp for $N=1^*,2,3,4$,
respectively. The Fricke involution is defined by
\begin{equation}
W_{N}: \tau\mapsto -{1\over N\tau}\,.
\end{equation}
It exchanges these two cusps and fixes the third branch point, see Fig.~\ref{funddomain}.\footnote{We point out that for the Seiberg-Witten curve family
given by $y^{2}=(x^{2}-u)^{2}-\Lambda^{4}$ and with monodromy group
$\Gamma_{0}(4)$, the Fricke involution acts as $2\tau\mapsto
-{1\over 2\tau}$. In the literature, see for example
~\cite{Huang:2006si}, by redefining $\tau$ as the above $2\tau$,
the Fricke involution is realized as the $S$-transformation. }

Recall that the modular curve $X_{0}(N)$ is the moduli space of enhanced
elliptic curves $(E,C)$, where $C$ is an order $N$ subgroup of the
torsion group $E_{N}\cong \mathbb{Z}_{N}\oplus \mathbb{Z}_{N}$. Using this
interpretation, the Fricke involution acts by sending $(E,C)$ to
$(E/C,E_{N}/C)$.

It turns out from e.g. ~\cite{Maier:2009} that the Fricke
involution maps the Hauptmodul
\begin{equation}
\alpha \quad  \textrm{to} \quad \beta:=1-\alpha\,.
\end{equation}
The Fricke involution acts on the ring of quasi modular forms according to
\begin{equation}
  \begin{aligned}
    \label{fricke}
    A &\mapsto \frac{\sqrt{N}}{i} \tau\, A\,, \\
    B&\mapsto \frac{\sqrt{N}}{i}\tau \,C\,,\\
    C&\mapsto\frac{\sqrt{N}}{i}\tau B\,,\\
    E&\mapsto  N\tau^{2}E+{12\over 2\pi i}{2N\tau\over N+1},\quad
    N=1^{*},2,3\,. \\
    E&\mapsto  N\tau^{2}E+{12\over 2\pi i}{2N\tau\over 6}\,,\;\;\quad
    N=4\,.
  \end{aligned}
\end{equation}
For all cases $N=1^{*},2,3,4$,
the non-holomorphic completion $\widehat{E}(\tau,\bar{\tau})$ transforms according to:
\begin{equation}
\widehat{E}\mapsto N\tau^{2}\widehat{E}\,.
\end{equation}

\subsection{Special geometry and modularity} \label{specialgeometryandmodularity}

\subsubsection{Argyres--Douglas $A_1$}

This theory is described by the curve:
\begin{equation}
\Sigma_{A_1} := \{y^2= z^2 - 4 u \in \mathbbm{C}^2 \}\,, \quad u\in \mathcal{M}=\mathbbm{C}^*
\end{equation}
which is given by a double cover of $\mathcal{C}=\mathbb{C}$, branched at $\pm 2\sqrt{u}$. The only compact cycle of $\Sigma$ is the lift of the cycle connecting the two  branch points in $\mathcal{C}$. We call this cycle $\gamma_1$. We can define a non-compact dual cycle $\gamma_2$, by introducing a parameter $\Lambda$ and consider the lift of the cycle in $\mathcal{C}$ connecting $2\sqrt{u}$ and $\Lambda$. The periods of the meromorphic differential $y\,dx$ can be obtained directly:

\begin{equation}
   a:=Z_{\gamma_1}(u)= \frac{1}{2\pi i}\int_{\gamma_1} y \, dx=  u\,,
\end{equation}
and the dual ($\Lambda$-)regularized period:
\begin{equation}
  2\pi i b:= 2\pi i Z_{\gamma_2}(u)= \int_{\gamma_2} y \, dx= u (\log u -1) + \frac{\Lambda^2}{2} -2 u \log \Lambda + \mathcal{O}(1/\Lambda^2)\,,
\end{equation}
these periods also define the special geometry of the Gaussian matrix model, see e.g. \cite{Marinowarwick}.

This gives the prepotential:
\begin{equation}
2\pi i F_0(a)=\frac{1}{2} a^2\log a -\frac{3}{4} a^2 -a^2 \log \Lambda + \frac{1}{2} a \Lambda^2  +  \mathcal{O}(1/\Lambda^2)\,.
    \end{equation}

The monodromy of the finite part of the periods around $u=0$ is given by:
\begin{equation}
    \left(\begin{array}{cc} a \\ b \end{array}\right) \rightarrow  \left(\begin{array}{cc} 1 &0 \\ 1&1 \end{array} \right) \left(\begin{array}{cc} a \\ b \end{array} \right)
\end{equation}

We furthermore introduce $\tau:= \frac{\partial b}{\partial a}= \frac{1}{2\pi i}\log u$, keeping only the finite parts. The special K\"ahler metric is now:
$$ g_{u\bar{u}}= 2 \textrm{Im} \tau\,.$$

The central charges $Z_{\gamma_{i}}(u)\, i=1,2$ for the other examples are found by solving the Picard-Fuchs equations for the periods of the corresponding meromorphic differential $\lambda$. From the basis of solutions we identify linear combinations which have integral monodromy and identify these with the central charges of a basis of BPS states. The bases are chosen such that these correspond to BPS states whose masses vanish at singular points in the moduli space and which correspond to stable BPS states everywhere in the moduli space. 

We will denote the monodromy around the singular point $u_*$ by $M_{u_*}$, i.e.:
\begin{equation}
    \left(\begin{array}{cc} Z_{\gamma_1}(u) \\ Z_{\gamma_2}(u) \end{array}\right) \rightarrow  M_{u_*} \left(\begin{array}{cc} Z_{\gamma_1}(u) \\ Z_{\gamma_2}(u) \end{array}\right) \,.
\end{equation}

\subsubsection{Argyres-Douglas $A_{2}$ realised on $\Sigma^{I}_{A_{2}}$}

The Picard-Fuchs operator derived in \ref{AD2firstPF} is:
$$  \mathcal{L}_m \varpi= \left((\Lambda^6-u^2) \partial^2_u -\frac{5}{36}\right) \varpi=0\,, $$
we can absorb the $\Lambda$ dependence into a redefinition of $u\rightarrow u/\Lambda^3$ (equivalent to setting $\Lambda$ to 1). The operator has three regular singular points at $\pm1,\infty$ and a basis of solutions 
is given by:
\begin{equation}
    \varpi_1= {}_2F_{1}\left(-\frac{5}{12},-\frac{1}{12},\frac{1}{2}, u^{2}\right)\,, \quad \varpi_2=u\,{}_2F_{1}\left(\frac{1}{12},\frac{5}{12},\frac{3}{2}, u^{2}\right)\,,
\end{equation}
where ${}_2F_1$ denotes the hypergeometric function. We identify the following linear combinations with the central charge:

\begin{align}
\label{A2secondrealZ1} Z_{\gamma_{1}}(u) &=  \frac{1}{1440 \pi ^{3/2}} \left(6 \Gamma\left(\frac{7}{12}\right) \Gamma\left(\frac{11}{12}\right) \varpi_1 
-\Gamma\left(\frac{1}{12}\right) \Gamma\left(\frac{17}{12}\right) \varpi_2 \right),
\\ \label{A2secondrealZ2}
 Z_{\gamma_{2}}(u)& = -\frac{i}{1440 \pi ^{3/2}} \left(6\Gamma\left(\frac{7}{12}\right) \Gamma\left(\frac{11}{12}\right) \varpi_1  + \Gamma\left(\frac{1}{12}\right) \Gamma\left(\frac{17}{12}\right)  \varpi_2\right).  
\end{align}

The combinations for the central charges are chosen such that $Z_{\gamma_1}$ corresponds to the central charge of the BPS state vanishing at $u=1$ and $Z_{\gamma_2}$ corresponds to the central charge of the BPS state vanishing at $u=-1$. The central charges correspond to integrations of the meromorphic one-form $\lambda$ over a dual basis of cycles of the family of curves which degenerate at $u=\pm1$. 

We call $a=Z_{\gamma_1}$ and $b=Z_{\gamma_2}$. In a local coordinate vanishing at $u=1$, for instance $v=\frac{1}{864}(1-u)$ we obtain the leading terms of the expansion of the exact solutions:\footnote{The numerical factor $1/864$ was chosen such that integral series expansions are obtained, it is however meaningless, the integral expansions in terms of the modular variables is however intrinsic.}
\begin{equation}
\begin{aligned}
a(v)&= v+30 v^2+ \mathcal{O}\left(v^3\right)\,\\
2\pi i b(v)&= \left(v+ 30 v^2 + \mathcal{O}(v^3)\right) \log (v)-v+\frac{1}{60} + 141 v^2 + \mathcal{O}(v^3)
\end{aligned}
\end{equation}

We can invert the series $a(v)$ to express the central charges as:

\begin{equation}
\begin{aligned}
Z_{\gamma_1}(a) &= a \,\\
Z_{\gamma_2}(a) &= b(a)= \frac{1}{2\pi i} \left( \frac{1}{60}+a (\log (a)-1)+141 a^2+O\left(a^3\right) \right) \,.
\end{aligned}
\end{equation}
Identifying $b(a)= \frac{\partial F_0(a)}{\partial a}$ gives the following expression for the prepotential $F_0(a)$:

\begin{equation}
2\pi i \cdot F_0(a)= \frac{a}{60}+\frac{1}{2} a^2 \left( \log (a)-\frac{3}{2}\right)+47
   a^3+O\left(a^4\right)
\end{equation}

The coordinate $\tau$ which maps $u$ to the upper half plane is obtained from the prepotential as:
$$\tau =\frac{\partial^2 F_0(a)}{\partial a^2}$$
\begin{equation}
2\pi i \tau(a)= \log (a)+282 a+46302 a^2+ \mathcal{O}(a^3)\,,
\end{equation}
we introduce $q:=\exp(2\pi i \tau)$ and obtain:

\begin{equation}
q(a)= a+282 a^2+ \mathcal{O}(a^3)\,,
\end{equation}

The monodromies of the central charges around the singular points are given by

\begin{align} \label{A2secondrealisationmonodromies}
 M_{+1} =
 \begin{pmatrix}
 1 &  0   \\
 1  &  1
 \end{pmatrix},
 \    \   \
 M_{-1} =
 \begin{pmatrix}
 1 &   -1   \\
 0  &  1
 \end{pmatrix},
\    \   \
 M_{\infty} = (M_{+1}\cdot M_{-1})^{-1}=
 \begin{pmatrix}
 0 &   1   \\
 -1  &  1
 \end{pmatrix}. 
 \end{align}

These are identified with the generators of the $\Gamma_0(1)^*$ subgroup of $SL(2,\mathbb{Z})$ which is reviewed in \ref{subsec:quasimodular}.

We note the following involution acting on the periods as $$u\rightarrow -u\,,$$
the two singular points $\pm1$ are interchanged and the action on the periods is:

$$Z_{\gamma_1}(-u)= -i Z_{\gamma_2}(u)\,, \quad Z_{\gamma_2}(-u)=i Z_{\gamma_1}(u)\,,$$

this involution is the Fricke involution acting on quasi modular forms of $\Gamma_0(1)^*$. We verify this by obtaining the expressions in terms of quasi modular forms of the central charges.

We recall the relations between the holomorphic differential on the curve $\omega_0$ and the meromorphic differential $\lambda$ derived using the Picard-Fuchs equations, we have on the one hand:
$$ \nabla_{\frac{\partial}{\partial u}} \lambda = \omega_0\,,$$
and on the other:
\begin{equation}
\lambda = \frac{36}{5} (\Lambda^6-u^2) \nabla_{\frac{\partial}{\partial u}} \omega_0\,.
\end{equation}

The periods of the holomorphic differential are directly related to modular forms using the relations between the Picard-Fuchs equation and modular forms \ref{quasiPF}, we find:

$$ \pi_0(\tau):=-\frac{1}{864} A(\tau)\,, \quad \pi_1(\tau)= -\frac{1}{864} A(\tau)\cdot \tau\,. $$

Using the relation of quasi modular forms to differential operators of Picard-Fuchs type as well as the relation between the periods of the meromorphic differential $\lambda$ and the holomorphic one \ref{relmeroholA2}, we obtain:
\begin{align}
u(\tau)&= -1 + \frac{2B(\tau)^6}{A(\tau)^6}\,,\\ 
\frac{d u(\tau)}{d\tau}&= \frac{ 2 B^6 (-A^6 + B^6)}{A^{10}} \,\, \\
Z_{\gamma_1}(\tau)&= \frac{A(\tau)^6 - 2 B(\tau)^6 + A(\tau)^4 E(\tau)}{720 A(\tau)^5}\,\\
Z_{\gamma_2}(\tau) &= \frac{\tau \left(A(\tau)^6-2
   B(\tau)^6+A(\tau)^4 E(\tau)\right)}{720 A(\tau)^5} -\frac{i}{120 \pi  A(\tau)}\,,
\end{align}
where:
\begin{equation}
A(\tau) = E_{4}(\tau)^{\frac{1}{4}}\quad B(\tau)=(\frac{E_{4}(\tau)^{\frac{3}{2}}+E_{6}(\tau)}{2})^{\frac{1}{6}} \quad C(\tau)=(\frac{E_{4}(\tau)^{\frac{3}{2}}-E_{6}(\tau)}{2})^{\frac{1}{6}}\,.
\end{equation}

The Fricke involution acts as:
\begin{align}
\mathcal{F}(\tau) &= -\frac{1}{\tau}\,,\\
\mathcal{F}(A)&=  -i A \cdot \tau \,\\
\mathcal{F}(B)&= - i C \cdot \tau \, \\
\mathcal{F}(C)&= - i B \cdot \tau \,\\
\mathcal{F}(E)&= E \cdot \tau^2 + \frac{6}{\pi i} \tau\,,
\end{align}
and we verify that indeed its action corresponds to:
\begin{align}
\mathcal{F}(u(\tau))&= - u(\tau)\,, \, \\
\mathcal{F}(Z_{\gamma_1}(\tau))&= - i Z_{\gamma_2}(\tau)\, , \\
\mathcal{F}(Z_{\gamma_2}(\tau))&=  i Z_{\gamma_1}(\tau)\, .
\end{align}

\begin{figure}[h!]
 	\centering
 	{\includegraphics[width=0.6\textwidth]{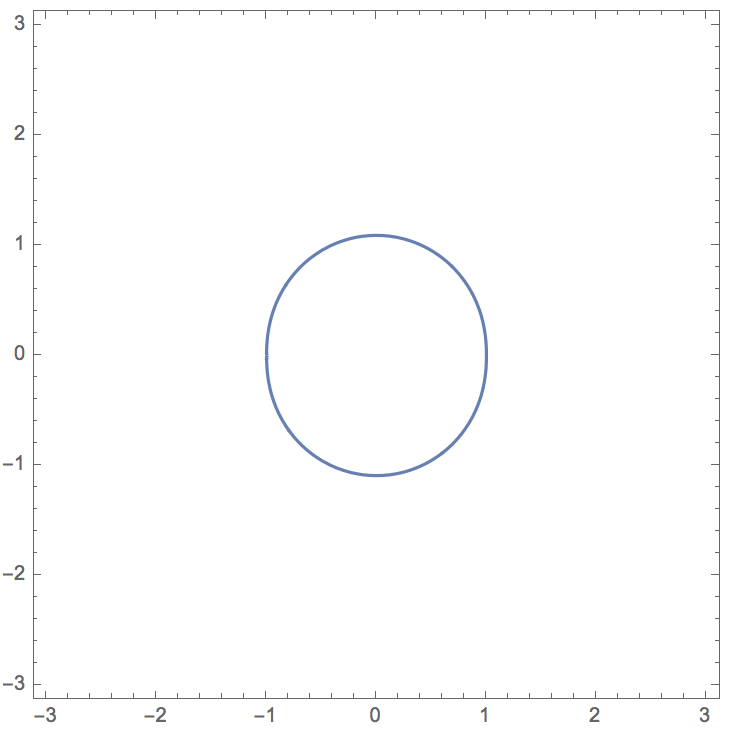}}
 	\caption{Wall of marginal stability}
 	\label{fig:walla2first}
 \end{figure}

Having found the exact solutions we can plot the wall of marginal stability 
\begin{equation}
    \mathcal{W}_{\gamma^1,\gamma^2} := \left\{ u \in \mathcal{B} \,|\, \hat{Z}_{\gamma^1}(u)/  \hat{Z}_{\gamma^2}(u)\in \mathbb{R}\right\}
\end{equation}
for this BPS structure is given in figure \ref{fig:walla2first}.

\subsubsection{Argyres-Douglas $A_{2}$ realised on $\Sigma^{II}_{A_{2}}$}

This parametrization of the curve for Argyres-Douglas theory is discussed for instance in \cite{Shapere:1999xr}:
   \begin{equation}
\Sigma_{A_2}^{II} := \{y^2= (z-\Lambda^2)(z+\lambda^2)(z-u) \in \mathbbm{C}^2 \}\,, \quad u\in \mathcal{B}=\mathbbm{P}^1\setminus \{\pm \Lambda^2,\infty\}\,.
\end{equation} 
We note that the combination of parameterizing the moduli space by $u$ and using the differential $\lambda$ does not yield the appropriate setting for the BPS-VHS introduced earlier. In particular the Gauss-Manin connection acting on $\lambda$ does not yield the holomorphic differential $\omega_0= \frac{dz}{y}$ of the curve but instead a combination of $\omega_0$ and $\nabla_{\frac{\partial}{\partial u}} \omega_0$ as determined in \ref{lambdainomega0AD2nd}.
\begin{equation}
    \nabla_{\frac{\partial}{\partial u}} \lambda =  \frac{2}{3} \left(u^2-1\right) \left(2 u \nabla_{\frac{\partial}{\partial u}} \omega_0+\omega_0\right)\,.
\end{equation}

We start by discussing the modularity of the periods of the holomorphic differential $\omega_0$, guided by the discussion in \ref{quasiPF}. Recall the Picard-Fuchs operator annihilating the periods of the holomorphic differential, \ref{holPF}:
\begin{equation}
    \mathcal{L}_h= (\Lambda^4-u^2) \partial^2_u - 2u \partial_u -\frac{1}{4}\,,
\end{equation}
we can redefine $u\rightarrow u/\Lambda^2$ to obtain:
\begin{equation}
    \mathcal{L}_h= (1-u^2) \partial^2_u - 2u \partial_u -\frac{1}{4}\,,
\end{equation}
and we identify the solutions:
\begin{equation}
    \pi_{\gamma_1}= \, _2F_1\left(\frac{1}{2},\frac{1}{2};1;\frac{1-u}{2}\right)\,, \quad  \pi_{\gamma_2}= \frac{-2 i Q_{-\frac{1}{2}}(u)}{\pi }\,,
\end{equation}
where $_2F_1$ denotes the hypergeometric function and  $Q_{-\frac{1}{2}}$ the Legendre function of the second kind. The solutions are chosen such they correspond up to normalization to the periods of the cycles $\gamma_1$ and $\gamma_2$ which vanish at $u=1,-1$ respectively. In terms of a local coordinate $v=\frac{1}{32}(1-u)$, the expansions of the periods read:
\begin{align}
    \pi_{\gamma_1}(v)&= 1+ 4v + 36 v^2 +\mathcal{O}(v^3)\,,\\
    \pi_{\gamma_2}(v) &= \frac{i}{\pi} \pi_{\gamma_1}(v) \log v + \frac{i}{\pi} \left( 8 v + 84 v^2+ \mathcal{O}(v^3)\right)\,,
\end{align}
the monodromies of this basis of solutions around the singular points is given by:
\begin{align} \label{AD2secondmonodromies}
 M_{+1} =
 \begin{pmatrix}
 1 &  0   \\
 -2  &  1
 \end{pmatrix},
 \    \   \
 M_{-1} =
 \begin{pmatrix}
 1 &   2   \\
 0  &  1
 \end{pmatrix},
\    \   \
 M_{\infty} = (M_{+1}\cdot M_{-1})^{-1}=
 \left(
\begin{array}{cc}
 -3 & -2 \\
 2 & 1 \\
\end{array}
\right)\,.
 \end{align}
The generate the subgroup $\Gamma(2)$ of $SL(2,\mathbb{Z})$ which is isomorphic to $\Gamma_0(4)$.
 The map from the moduli space of the family of curves parameterized by $u$ to the upper half plane near $u=1 (v=0)$ is given by:
 \begin{equation}
     \tau (v) = -\pi_{\gamma_2}(v) /\pi_{\gamma_1}(v) \,.
 \end{equation}

Defining $q:= \exp(2\pi i \tau)$ and inverting the series we obtain:
$$v(q)=\sqrt{q}-8 q+44 q^{3/2}+\mathcal{O}\left(q^{2}\right)\,,$$
to use the quasi-modular forms of $\Gamma_0(4)$ discussed earlier we use the isomorphism of $\Gamma(2)$ and $\Gamma_0(4)$ and redefine $\tilde{\tau}=\tau/2$ and $\tilde{q}=\sqrt{q}$, dropping the tilde we henceforth obtain:
$$v(q)=q-8 q^2+44 q^3+\left(q^{4}\right)\,,$$

We use the generators $A,B,C$ of quasi modular forms discussed in \ref{subsec:quasimodular}, for the subgroup $\Gamma(2)$ of $SL(2,\mathbb{Z})$  which is isomorphic to $\Gamma_0(4)$ these generators are:

$$A(\tau)=\theta_3^2(2\tau)\,, \quad B(\tau)=\theta_4^2(2\tau)\,, \C(\tau)=\theta_2^2(2\tau)\,.$$

$$ \pi_{\gamma_1}(\tau):= A(\tau)\,, \quad \pi_{\gamma_2}(\tau)= -A(\tau)\cdot 2 \tau\,, $$

We furthermore obtain:
\begin{align}
u(\tau)&= -1 + \frac{2B(\tau)^2}{A(\tau)^2}\,,\\ 
\frac{d u(\tau)}{d\tau}&=  -2 B^2 + \frac{2 B^4}{A^2} \,.
\end{align}

We now come to the periods of the meromorphic differential $\lambda$ which are annihilated by the Picard-Fuchs operators.

We choose as solutions:
\begin{align}
 \label{A2firstrealZ1}   Z_{\gamma_1}(u) &= \frac{1}{480} \left(u^2-1\right) P_{\frac{1}{2}}^2(u) \,, \\ 
\label{A2firstrealZ2} Z_{\gamma_2}(u) &=-\frac{i \left(u^2-1\right) Q_{\frac{1}{2}}^2(u)}{240 \pi }\,,
\end{align}

where $P^{2}_{\frac{1}{2}}(u)$ and $Q^{2}_{\frac{1}{2}}(u)$ are associated Legendre polynomials. These correspond to the central charges of the elements of the lattice which correspond to the vanishing cycles at $u=1,-1$ respectively and they obey the same monodromies as $\pi_{\gamma_1}$ and $\pi_{\gamma_2}$ discussed earlier.

Using the normalizations chosen here as well as the relations between the mermorphic differential and the holomorphic one determined earlier:

\begin{equation}
    \lambda = \frac{1}{1920} \left(u^2-1\right) \left(2 \left(u^2+3\right) \nabla_{\frac{\partial}{\partial u}}\omega_0+u\, \omega_0\right)\, , 
\end{equation}

we find the following expressions in terms of quasi-modular forms of the central charges:
\begin{align}
    Z_{\gamma_1}(\tau) &= \frac{A(\tau)^6+A(\tau)^4 \left(E(\tau)-2
   B(\tau)^2\right)-A(\tau)^2 B(\tau)^2
   E(\tau)+B(\tau)^4 E(\tau)}{480 A(\tau)^5}\,, \\
    Z_{\gamma_2}(\tau) &= \frac{\tau \left(-A(\tau)^6-A(\tau)^4
   \left(E(\tau)-2 B(\tau)^2\right)+A(\tau)^2
   B(\tau)^2 E(\tau)-B(\tau)^4 E(\tau)\right)}{240
   A(\tau)^5}  \nonumber \\  &+\frac{i \left(A(\tau)^4-A(\tau)^2
   B(\tau)^2+B(\tau)^4\right)}{120 \pi  A(\tau)^5}\,.
\end{align}

\begin{figure}[h!]
 	\centering
 	{\includegraphics[width=0.6\textwidth]{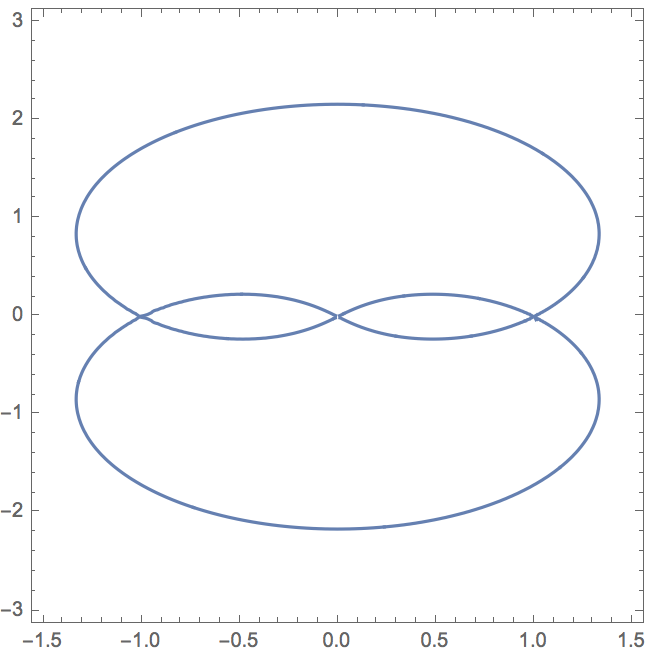}}
 	\caption{Wall of marginal stability, for $u \rightarrow \frac{1}{u}$.}
 	\label{fig:walla2second}
 \end{figure}

Having found the exact solutions we can plot the wall of marginal stability 
\begin{equation}
    \mathcal{W}_{\gamma^1,\gamma^2} := \left\{ u \in \mathcal{B} \,|\, \hat{Z}_{\gamma^1}(u)/  \hat{Z}_{\gamma^2}(u)\in \mathbb{R}\right\}
\end{equation}
for this BPS structure, it is given in figure \ref{fig:walla2second}.


\subsubsection{Seiberg-Witten SU(2) theory}

The Picard-Fuchs operator derived in \ref{SWPF} is:
$$    \mathcal{L}_m =  \left((\Lambda^4-u^2) \partial^2_u -\frac{1}{4}\right)\,,  $$
we can absrob the $\Lambda$ dependence into a redefinition of $u\rightarrow u/\Lambda^2$ (equivalent to setting $\Lambda$ to 1). The operator has three regular singular points at $\pm1,\infty$ and a basis of solutions 
is given by:
\begin{equation}
    \varpi_1= {}_2F_{1}\left(-\frac{1}{4},-\frac{1}{4},\frac{1}{2}, u^{2}\right)\,, \quad \varpi_2=u\,{}_2F_{1}\left(\frac{1}{4},\frac{1}{4},\frac{3}{2}, u^{2}\right)\,,
\end{equation}
where ${}_2F_1$ denotes the hypergeometric function. We identify the following linear combinations with the central charge:

\begin{align}
\label{swcharge1}Z_{\gamma_{1}}(u) &=  -\frac{ \Gamma \left(-\frac{1}{4}\right) \Gamma
   \left(\frac{3}{4}\right)\varpi_1+2 \varpi_2 \Gamma
   \left(\frac{1}{4}\right) \Gamma
   \left(\frac{5}{4}\right)}{32 \pi ^{3/2}}
\\ 
 \label{swcharge2}Z_{\gamma_{2}}(u)& =\frac{1}{32} \left(\frac{4 i \sqrt{\frac{2}{\pi }} \Gamma
   \left(\frac{3}{4}\right) \,
   \varpi_1}{\Gamma \left(\frac{1}{4}\right)}+\frac{i  \Gamma
   \left(\frac{1}{4}\right) \,
   \varpi_2}
   {\sqrt{2 \pi } \Gamma \left(\frac{3}{4}\right)}\right).  
\end{align}

The combinations for the central charges are chosen such that $Z_{\gamma_1}$ corresponds to the central charge of the BPS state vanishing at $u=1$ and $Z_{\gamma_2}$ corresponds to the central charge of the BPS state vanshing at $u=-1$, these correspond to the magnetic monopole and the dyon of Seiberg-Witten theory \cite{SW}. The central charges correspond to integrations of the meromorphic one-form $\lambda$ over a dual basis of cycles of the family of curves which degenerate at $u=\pm1$. 

We call $t_1=Z_{\gamma_1}$ and $t_2=Z_{\gamma_2}$. In a local coordinate vanishing at $u=1$, for instance $v=\frac{1}{32}(1-u)$ we obtain the leading terms of the expansion of the exact solutions:
\begin{equation}
\begin{aligned}
t_1(v)&= v+2 v^2+ \mathcal{O}\left(v^3\right)\,\\
\pi i t_2(v)&= \left(-v- 2 v^2 + \mathcal{O}(v^3)\right) \log (v)
-\frac{1}{4}+v-3 v^2 +\mathcal{O}(v^3)
\end{aligned}
\end{equation}

Inverting the series $t_1(v)$ and
identifying $t_2(t_1)= \frac{\partial F_0(t_1)}{\partial t_1}$ gives the following expression for the prepotential $F_0(a)$:

\begin{equation}
\pi i \cdot F_0(t_1)= -\frac{t_1}{4}+\frac{1}{4} t_1^2 (3-2 \log
   (t_1))+ \mathcal{O}\left(t_1^3\right)
\end{equation}

The coordinate $\tau$ which maps $u$ to the upper half plane is obtained from the prepotential as:
$$\tau =\frac{\partial^2 F_0(t_1)}{\partial t_1^2}$$
\begin{equation}
\pi i \tau(t_1)= -\log (t_1)-6 t_1-30 t_1^2+ \mathcal{O}\left(t_1^3\right)
\end{equation}
we introduce $q:=\exp(-\pi i \tau)$ .

The monodromies of the central charges around the singular points are given by

\begin{align} \label{SWmonodromies}
 M_{+1} =
 \begin{pmatrix}
 1 &  0   \\
 -2  &  1
 \end{pmatrix},
 \    \   \
 M_{-1} =
 \begin{pmatrix}
 1 &   2   \\
 0  &  1
 \end{pmatrix},
\    \   \
 M_{\infty} = (M_{+1}\cdot M_{-1})^{-1}=
 \left(
\begin{array}{cc}
 -3 & -2 \\
 2 & 1 \\
\end{array}
\right)\,.
 \end{align}

These are identified with the generators of the $\Gamma(2)$ subgroup of $SL(2,\mathbb{Z})$ which is isomorphic to $\Gamma_0(4)$ which is reviewed in \ref{subsec:quasimodular}.

We note the following involution acting on the periods as $$u\rightarrow -u\,,$$
the two singular points $\pm1$ are interchanged and the action on the periods is:

$$Z_{\gamma_1}(-u)= i Z_{\gamma_2}(u)\,, \quad Z_{\gamma_2}(-u)=-i Z_{\gamma_1}(u)\,,$$

this involution is the Fricke involution acting on quasi modular forms of $\Gamma(2)$. We verify this by obtaining the expressions in terms of quasi modular forms of the central charges, for the Seiberg-Witten geometry, quasi-modular expressions for the central charges have been previously obtained in \cite{Huang:2006si}, following modularity discussion in \cite{Klemm:1995wp}, see also \cite{Lerche:1996xu}.
and references therein.

We recall the relations between the holomorphic differential on the curve $\omega_0$ and the meromorphic differential $\lambda$ derived using the Picard-Fuchs equations, we have on the one hand:
$$ \nabla_{\frac{\partial}{\partial u}} \lambda = \omega_0\,,$$
and on the other hand:
$$ \lambda = 4 (1-u^2) \nabla_{\frac{\partial}{\partial u}} \omega_0\,.$$

We use the generators $A,B,C$ of quasi modular forms discussed in \ref{subsec:quasimodular}, for the subgroup $\Gamma(2)$ of $SL(2,\mathbb{Z})$  which is isomorphic to $\Gamma_0(4)$ these generators are:

$$A(\tau)=\theta_3^2(2\tau)\,, \quad B(\tau)=\theta_4^2(2\tau)\,, \C(\tau)=\theta_2^2(2\tau)\,.$$
The periods of the holomorphic differential are directly related to modular forms using the relations between the Picard-Fuchs equation and modular forms \ref{quasiPF}, we find:

$$ \pi_0(\tau):= A(\tau)\,, \quad \pi_1(\tau)= A(\tau)\cdot \tau\,, $$

where $$\partial_u Z_{\gamma_1} (u)= -\frac{1}{32} \pi_0(u)\,, \quad \partial_u Z_{\gamma_2} (u)= \frac{1}{32} \pi_1(u)$$

Using the relation of quasi modular forms to differential operators of Picard-Fuchs type as well as the relation between the periods of the meromorphic differential $\lambda$ and the holomorphic one, we obtain:
\begin{align}
u(\tau)&= -1 + \frac{2B(\tau)^2}{A(\tau)^2}\,,\\ 
\frac{d u(\tau)}{d\tau}&=  -2 B^2 + \frac{2 B^4}{A^2} \, \\
Z_{\gamma_1}(\tau)&= \frac{A(\tau)^2-2 B(\tau)^2+E(\tau)}{16 A(\tau)}\\
Z_{\gamma_2}(\tau) &= \frac{\pi  A(\tau)^2 \tau-2 \pi  B(\tau)^2
   \tau+\pi  E(\tau) \tau-2 i}{16 \pi 
   A(\tau)}\,.
\end{align}

The Fricke involution acts as:
\begin{align}
\mathcal{F}(\tau) &= -\frac{1}{\tau}\,,\\
\mathcal{F}(A)&=  -i A \cdot \tau \,\\
\mathcal{F}(B)&= - i C \cdot \tau \, \\
\mathcal{F}(C)&= - i B \cdot \tau \,\\
\mathcal{F}(E)&= E \cdot \tau^2 + \frac{6}{\pi i} \tau\,,
\end{align}
and we verify that indeed its action corresponds to:
\begin{align}
\mathcal{F}(u(\tau))&= - u(\tau)\,, \, \\
\mathcal{F}(Z_{\gamma_1}(\tau))&= - i Z_{\gamma_2}(\tau)\,, \\
\mathcal{F}(Z_{\gamma_2}(\tau))&=  i Z_{\gamma_1}(\tau)\, .
\end{align}

\begin{figure}[h!]
 	\centering
 	{\includegraphics[width=0.6\textwidth]{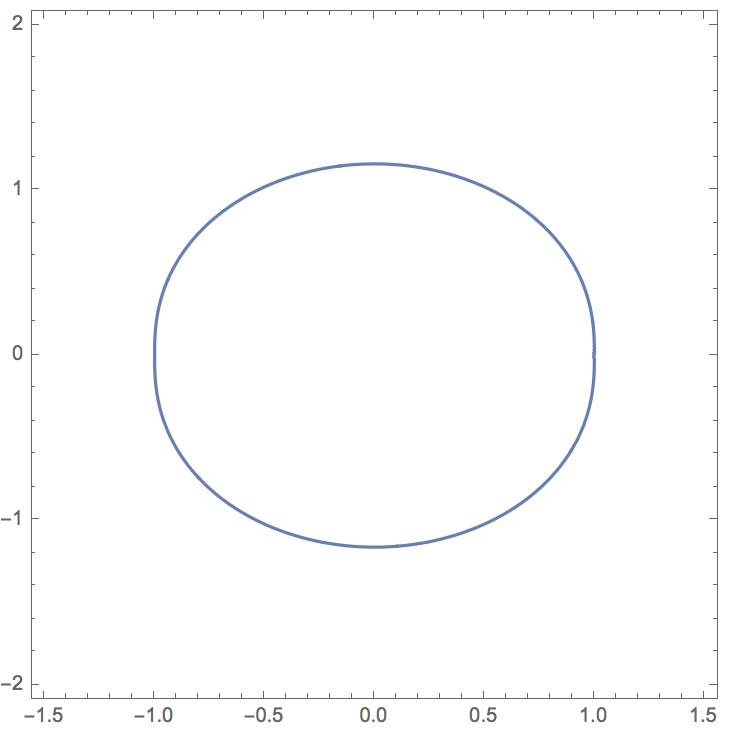}}
 	\caption{Wall of marginal stability, for $u \rightarrow \frac{1}{u}$.}
 	\label{fig:swwall}
 \end{figure}

Having found the exact solutions we can plot the wall of marginal stability 
\begin{equation}
    \mathcal{W}_{\gamma^1,\gamma^2} := \left\{ u \in \mathcal{B} \,|\, \hat{Z}_{\gamma^1}(u)/  \hat{Z}_{\gamma^2}(u)\in \mathbb{R}\right\}
\end{equation}
for this BPS structure, it is given in figure \ref{fig:swwall}.


\section{BPS spectrum from attractor flow} \label{section:attractorflowproject2examplesdiagrams}

Now the Picard-Fuchs equations are known and the solutions and their monodromies have been found and matched with the moduli dependent central charges of BPS states in theories with complex 1-dimensional moduli spaces. This now allows us to apply the method of steepest decent from \cite{Denef:2001xn} described in subsections \ref{ss:specialgeom}-\ref{One-dimensional case} and developed in \cite{Denef:1998sv} to theories described in the literature \cite{KLMVW,KKV,KlemmEffective,Billo:1997mi,Billo:1998yr} on type II effective field theories. We use this to find and plot the attractor flow lines (for the spherically symmetric approximation \footnote{This is the approximation for a supergravity solution of a spherically symmetric 4d metric \cite{Denef:2000nb} needed to derive the equations of motion.}) iteratively (on \textsc{Mathematica}) for all the cases mentioned above including the Argyres-Douglas $A_{1},A_{2}$ theories \cite{Argyres:1995jj,Shapere:1999xr} as well as Seiberg-Witten theory \cite{SW}. In the cases with a wall of marginal stability (see subsection \ref{sec:wall}) between the weak and strong coupling regions, this is also plotted. The existence conditions on the endpoint of the flow \cite{Moore:1998pn,Denef:2001xn,Denef:2000nb} from subsection \ref{sec:existenceconditions} are used to determine which BPS states exist in which chambers in the moduli space. Any split attractor flow lines are plotted. The branch cuts are also plotted and when the flow lines enter or leave a branch cut they are continued through the branch cut by taking a loop around a singular point and acting on them with the appropriate monodromies from subsection \ref{specialgeometryandmodularity} around this loop.

\subsection{Attractor flow for $A_{1}$-theory}
We proceed to derive the attractor flow for the $A_{1}$-model. We are using the path of steepest descent method described in subsection \ref{One-dimensional case} to derive the flow lines.
In this case the central charge is simply $Z_{\gamma_{1}}(u) = u$.  

\begin{figure}[h!]
 	\centering
 	{\includegraphics[width=0.6\textwidth]{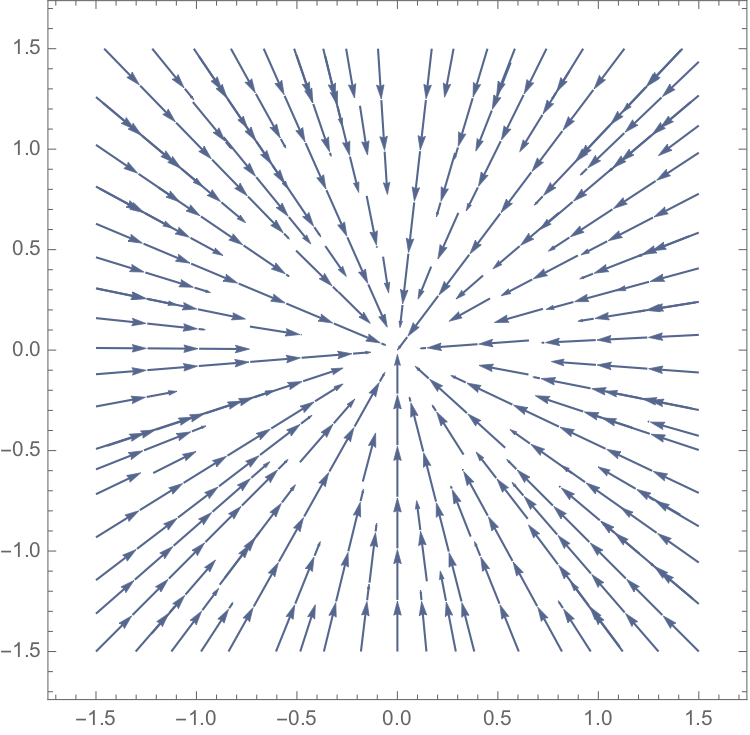}}
 	\caption{This diagram shows the attractor flow lines of $A_{1}$ flowing to its attractor point at the origin.}
 	\label{fig:a1flow}
 \end{figure}

\begin{enumerate}[label = \alph*.)]

	\item We let $u := x+iy$ such that $|Z_{\gamma_{i}}(u)| := |Z_{\gamma_{i}}(x+iy)| \in \mathbb{R}^{+}$.
	
	\item For $Z_{\gamma_{1}}(u) = u$ we have $|Z_{\gamma_{1}}(u)| = |u| = |x+iy|= \sqrt{x^{2}+y^{2}}$. 
	
	\item We substitute this expression into the gradient flow equations to derive the attractor flow lines
	\begin{align} \label{diffeq}
	\frac{dy}{dx} =  \frac{\Big(\frac{\partial |Z_{\gamma_{i}}(x,y)|}{\partial y}\Big)}{\Big(\frac{\partial |Z_{\gamma_{i}}(x,y)|}{\partial x}\Big)} =\frac{\frac{y}{\sqrt{x^{2}+y^{2}}}}{\frac{x}{\sqrt{x^{2}+y^{2}}}} = \frac{y}{x}.
	\end{align}
  \item   The equation can be solved as
\begin{align} \label{straightline}
    \frac{1}{y}dy = \frac{1}{x}dx \rightarrow y=Ax.
\end{align}
\end{enumerate}	
Hence the equations $y = Ax$ from(\ref{diffeq}-\ref{straightline}) describe the set of all straight lines $L$ passing through the origin. The attractor flow lines then correspond to straight lines flowing to the origin. Now we can also look at this in terms of the time parameter $\tau$. In this form, the attractor flow equation can be written as 
\begin{align}
   \frac{dx}{d \tau} =  - \frac{\partial |Z_{\gamma_{i}}(x,y)|}{\partial x} = - \frac{x}{\sqrt{x^{2}+y^{2}}} = - \frac{x}{\sqrt{(1+A^{2})x^{2}}} = -\frac{1}{\sqrt{(1+A^{2})}} = \alpha. 
\end{align}
Therefore the $x$ coordinate is linear in the time coordinate such that:
\begin{align}
  x = \alpha \tau + \beta.
\end{align}
One can plot these attractor flow lines (see Fig. \ref{fig:a1flow}) on the moduli space as a radial flow flowing into the origin.

\subsubsection*{Final state in chamber}

Hence we can see that in this theory there is only 1 BPS state flowing to the attractor point at the origin. There are no walls of marginal stability or jumps in the number of BPS states.

\begin{table}[h!]
\begin{center}
	\begin{tabular}{ | l | l | l | l | p{5cm} |}
		\hline
		Chamber & Existing charges   & Count \\ \hline
		1: All space & $\gamma_{1}$ & 1 \\ \hline
	\end{tabular}
\end{center}   
\caption{Single BPS state that exists everywhere in the moduli space.}
\label{tab:A1singleBPS}
\end{table}

\subsection{Attractor flow for Argyres--Douglas $A_{2}$ theory}

We now repeat this process for $A_{2}$ theory \cite{Argyres:1995jj, Shapere:1999xr}. We take the exact expressions for the all possible linear combinations of the central charges for each realisation of the theory $Z_{\gamma_{i}}(u)$ and derive the attractor flow lines from the gradient flow (see section \ref{One-dimensional case}) using \textsc{Mathematica}. We let $u \rightarrow \frac{1}{u}$ to have the origin of the complex plane at infinity. We proceed in the following way;  we first let $u:= x+iy \in \mathbbm{P}^1 $ such that the modulus of the central charge $|Z_{\gamma_{i}}(u)| := |Z_{\gamma_{i}}(x+iy)| \in \mathbb{R}^{+} $.
	As before, we substitute this expression into the gradient flow equations to derive the attractor flow lines
	\begin{align}
	\text{gradient flow:} \ \ \ \frac{dy}{dx} =  \frac{\Big(\frac{\partial |Z_{\gamma_{i}}(x,y)|}{\partial y}\Big)}{\Big(\frac{\partial |Z_{\gamma_{i}}(x,y)|}{\partial x}\Big)}.
	\end{align}

\subsubsection{Walls, branch cuts and flow lines}

\subsubsection*{Wall of marginal stability and existence}

We take into consideration the wall crossing phenomena in which a BPS exists (is stable) in one region of the $u$-plane, but is excluded by the existence conditions from \cite{Moore:1998pn,Denef:2001xn,Denef:2000nb} in another region (see section \ref{sec:existenceconditions}). Physically this corresponds to a region of the moduli space in which the composite BPS particle is unstable and  decays into a combination of its constituents. For this we also plot the wall of marginal stability $MS_{\gamma_{1}, \gamma_{2}}$ that bounds the stable and unstable regions on the $u$-plane: for the meromorphic differential $\tilde{y}dz$ from section \ref{subsec:examplecurves} this is given by the locus of real ratio of the periods
\begin{align}
	MS_{\gamma_{1}, \gamma_{2}}: \  \ \frac{Z_{\gamma_{1}}(u)}{Z_{\gamma_{2}}(u)} \in \mathbb{R}.
\end{align}
The decay of a BPS state in a chamber is represented diagrammatically by split attractor flow lines: a flow line enters a chamber and would end at a regular point in the moduli space. Therefore it is excluded in the chamber but still existed before it crossed the wall of marginal stability $MS_{\gamma_{1}, \gamma_{2}}$. In this case the flow line hits the wall of marginal stability and splits into constituent BPS flow lines corresponding to BPS states that are stable within the region.

   \begin{figure}[h!]
	\centering
	{\includegraphics[width=0.7\textwidth]{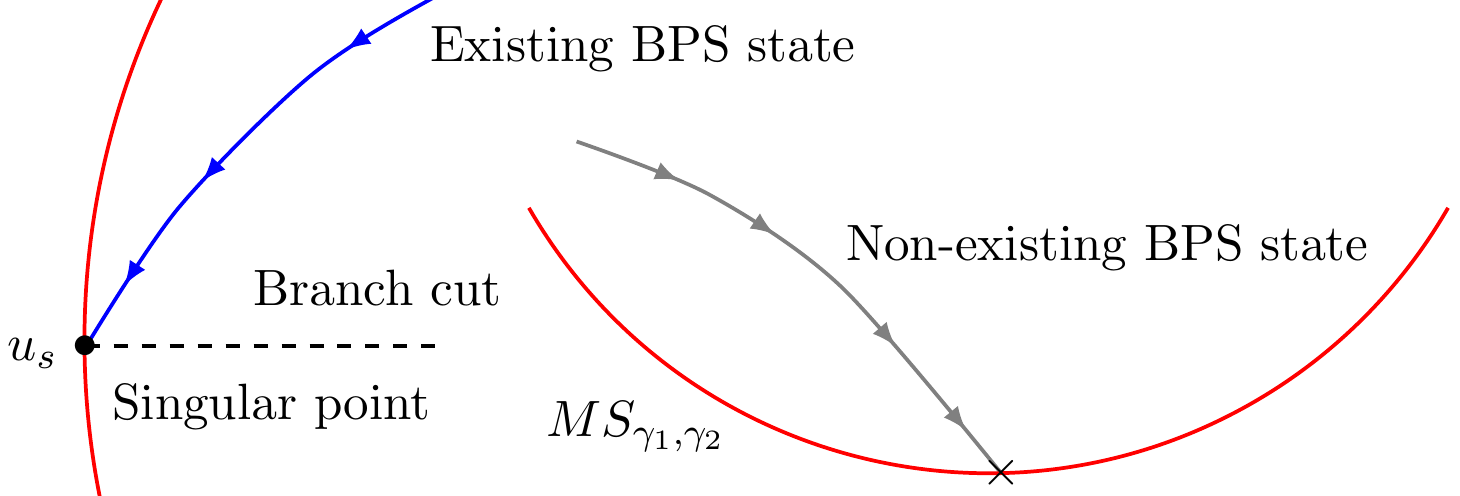}}
	
\caption{BPS state only exists if flow terminates at singular point $u_{s}$.}

\end{figure}

\subsubsection*{Comment on branch cuts} \label{subsubcontinuationthroughcuts}
	
\begin{figure}[h!]
	\centering
	{\includegraphics[width=0.7\textwidth]{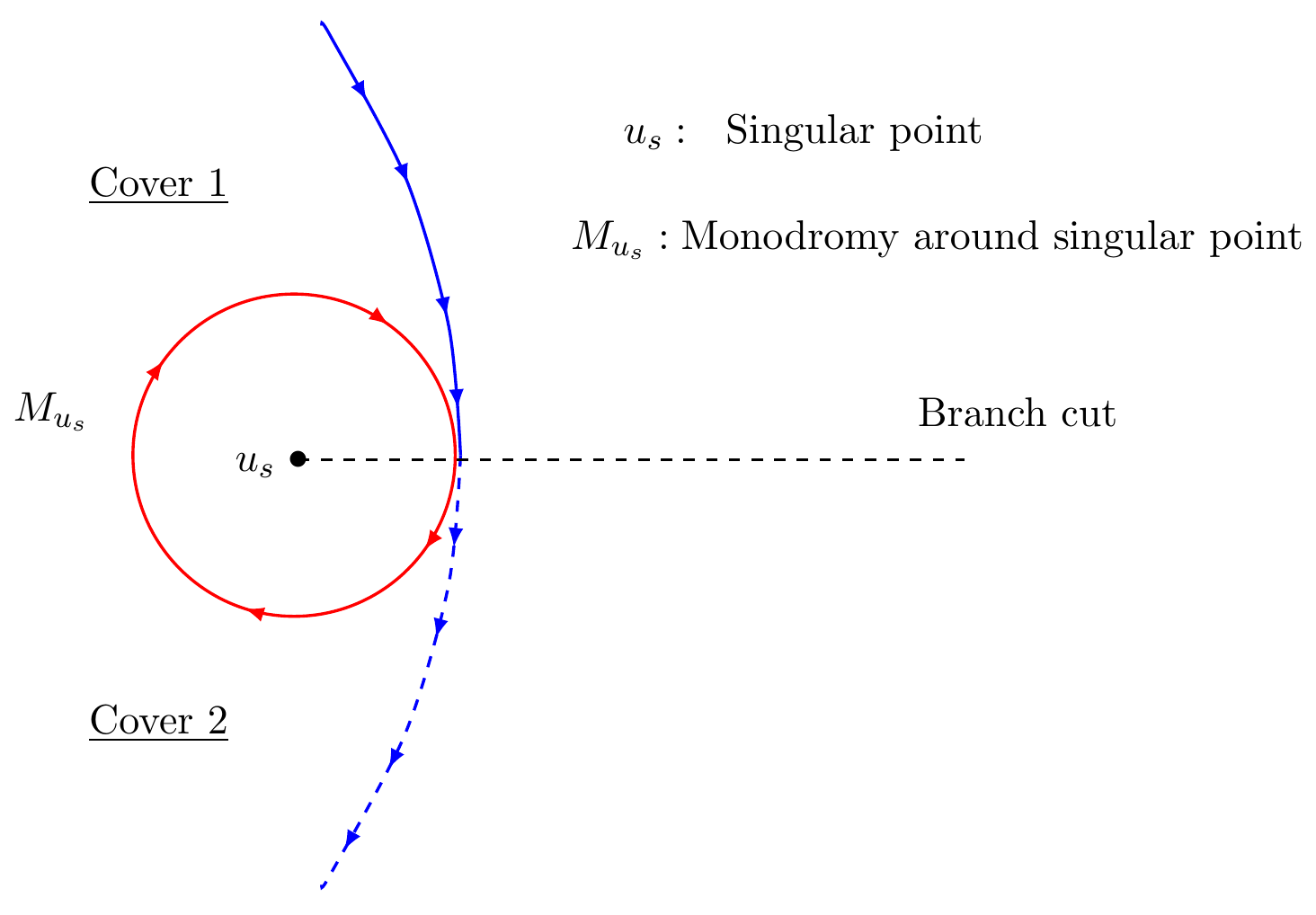}}
	\caption{Single flow line passing through branch cut after acting with monodromy around singular point.}
	\label{fig:monodromybranchcuts}
\end{figure}

	We also consider the branch cuts of the $Z_{\gamma_{i}}(u)$ in the $u$-plane. We find the branch cuts and plot the segments which affect the attractor flow lines.
	When an attractor flow line enters a branch cut, we follow \cite{Denef:2001xn, Denef:2000nb} and find a path in the $u$-plane which connects the point at which the attractor flow line enters the branch cut (shown from above in Fig. \ref{fig:monodromybranchcuts}) and leaves it (shown from below). We determine which singular points this path encloses and act on $Z_{\gamma_{i}}(u)$ with the monodromies $M_{u_{s}}$ associated to the corresponding singular points $u_{s} \in \mathcal{B}$. We write $u = u_{s}+\epsilon e^{-i \theta}, \ \epsilon \in \mathbb{R}^{+}, \ \theta \in \{0,2 \pi \}$ and let $\theta: 0   \mapsto  2\pi$. The monodromies  from section \ref{specialgeometryandmodularity} have been be read off from the expansions around the singular points provided there.

We then continue this modified central charge through the branch cut. The same existence conditions then apply on the other side of the branch cut. If the state is excluded on the other side of the branch cut by a regular termination point, then it also didn't exist before it entered the branch cut - up until the last point after it crossed the wall of marginal stability $MS_{\gamma_{1}, \gamma_{2}}$ where it decayed into its constituent BPS states.

\subsubsection{Attractor flow first realisation}

\begin{figure}[h!]
	\centering
    {\includegraphics[width=0.7\textwidth]{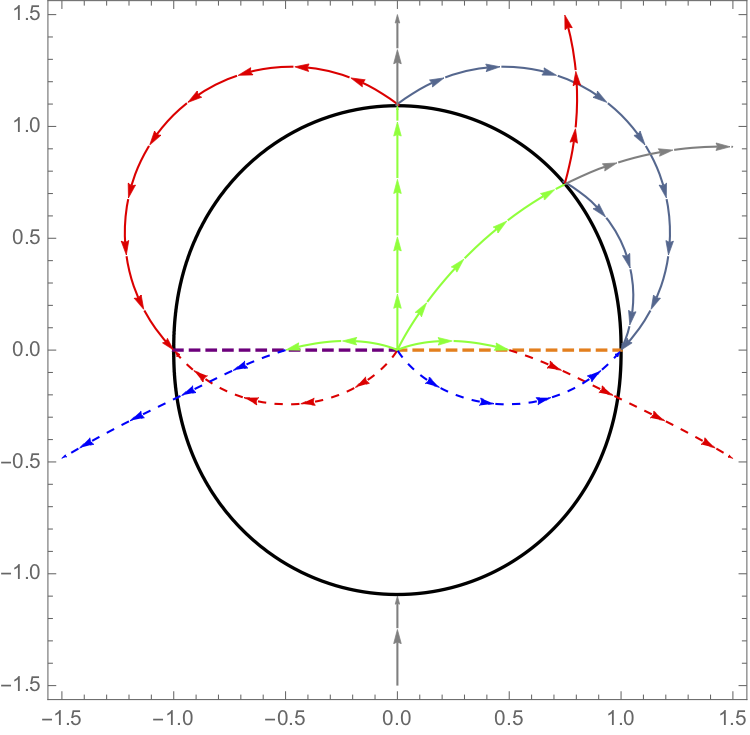}}
	
	\vspace{-30ex}
	\begin{center}
	B2  $\   \  \ \  \  \  \  \ \ $  
	\end{center}
	\vspace{20ex}
	\vspace{-50ex}
	\begin{center}
	B1  $\   \  \ \  \  \  \  \ $  
	\end{center}
	\vspace{50ex}
	
	\vspace{-23ex}
	\begin{center}
	A  $\   \  \ \  \  \  \  \ \  \   \  \ \  \  \  \  \ \  \   \  \ \  \  \  \  \ \ \   \  \ \  \  \  \  \ \  \   \  \ \  \  \  \  \ \ $  
	\end{center}
	\vspace{10ex}
	
	\captionsetup{singlelinecheck=off}
	\caption[flowline2]{As before the black line is the wall of marginal stability. 
	
	\begin{enumerate}[label = \alph*.)]
	\item
	The purple line represents the $Z_{\gamma_{1}}(u)$ branch cut and the orange line is that of $Z_{\gamma_{2}}(u)$. The blue and red flow lines represent $Z_{\gamma_{1}}(u)$ and  $Z_{\gamma_{2}}(u)$ respectively. 
	\item
	The solid green line represents the sum of the basis states $Z_{\gamma_{1}}(u)+Z_{\gamma_{2}}(u)$. The dashed blue and red lines represent the analytic continuation of $Z_{\gamma_{1}}(u)+Z_{\gamma_{2}}(u)$ through the branch cuts of $Z_{\gamma_{1}}(u)$ and $Z_{\gamma_{2}}(u)$ respectively. 
	
	\end{enumerate}
	
	In this case, when the sum is analytically continued through the branch cuts, it becomes $Z_{\gamma_{1}}(u)$ around the left cut and $Z_{\gamma_{2}}(u)$ around the right cut. Again the gray lines represents the unstable continuation of $Z_{\gamma_{1}}(u)+ Z_{\gamma_{2}}(u)$ flow in the outer region A where it crashes at a regular point.}
	
	\label{fig:1chamberinf2b}
\end{figure}

We will start by using the curve $\Sigma^{I}_{A_{2}}$ for the first realisation and consider flow lines of all possible $Z_{\gamma_{i}}(u)$. In this section we use the basis $Z_{\gamma_{1}}(u),  \ Z_{\gamma_{2}}(u) \rightarrow - Z_{\gamma_{2}}(u)$ to better visualise and position the split flow lines and attractor points. We plot a diagram Fig. \ref{fig:1chamberinf2b} with the existing flow lines and the split attractor flow. When we compute the central charges and take the ratio we just obtain a wall separating 2 chambers: an outer and inner region (see Fig. \ref{fig:1chamberinf2b}) labelled as A and B respectively. This is the standard realisation of the wall used the literature e.g. \cite{GMN1} and can be obtained by taking a slice in the complex 2d moduli space of $SU(3)$ \cite{Argyres:1995jj} containing the Argyres-Douglas points \cite{DDN}. In their parameterisation there are also 2 regions but with the wall shifted along the $y$-axis.

\begin{table}[h!]

  \begin{center}
	\begin{tabular}{ | l | l | l | l | l | l | l | p{5cm} |}
		\hline
	\multicolumn{6}{|c|}{Existing BPS states} & Non-existing BPS states\\
	\hline
		Flow line & \begin{tikzpicture}
\text{The blue line}
    \draw [blue,thick,-{Stealth}] (0,-1)--(1,-1); 
\end{tikzpicture}   & \begin{tikzpicture}
\text{The blue line}
    \draw [red,thick,-{Stealth}] (0,-1)--(1,-1); 
\end{tikzpicture}   & \begin{tikzpicture}
\text{The blue line}
    \draw [green,thick,-{Stealth}] (0,-1)--(1,-1); 
\end{tikzpicture} &  \begin{tikzpicture}
\text{The blue line}
    \draw [blue,thick,dashed,-{Stealth}] (0,-1)--(1,-1); 
\end{tikzpicture} & \begin{tikzpicture}
\text{The blue line}
    \draw [red,thick,dashed,-{Stealth}] (0,-1)--(1,-1); 
\end{tikzpicture} & \begin{tikzpicture}
\text{The blue line}
    \draw [gray,thick,-{Stealth}] (0,-1)--(1,-1); 
\end{tikzpicture} \\ \hline
		Charges & $\gamma_{1}$ & $\gamma_{2}$ & $\gamma_{1}+\gamma_{2}$& $ \gamma_{1}$ continuation & $ \gamma_{2}$ continuation & $\gamma_{1}+\gamma_{2} \ \ $ outside wall\\ \hline
	\end{tabular}
\end{center}

    \caption{Flow lines in Fig. \ref{fig:1chamberinf2b}.}
    \label{tab:1chamberinf2b}
\end{table}

\subsubsection*{Description of flow lines} \label{firstrealisationdescriptionofflowlines}

$Z_{\gamma_{1}}(u)$ and  $Z_{\gamma_{2}}(u)$ (\ref{A2secondrealZ1}-\ref{A2secondrealZ2}) are, for a particular cover, defined for  $u \in \mathbbm{P}^1\setminus  [-1, \infty)$ and $u \in \mathbbm{P}^1\setminus  [1, \infty)$ respectively.
They have $u^{b} \log u$ branch cuts $[-1, \infty)$ and  $(\infty, 1]$ represented by \begin{tikzpicture}
\text{The blue line}
    \draw [purple,thick,dashed] (0,0)--(1,0); 
\end{tikzpicture}
and \begin{tikzpicture}
\text{The blue line}
    \draw [orange,thick,dashed] (0,0)--(1,0); 
\end{tikzpicture} on Fig. \ref{fig:1chamberinf2b}  respectively, separating the regions B1 and B2. This allows for the analytic continuation of the central charges onto new covers. Their flow lines are represented on this diagram by blue \begin{tikzpicture}
\text{The blue line}
    \draw [blue,thick,-{Stealth}] (0,-1)--(1,-1); 
\end{tikzpicture} and red \begin{tikzpicture}
\text{The blue line}
    \draw [red,thick,-{Stealth}] (0,-1)--(1,-1); 
\end{tikzpicture} lines, and flow to $+1$ and $-1$ respectively. Both are singular points, hence $Z_{\gamma_{1}}(u)$ and  $Z_{\gamma_{2}}(u)$ exist everywhere. This holds in both the inner region B and outer region A.

\begin{align*}
 \text{\textit{Wall crossing of dyon} } 
 \end{align*}
 \\
The sum, represented by \begin{tikzpicture}
\text{The blue line}
    \draw [green,thick,-{Stealth}] (0,-1)--(1,-1); 
\end{tikzpicture}, and written as  $Z_{\gamma_{1}}(u)+Z_{\gamma_{2}}(u)$, has a more complicated behavior. Its source is at infinity. It exists within the inner region B. However, outside this region, where it is represented by the gray line \begin{tikzpicture}
\text{The blue line}
    \draw [gray,thick,-{Stealth}] (0,-1)--(1,-1); 
\end{tikzpicture}, it flows to a regular point on the wall, is hereby excluded, and must therefore split into its constituent BPS states $Z_{\gamma_{1}}(u), Z_{\gamma_{2}}(u)$, as shown on the diagram above.
The situation with the sum $Z_{\gamma_{1}}(u)+Z_{\gamma_{2}}(u)$ in the inner chamber B is more involved: it first appears that the attractor flow lines terminate at the same regular point in the moduli space as they do from the outside. However, in this case we must take the branch cuts of the basis charges into account.
\\
\\
We must now use the method introduced in \ref{subsubcontinuationthroughcuts}. This means we analytically continue the flow of $Z_{\gamma_{1}}(u)+Z_{\gamma_{2}}(u)$ through the branch cuts between B1 and B2 by taking paths around the singular points at $\pm 1$ and acting with the associated monodromy matrix, from (\ref{A2secondrealisationmonodromies}), in the right direction. We look at 2 cases:

\begin{enumerate}[label= (\roman*)]
\item
In the first case, represented by \begin{tikzpicture}
\text{The blue line}
    \draw [blue,thick, dashed, -{Stealth}] (0,-1)--(1,-1); 
\end{tikzpicture}, the analytic continuation of the central charges acts in a clockwise direction around $-1$ as \footnote{We must remember here that the monodromies $M_{u_{s}}$ are now acting on the basis $Z_{\gamma_{1}}(u), \ Z_{\gamma_{2}}(u) \rightarrow -Z_{\gamma_{2}}(u)$.} 
\begin{align} \label{transformation1}
(M_{-1})^{-1}: \  \  & Z_{\gamma_{1}}(u) \longmapsto Z_{\gamma_{1}}(u)- Z_{\gamma_{2}}(u), \\ \nonumber
\\ \nonumber  \text{such that}  \  \  \ & Z_{\gamma_{1}}(u)+Z_{\gamma_{2}}(u) \longmapsto Z_{\gamma_{1}}(u).
\end{align}  
This then leaves the branch cut in B2 as the dashed blue line.

\item
In the second case, represented by \begin{tikzpicture}
\text{The blue line}
    \draw [red,thick, dashed, -{Stealth}] (0,-1)--(1,-1); 
\end{tikzpicture}, the continuation acts in a counter-clockwise direction around $+1$:
\begin{align} \label{transformation2}
M_{+1}: \ \ & Z_{\gamma_{2}}(u)\longmapsto Z_{\gamma_{2}}(u)- Z_{\gamma_{1}}(u), \\ \nonumber
\\ \nonumber  \text{such that}  \  \  \ & Z_{\gamma_{1}}(u)+Z_{\gamma_{2}}(u)\longmapsto Z_{\gamma_{2}}(u),
\end{align}
and leaves the branch cut in B2 as the dashed red line.

Hence, one can see from (\ref{transformation1}-\ref{transformation2}) when the sum in the upper half plane is continued through the branch cuts, where it flows in, it subsequently flows out in the lower half plane as one of the basis states. This then flows to the singular points $\pm 1$ and therefore exists.

\end{enumerate}

One can use the analytic continuation of the central charges to find a region in which the sum of a particle and an antiparticle, e.g. $\gamma_{1}-\gamma_{2}$, can exist. In the region B2 on the diagram Fig. \ref{fig:1chamberinf2b}, taken below the 2 branch cuts and above the outer lower arc \footnote{This is the lower part of the central chamber.}, the central charge of the basis states $Z_{\gamma_{1}}(u)$, $Z_{\gamma_{2}}(u)$ becomes either 
$Z_{\gamma_{1}}(u)-Z_{\gamma_{2}}(u)$ or $-Z_{\gamma_{1}}(u)+Z_{\gamma_{2}}(u)$. This depends on which branch cut the analytic continuation is done through, the sign of the basis state before the continuation. There are now 2 possible covers one must consider for the central charges with either
$\gamma_{1}-\gamma_{2}$ or $-\gamma_{1}+\gamma_{2}$ existing on it.
.

\subsubsection*{Exclusion of higher linear combinations $nZ_{\gamma_{1}}(u)+mZ_{\gamma_{2}}(u)$}
\label{higherlinearfirstrealisation}

\begin{figure}[h!]
	\centering
	{\includegraphics[width=0.9\textwidth]{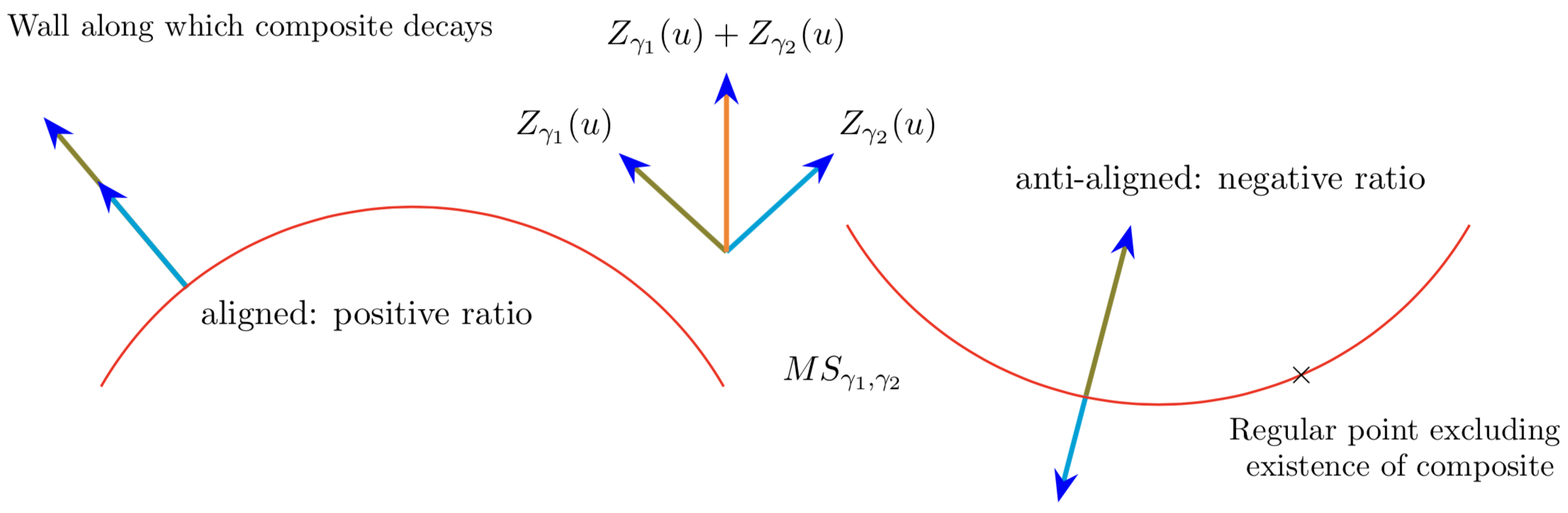}}
	
	\caption{Alignment and anti-alignment of central charges on different general segments of a wall of marginal stability.}
 
 \label{fig:alignmentofcentralcharges}
\end{figure}

After this we consider the general state $n\gamma_{1}+m\gamma_{2}$. For this, one needs to consider termination points corresponding to 
\begin{align} \label{eq:higherfirstrealisation}
n Z_{\gamma_{1}}(u)+m Z_{\gamma_{2}}(u) =0,
\end{align}
for all $(n,m)$. If the point is a regular point on the wall, the state doesn't exist. In the discussion below we will show that all states other than $(0,1), (1,0) $ and $(1,1)$ are excluded by regular termination points on segments of the wall. This is done by considering the alignment or anti-alignment of the central charges (described in Fig. \ref{fig:alignmentofcentralcharges}) on the wall
and the range of the ratio on paths between singular points. If there is a change in sign along the path then equation (\ref{eq:higherfirstrealisation}) has a solution and a combination is excluded.

From this one can see that in the outer region A these higher linear combinations flow to and are excluded by a regular point on the lower segment of the wall, like the sum, but this time shifted according to the ratio $\frac{n}{m}$. This is because of the anti-alignment of the central charges on this segment. We present a table (\ref{tab:ratios2}) here representing the ratios of the central charges along a path (\ref{path2}) on the wall from 
\begin{align} \label{path2}
1 \longrightarrow -1 \longrightarrow 1,
\end{align}
along the lower segment then the upper segment respectively. 
\begin{table}[h!]
    
\begin{center}
	\begin{tabular}{ | l | l | l | l | p{2.5cm} |}
		\hline
		 \multicolumn{5}{|c|}{\vphantom{\LARGE{H}} ratios along path}            \\[4pt] \hline
		paths between singular points & $Z_{\gamma_{1}}(u)$ & $Z_{\gamma_{2}}(u)$ & $ Z_{\gamma_{1}}(u)/ Z_{\gamma_{2}}(u)$ &  $Z_{\gamma_{2}}(u) / Z_{\gamma_{1}}(u)$ \\ \hline
		$1$ & $  0$ & $0.0027i$ & $+\infty$ &  0 \\ \hline
		$-1$ & $0.0027$ & $ 0$  & $0$ &  $\pm \infty$ \\ \hline
		$1$ &  $ 0$ &  $0.0027i$ &   $-\infty$  &  $0$ \\ \hline
	\end{tabular}
\end{center}    
    
    \caption{Ratio of central charges at singular points along wall.}
    \label{tab:ratios2}
\end{table} 
Within the inner region B things become more involved and one must consider the logarithmic branch cuts $[\pm 1, \infty)$. In this case we find that for 
$n \gamma_{1}+m \gamma_{2}$ within the inner chamber the flow line for $n Z_{\gamma_{1}}(u)+m Z_{\gamma_{2}}(u), \  \   \forall n,m \geq 1$
also ends at a regular point on the lower segment of the wall. 
\\
However, as with the case for $ Z_{\gamma_{1}}(u)+ Z_{\gamma_{2}}(u)$, it flows through the branch cuts before it can reach the point. Again we must act with the corresponding monodromies from (\ref{A2secondrealisationmonodromies}) around the singular points to transform the central charges and then analytically continue through the branch cut.

\begin{align*}
 \text{\textit{Excluding higher combinations using range of ratio} } \\
 \end{align*} 
Unlike $ Z_{\gamma_{1}}(u)+ Z_{\gamma_{2}}(u)$ the higher $n,m > 1, n \neq m$ do not become the basis states when flowing through the branch cuts, instead they remain states with $m, n >1, \  n \neq m$, and the ratio $\frac{n}{m}$ keeps the same sign. This means the states will continue to flow and terminate at another regular point on the lower segment of the wall. Therefore, as with the previous parameterisation states of the form, $n \gamma_{1}+m \gamma_{2} \ \ n,m \geq 1,  \frac{n}{m}, \frac{m}{n} > 1$ are excluded in the inner region B as well and hence do not exist/ are unstable anywhere in the moduli space.

\subsubsection*{Example}

\begin{table}[h!]

 \begin{center}
	\begin{tabular}{ | l | l | l | l | l | p{5cm} |}
		\hline
	\multicolumn{5}{|c|}{Non - existing BPS states} \\
	\hline
		Flow line & \multicolumn{2}{|c|}{\begin{tikzpicture}
\text{The blue line}
    \draw [black,thick,dashed,-{Stealth}] (0,-1)--(1,-1); 
\end{tikzpicture} }  & \multicolumn{2}{|c|}{\begin{tikzpicture}
\text{The blue line}
    \draw [brown,thick,-{Stealth}] (0,-1)--(1,-1); 
\end{tikzpicture}} \\ \hline
		Charges & \multicolumn{2}{|c|}{$3\gamma_{1}+2\gamma_{2}$} &    \multicolumn{2}{|c|}{$\gamma_{1}+2\gamma_{2}$} \\ \hline
		Forked flow lines & \begin{tikzpicture}
\text{The blue line}
    \draw [green,thick,dashed,-{Stealth}] (0,-1)--(1,-1); 
\end{tikzpicture}  & \begin{tikzpicture}
\text{The blue line}
    \draw [darkgray,thick,dashed,-{Stealth}] (0,-1)--(1,-1); 
\end{tikzpicture} & \begin{tikzpicture}
\text{The blue line}
    \draw [red,thick,-{Stealth}] (0,-1)--(1,-1); 
\end{tikzpicture}  &     \begin{tikzpicture}
\text{The blue line}
    \draw [green,thick,-{Stealth}] (0,-1)--(1,-1); 
\end{tikzpicture}                                \\ \hline
		Charges & $\gamma_{1}+\gamma_{2}$ & $2\gamma_{1}+\gamma_{2}$ &  $\gamma_{2}$ &     $\gamma_{1}+\gamma_{2}$                           \\ \hline
	\end{tabular}
\end{center}

    \caption{Forked flow lines of non-existing BPS states on Fig. \ref{fig:higherstates}.}
    \label{tab:higherstates}
\end{table}

\begin{figure}[h!]
	\centering
	{\includegraphics[width=0.6\textwidth]{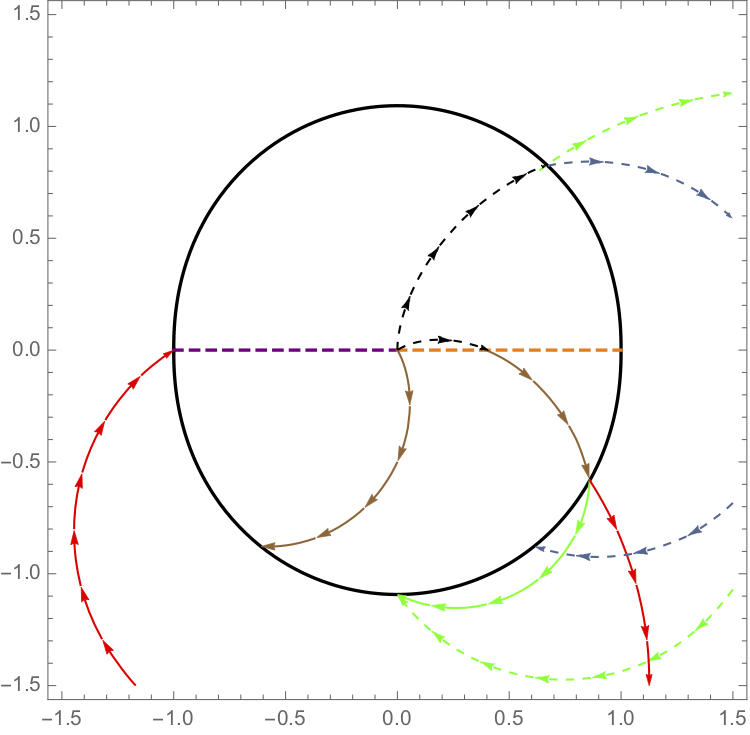}}

	\vspace{-42.5ex}
	\begin{center}
	B1  $\   \  \ \  \  \  \  \ $  
	\end{center}
	\vspace{40ex}
	
	\vspace{-35ex}
	\begin{center}
	B2  $\   \  \ \  \  \  \  \ \ $  
	\end{center}
	\vspace{23ex}

	\vspace{-23ex}
	\begin{center}
	A  $\   \  \ \  \  \  \  \ \  \   \  \ \  \  \  \  \ \  \   \  \ \  \  \  \  \ \ \   \  \ \  \  \  \  \ \  \   \  \ \  \  \  \  \ \ $  
	\end{center}
	\vspace{13ex}

	\caption{The black dotted line in B1 represents the flow of the charge $ 3Z_{\gamma_{1}}(u)+ 2Z_{\gamma_{2}}(u)$. Its flow through the branch cut along $(\infty,1]$ is shown. In this case it becomes $ Z_{\gamma_{1}}(u)+ 2Z_{\gamma_{2}}(u)$ in B2, represented by the brown line, which can terminate at a regular point on the wall. The black dotted line splits into the green and blue dashed lines, representing the sum   $Z_{\gamma_{1}}(u)+ Z_{\gamma_{2}}(u)$ and $2Z_{\gamma_{1}}(u)+ Z_{\gamma_{2}}(u)$ respectively. Similarly the diagram shows the splitting of the brown line into green and red lines,  representing $Z_{\gamma_{1}}(u)+ Z_{\gamma_{2}}(u)$ and $Z_{\gamma_{2}}(u)$ on the first cover.}
	
	\label{fig:higherstates}
\end{figure}

An example flow \begin{tikzpicture}
\text{The blue line}
    \draw [black,thick,dashed,-{Stealth}] (0,-1)--(1,-1); 
\end{tikzpicture} in chamber B (the dotted black line) is given by $ 3Z_{\gamma_{1}}(u)+ 2Z_{\gamma_{2}}(u)$ and is shown in the diagram Fig. \ref{fig:higherstates} above. As it passes through the branch cut between B1 and B2, we act with an $M_{+1}$ (from (\ref{A2secondrealisationmonodromies})) in a counter-clockwise direction and transform
\begin{align} 
3Z_{\gamma_{1}}(u)+ 2Z_{\gamma_{2}}(u) \longmapsto 3Z_{\gamma_{1}}(u)+ 2(Z_{\gamma_{2}}(u)-Z_{\gamma_{1}}(u)) \longmapsto Z_{\gamma_{1}}(u)+ 2Z_{\gamma_{2}}(u).
\end{align}

This leaves the branch cut in B2 as \begin{tikzpicture}
\text{The blue line}
    \draw [brown,thick,-{Stealth}] (0,-1)--(1,-1); 
\end{tikzpicture} (the brown line). This terminates at a regular point in the lower half plane and is therefore excluded. This then excludes the higher linear combinations in the inner chamber B around infinity. \footnote{They are excluded outside in A as they simply flow to another regular point on the lower segment of the wall without encountering a branch cut.}

\subsubsection*{Final existing states in each chamber}

 The complete tabulation (\ref{finalexistingstatestable}) for the existing states on the 2 covers discussed in section \ref{firstrealisationdescriptionofflowlines} in this parameterisation is as follows:
 
 \begin{table}[h!]
     
   \begin{center} 
	\begin{tabular}{ | l | l | l | l | p{5cm} |}
		\hline
		Chamber & Existing charges cover 1  & Existing charges cover 2    & Count \\ \hline
		B1: Central region upper half & $\gamma_{1} , \ \  \gamma_{2}, \ \ \gamma_{1}+\gamma_{2}$ & $\gamma_{1} , \ \  \gamma_{2}, \ \ \gamma_{1}+\gamma_{2}$  & 3 \\ \hline
		B2: Central region lower half & $\gamma_{1} , \ \  \gamma_{2}, \ \ \gamma_{1}-\gamma_{2}$ & $\gamma_{1} , \ \  \gamma_{2}, \ \ -\gamma_{1}+\gamma_{2}$  & 3 \\ \hline
		A: Outer region & $\gamma_{1},  \ \ \gamma_{2}$ & $\gamma_{1},  \ \ \gamma_{2}$ &2 \\ \hline
		
	\end{tabular}
\end{center}  
     
     \caption{Existing BPS states on both covers, in all chambers.}
     \label{finalexistingstatestable}
 \end{table}
 
\newpage 

\subsubsection{Attractor flow second realisation} \label{section:attractorflowfirstrealisation}

\begin{figure}[h!]
	\centering
	{\includegraphics[width=0.7\textwidth]{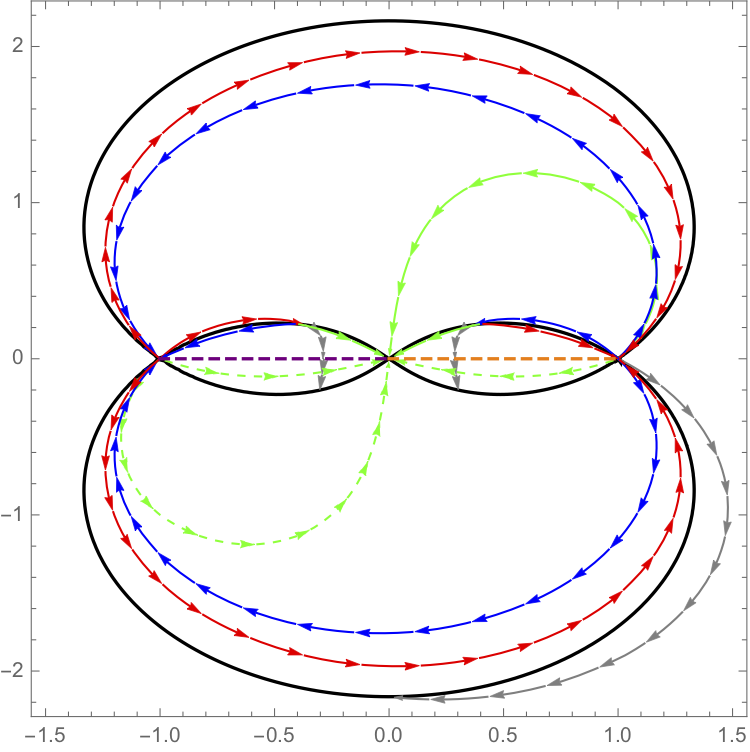}}
	
\vspace{-20ex}
	\begin{center}
	$\   \  \ \  \  \  $   C    
	\end{center}
	\vspace{28ex}
	\vspace{-78ex}
	\begin{center}
	B  $\   \  \ \  \  \  \  \ $  
	\end{center}
	\vspace{70ex}
	
	\vspace{-86ex}
	\begin{center}
	A  $\   \  \ \  \  \  \  \ \  \   \  \ \  \  \  \  \ \  \   \  \ \  \  \  \  \ \ \   \  \ \  \  \  \  \ \  \   \  \ \  \  \  \  \ \ \  \   \  \ \  \  \  \  \ \ \   \ \ \  \  \ \ \   \ \ \  $  
	\end{center}
	\vspace{64ex}	

\vspace{-49.5ex}
	\begin{center}
	D1  $\   \  \ \  \  \  \  \  \ \   \ \ \  \  \ \ \   \ \ \ \  \  \  \  \  \ \   \ \ \  \  \ \ \   \ \ \  $  
	\end{center}
	\vspace{40ex}

   \vspace{-37ex}
	\begin{center}
	D2  $\   \  \ \  \  \  \  \  \ \   \ \ \  \  \ \ \   \ \ \ \  \  \  \  \  \ \   \ \ \  \  \ \ \   \ \ \  $  
	\end{center}
	\vspace{30ex}

\vspace{-49.5ex}
	\begin{center}
	 $\   \  \ \  \  \  \  \  \ \   \ \ \  \  \ \ \   \ \ \ \  \  \  \  \  \ \   \ \ \  \  \ \ \   \ \ \  $  E1    
	\end{center}
	\vspace{40ex}

   \vspace{-37ex}
	\begin{center}
	 $\   \  \ \  \  \  \  \  \ \   \ \ \  \  \ \ \   \ \ \ \  \  \  \  \  \ \   \ \ \  \  \ \ \   \ \ \  $    E2    
	\end{center}
	\vspace{29ex}

	\caption{Wall of marginal stability in black, the purple and orange lines correspond to branch cuts of $Z_{\gamma_{1}}(u)$ and $Z_{\gamma_{2}}(u)$ respectively.  The red and blue lines correspond to sample attractor flow lines of $Z_{\gamma_{1}}(u)$ and $Z_{\gamma_{2}}(u)$ . The green line is a sample flow line for the sum  $Z_{\gamma_{1}}(u)+ Z_{\gamma_{2}}(u)$. It appears as a dashed line on the other side of the branch cuts. The grey lines represent the flow lines continued into unstable regions: $Z_{\gamma_{1}}(u)$ and $Z_{\gamma_{2}}(u)$ in the left and right central chambers respectively, and $Z_{\gamma_{1}}(u)+Z_{\gamma_{2}}(u)$ in the outer region.}

	\label{fig:Z1flow1}
\end{figure}

Now we repeat the attractor flow analysis for the second realisation which is described by the curve $\Sigma^{II}_{A_{2}}$. When one again computes the ratio of the central charges one finds a wall of marginal stability with 5 chambers analogous to that in \cite{Shapere:1999xr} ( see Fig. \ref{fig:Z1flow1}). This time this includes a center right and center left inner chamber, an outer chamber as well as 2 chambers below the upper arc and above the lower arc. 

\begin{enumerate}[label = \alph*.)]

\item We continue with the curve $\Sigma^{II}_{A_{2}}$ and consider flow lines of all possible $Z_{\gamma_{i}}(u)$.

\item We normalise the central charges to $Z_{\gamma_{i}}(u) \rightarrow \frac{1}{u^{2}-1}Z_{\gamma_{i}}(u) $ before plotting the attractor flow lines. This is to produce symmetric results such that all attractor points are on equal footing. This means that, for each existing BPS state, each flow line flows from an infinity of the central charge at 2 singular starting points to a 0 at the third singular point.

\end{enumerate}

The existing flow lines in the inner chambers are shown on the Figure \ref{fig:Z1flow1} below:

 \begin{table}[h!]
     
     \begin{center}
	\begin{tabular}{ | l | l | l | l | l | p{5cm} |}
		\hline
	\multicolumn{5}{|c|}{Existing BPS states} \\
	\hline
		Flow line & \begin{tikzpicture}
\text{The blue line}
    \draw [red,thick,-{Stealth}] (0,-1)--(1,-1); 
\end{tikzpicture}   & \begin{tikzpicture}
\text{The blue line}
    \draw [blue,thick,-{Stealth}] (0,-1)--(1,-1); 
\end{tikzpicture}   & \begin{tikzpicture}
\text{The blue line}
    \draw [green,thick,-{Stealth}] (0,-1)--(1,-1); 
\end{tikzpicture} &  \begin{tikzpicture}
\text{The blue line}
    \draw [green,thick,dashed,-{Stealth}] (0,-1)--(1,-1); 
\end{tikzpicture} \\ \hline
		Charges & $\gamma_{1}$ & $\gamma_{2}$ & $\gamma_{1}+\gamma_{2}$& $\gamma_{2}-\gamma_{1}$\\ \hline
	\end{tabular}
\end{center}

\begin{center}
	\begin{tabular}{ | l | l | l | l | l | p{5cm} |}
		\hline
	\multicolumn{2}{|c|}{Non-existing BPS states: \begin{tikzpicture}
\text{The blue line}
    \draw [gray,thick,-{Stealth}] (0,-1)--(1,-1); 
\end{tikzpicture}} \\
	\hline
		Chamber & Charge      \\ \hline
		outer, lower half plane  & $\gamma_{1}+\gamma_{2}$   \\ \hline
		center left, above cut & $\gamma_{1}$ \\ \hline
		center left, below cut & $\gamma_{1}-2 \gamma_{2}$ \\ \hline
		center right, above cut & $\gamma_{2}$ \\ \hline
		center right, below cut &  $\gamma_{2}-2 \gamma_{1}$        \\ \hline
	\end{tabular}
\end{center}

     \caption{Flow lines on Fig. \ref{fig:Z1flow1}.}
     
 \end{table}
 
\subsubsection*{Description of each flow line in outer chambers}

The central charges $Z_{\gamma_{1}}(u), Z_{\gamma_{2}}(u)$ (\ref{A2firstrealZ1}-\ref{A2firstrealZ2}) are again defined on a particular cover for $u \in \mathbbm{P}^1\setminus  [-1, \infty)$ and $u \in \mathbbm{P}^1\setminus  [1, \infty)$ respectively. As with the first realisation there are logarithmic branch cuts arising from the $u^{a} \log u$ terms in the expansion around the singular points. These can again be taken from $[\pm 1, \infty)$ and are represented by the lines \begin{tikzpicture}
\text{The blue line}
    \draw [purple,thick,dashed] (0,0)--(1,0); 
\end{tikzpicture}
and \begin{tikzpicture}
\text{The blue line}
    \draw [orange,thick,dashed] (0,0)--(1,0); 
\end{tikzpicture}
respectively. The flow lines can be continued through the branch cuts onto a new cover. 

\begin{enumerate}[label = (\roman*)]

\item The blue line  
\begin{tikzpicture}
\text{The blue line}
    \draw [blue,thick,-{Stealth}] (0,-1)--(1,-1); 
\end{tikzpicture}
corresponds to a sample flow of $Z_{\gamma_{2}}(u)$: in the 2 chambers B,C within the outer arc, this charge flows from the singular points at $u=+1$ and $u= \infty$ to the termination point at $u= -1$. Therefore, because $-1$ is a singular point, $Z_{\gamma_{2}}(u)$ exists within these chambers. The blue line can also be taken in the outer region A just above the outer arc and in this case will flow from $+1$ to $-1$. So, again, $Z_{\gamma_{2}}(u)$ exists in this outer region.

\item The red line \begin{tikzpicture}
\text{The blue line}
    \draw [red,thick,-{Stealth}] (0,-1)--(1,-1); 
\end{tikzpicture}
corresponds to a sample flow of $Z_{\gamma_{1}}(u)$: This follows the same flow pattern with the direction of flow reversed: in the regions B,C just below the outer arc, $Z_{\gamma_{1}}(u)$ flows from $u= -1$ and $u= \infty$ to $u=+1$. Again, $+1$ is a singular point and therefore again $Z_{\gamma_{1}}(u)$ exists in these chambers as well as the outer region.

\item Finally, the green line \begin{tikzpicture}
\text{The blue line}
    \draw [green,thick,-{Stealth}] (0,-1)--(1,-1); 
\end{tikzpicture}
corresponds to a sample attractor flow line of the sum $Z_{\gamma_{3}}(u) =Z_{\gamma_{1}}(u)+Z_{\gamma_{2}}(u)$: in the chamber B below the upper outer arc, as well as above the branch cuts in the 2 central chambers, D1 and E1, this state flows from $\pm1$ to its termination point at $u = \infty$ and therefore exists, because $u=\infty$ is a singular point.

\item
The gray line \begin{tikzpicture}
\text{The blue line}
    \draw [gray,thick,-{Stealth}] (0,-1)--(1,-1); 
\end{tikzpicture}
represents the sum $Z_{\gamma_{1}}(u)+Z_{\gamma_{2}}(u)$ in chamber A outside the wall: here it doesn't exist and terminates at a regular point on the lower arc of the wall. This means that the third state decays across the outer wall in the upper half plane. We have:
\begin{align}
\text{decay pathway outer chamber} \  \  \  \ Z_{\gamma_{3}}(u) \ \ \begin{tikzpicture}
\draw[-{>[scale=2.5,
          length=2,
          width=3]},line width=0.4pt] (0,0) to (1,0);
\end{tikzpicture} \  & \  \ Z_{\gamma_{1}}(u)+ Z_{\gamma_{2}}(u), \\ \nonumber
\text{in terms of charges} \  \ \gamma_{3} \ \begin{tikzpicture}
\draw[-{>[scale=2.5,
          length=2,
          width=3]},line width=0.4pt] (0,0) to (1,0);
\end{tikzpicture} & \ \ \gamma_{1}+ \gamma_{2}.
\end{align}

\end{enumerate}

\subsubsection*{Analytic continuation of $Z_{\gamma_{1}}(u)+Z_{\gamma_{2}}(u)$ dyon through the branch cuts}

The sum $Z_{\gamma_{1}}(u)+Z_{\gamma_{2}}(u)$ also wouldn't exist when evaluated in the chamber C just above the lower arc because it would flow to the same regular termination point.  However, we apply the same method as for the previous parameterisation $\Sigma^{I}_{A_{2}}$ by taking into account the branch cuts of the basis charges. The sum represented by the green line \begin{tikzpicture}
\text{The blue line}
    \draw [green,thick,-{Stealth}] (0,-1)--(1,-1); 
\end{tikzpicture} in B can be analytically continued, using the monodromies in (\ref{AD2secondmonodromies}), into the lower half plane. Both $Z_{\gamma_{1}}(u)$ and $Z_{\gamma_{2}}(u)$ have logarithmic branch cuts in the intervals $(\infty, -1]$ and $[+1,\infty )$ respectively. In these cases we took the paths around the singular points. The one around $-1$ is clockwise and we act with 
\begin{align} \label{monodromyaction1}
(M_{-1})^{-1}: Z_{\gamma_{1}}(u) \longmapsto Z_{\gamma_{1}}(u) -2Z_{\gamma_{2}}(u).	
\end{align}
For $+1$ we must consider $M_{+1}$. However, this time we rotate in a anti-clockwise direction such that 
\begin{align} \label{monodromyaction2}
M_{+1}: Z_{\gamma_{2}}(u) \longmapsto Z_{\gamma_{2}}(u) -2Z_{\gamma_{1}}(u).	
\end{align}
Therefore the sum becomes 
\begin{align} \label{sumanalyticcontinuation}
Z_{\gamma_{1}}(u)+Z_{\gamma_{2}}(u) & \longmapsto -Z_{\gamma_{2}}(u)+Z_{\gamma_{1}}(u) \  \  \ \text{around $-1$ and} \\ \nonumber 
\\ \nonumber
Z_{\gamma_{1}}(u)+Z_{\gamma_{2}}(u) & \longmapsto Z_{\gamma_{2}}(u)-Z_{\gamma_{1}}(u)	\  \  \ \text{around $+1$.}
\end{align}
These combinations are represented on Fig. \ref{fig:Z1flow1} as dotted green lines \begin{tikzpicture}
\text{The blue line}
    \draw [green,thick,dashed,-{Stealth}] (0,0)--(1,0); 
\end{tikzpicture}
. They also flow from $\pm 1$ to $\infty$ in the chamber C just above the lower arc, and also in the 2 central chambers, D2 and E2, below the branch cut. Therefore the analytic combination of the sum (\ref{sumanalyticcontinuation}) can be taken to exist there.

We remember that the sum and its analytic continuation do not exist in the outer region A. Here, on both sides of the branch cuts, the flow lines terminate at a regular point on the wall bounding the outer region. This also means that as with the first realisation  $\Sigma^{I}_{A_{2}}$ in section  \ref{firstrealisationdescriptionofflowlines}, there are again 2 covers on which either $\gamma_{1}-\gamma_{2}$ or $-\gamma_{1}+\gamma_{2}$ and their antiparticles can exist.

\subsubsection*{Split flow of $Z_{\gamma_{1}}(u)$, $Z_{\gamma_{2}}(u)$ in central two regions}

We next consider existence of the basis charges $Z_{\gamma_{1}}(u)$, $Z_{\gamma_{2}}(u)$ (\ref{A2firstrealZ1}-\ref{A2firstrealZ2}) in the 2 central regions D and E and their analytic continuation through the branch cuts. Each basis charge has a chamber in the central region near its source point with its logarithmic branch cut passing through it. In these chambers the attractor flow flows into the branch cut. We proceed as before in (\ref{monodromyaction1}-\ref{monodromyaction2}), by analytically continuing the basis charge through the branch cut by taking a path around the singular point and acting with the corresponding monodromy from (\ref{AD2secondmonodromies}). In this case, depending on whether one considers the flow flowing into the branch cut from above or below, the central charges get mapped to 
\begin{align}
Z_{\gamma_{1,2}}(u) \longmapsto Z_{\gamma_{1,2}}(u) \pm 2Z_{\gamma_{1,2}}(u).
\end{align}
This always leads to the flow terminating at a regular point on the wall just on the other side of the branch cut, excluding the basis state from existing within the smaller chamber next to its source point. In this case the central charges $Z_{\gamma_{1}}(u)$, $Z_{\gamma_{2}}(u)$ can no longer be considered basis states. Instead, at the wall of marginal stability surrounding the 2 small central chambers D, E these BPS states decay into the other (now constituent states) that are stable within the chamber:

\begin{figure}[h!]
	\centering
	{\includegraphics[width=0.7\textwidth]{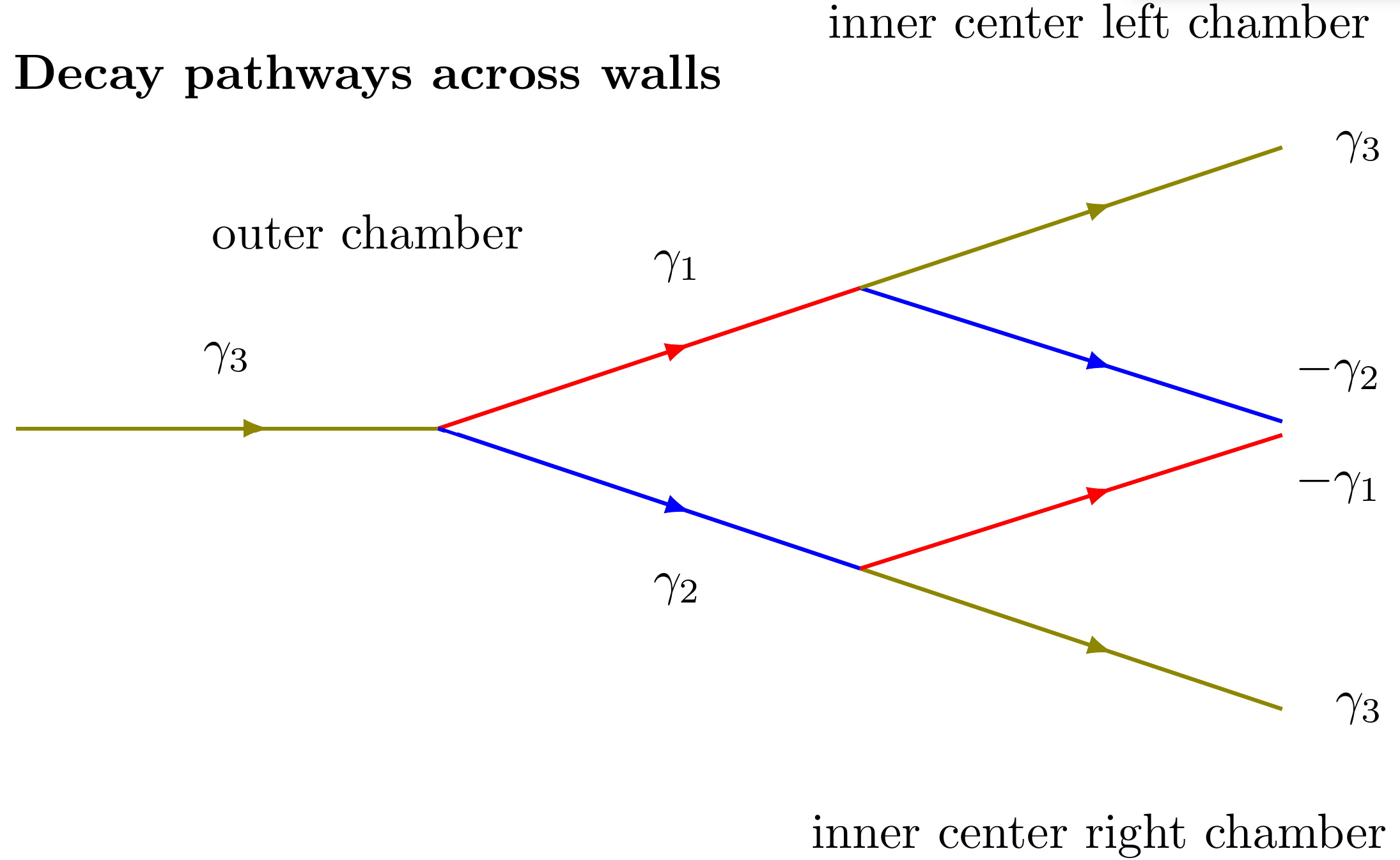}}
	
	\label{fig:decaypathwaysallchambers}
\end{figure}
\begin{align} \label{decaypathway1}
 \text{decay pathway center left chamber D} \ \  \  \ Z_{\gamma_{1}}(u) \ \begin{tikzpicture}
\draw[-{>[scale=2.5,
          length=2,
          width=3]},line width=0.4pt] (0,0) to (1,0);
\end{tikzpicture} \  & -Z_{\gamma_{2}}(u)+  Z_{\gamma_{3}}(u), \\ \nonumber
 \text{in terms of charges} \ \  \  \ \gamma_{1} \  \begin{tikzpicture}
\draw[-{>[scale=2.5,
          length=2,
          width=3]},line width=0.4pt] (0,0) to (1,0);
\end{tikzpicture} \ &  -\gamma_{2}+ \gamma_{3},
 \\ \nonumber
 \\ \nonumber
 \text{and} \  \  \  \
\end{align}
\begin{align} \label{decaypathway2}
\text{decay pathway center right chamber E} \  \  \  \ Z_{\gamma_{2}}(u) \ \ \begin{tikzpicture}
\draw[-{>[scale=2.5,
          length=2,
          width=3]},line width=0.4pt] (0,0) to (1,0);
\end{tikzpicture} & \  -Z_{\gamma_{1}}(u)+ Z_{\gamma_{3}}(u), \\ \nonumber
\  \  \  \ \gamma_{2} \ \ \begin{tikzpicture}
\draw[-{>[scale=2.5,
          length=2,
          width=3]},line width=0.4pt] (0,0) to (1,0);
\end{tikzpicture} \ \  & -\gamma_{1}+ \gamma_{3},
\end{align}
These split attractor flow processes (\ref{decaypathway1}-\ref{decaypathway2}) can be seen in the right and left central chambers, D, E, of Fig. \ref{fig:Z1flow1} above respectively. The Fig. \ref{fig:Z1flow2} below shows a zoomed-in version of these chambers:

\begin{center}

\begin{table}[h!]
    \centering
    
 \begin{tabular}{ | l | l | l | p{5cm} |}
		\hline
	\multicolumn{3}{|c|}{Existing BPS states} \\
	\hline
		Flow line & \multicolumn{2}{|c|}{\begin{tikzpicture}
\text{The blue line}
    \draw [red,thick,-{Stealth}] (0,-1)--(1,-1); 
\end{tikzpicture} }   \\ \hline
		Charges & \multicolumn{2}{|c|}{$\gamma_{1}$}  \\ \hline
		Split flow lines & \begin{tikzpicture}
\text{The blue line}
    \draw [blue,thick,-{Stealth}] (0,-1)--(1,-1); 
\end{tikzpicture}  & \begin{tikzpicture}
\text{The blue line}
    \draw [green,thick,-{Stealth}] (0,-1)--(1,-1); 
\end{tikzpicture}      \\ \hline
		Charges & $-\gamma_{2}$ & $\gamma_{1}+\gamma_{2}$                            \\ \hline
		\multicolumn{3}{|c|}{Non-existing BPS states}  \\ \hline
	Flow line	&   \multicolumn{2}{|c|}{\begin{tikzpicture}
\text{The blue line}
    \draw [gray,thick,-{Stealth}] (0,-1)--(1,-1); 
\end{tikzpicture} }      \\ \hline
 Charges      &    \multicolumn{2}{|c|}{$ \  \  \ \gamma_{1}  \  \  \ \  \ \ $ above cut}            \\ \cline{2-3}
     &   \multicolumn{2}{|c|}{$\gamma_{1}-2\gamma_{2} \ $ below cut}   \\ \hline
	\end{tabular}   
 
 \caption{Split flow lines of $\gamma_{1}$ on Fig. \ref{fig:Z1flow2}.}

\end{table}

\begin{figure}
	\centering
	{\includegraphics[width=0.7\textwidth]{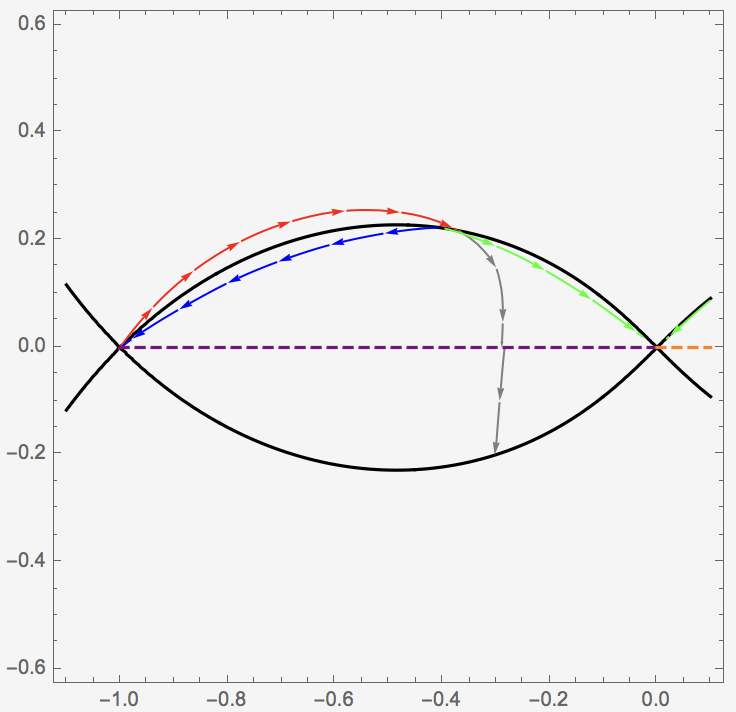}}

	\vspace{-12ex}
	\begin{center}
	   C   $\   \  \ \  \  \  \  \ $ 
	\end{center}
	\vspace{17ex}
	\vspace{-75ex}
	\begin{center}
	B  $\   \  \ \  \  \  \  \ $  
	\end{center}
	\vspace{70ex}

\vspace{-57ex}
	\begin{center}
	D1   $\   \  \ \  \  \  \  \ $ 
	\end{center}
	\vspace{40ex}

   \vspace{-37ex}
	\begin{center}
	D2   $\   \  \ \  \  \  \  \ $ 
	\end{center}
	\vspace{27ex}

	\captionsetup{singlelinecheck=off}
	\caption[flowlinecloseup]{Zoomed version of left central chamber D. 
	
	\begin{enumerate}[label = \alph*.)]
	\item
	This diagram showing split flow corresponding to $Z_{\gamma_{1}}(u)$ splitting into $-Z_{\gamma_{2}}(u)+  Z_{\gamma_{3}}(u)$, where $Z_{\gamma_{2}}(u)$ in blue exists in the chamber and flows to $u=-1$. Similarly $Z_{\gamma_{3}}(u)$ in green exists by flowing to $u \rightarrow \infty$. 
	\item
	The gray line shows the flow line of $Z_{\gamma_{1}}(u)$ in D1 continued into its non-existing chamber D2. 
	This gray flow line is analytically continued through the branch cut by mapping $Z_{\gamma_{1}}(u) \longmapsto Z_{\gamma_{1}}(u) - 2Z_{\gamma_{2}}(u) $, which is represented by the dashed line. 
	\item
	This flow then crashes at a regular point on the lower wall, hence excluding the state. 
	\end{enumerate}
	}

	\label{fig:Z1flow2}
\end{figure}

\end{center}

\subsubsection*{Exclusion of general $m \gamma_{1}+n \gamma_{2}$} \label{section:generallinearcombinations1}

 Now we consider again, as in \ref{higherlinearfirstrealisation}, the general states  of the form 
\begin{align} \label{states2}
m \gamma_{1}+n \gamma_{2}\ \ \text{for} \ \ n > 1, \ m \neq 0,1 \ \ \text{or} \ \  m > 1, \ n \neq 0, 1.
\end{align}
 
Such states in (\ref{states2}) are again found not to exist. They flow to a regular point, analogous to a solution of (\ref{eq:higherfirstrealisation}), but for this realisation of the curve. The equation for this point is again 
\begin{align} 
mZ_{\gamma_{1}}(u)+nZ_{\gamma_{2}}(u)= 0.
\end{align}

To proceed, we again consider the alignment of the central charges along all segments of the wall and determine the sign of the ratio of the central charges, as well as its range of values between the singular points. As with the first parameterisation we tabulate these data in a table (\ref{tableratio}) below. We find that the alignment reverses discontinuously at the point at infinity. In particular, the alignment of $Z_{\gamma_{2}}(u)$ reverses as the wall passes through the point at infinity.

The final table (\ref{tableratio}) shows the ratios and the normalised central charges along the path on the wall: \footnote{The double tabulated singular points represent the different paths between the singular points along different arcs on the wall to show the range of the ratios along these arcs.} 
 \begin{align}
1 \longrightarrow -1 \longrightarrow \infty \longrightarrow  1 \longrightarrow -1 \longrightarrow \infty.
\end{align}
\begin{table}[h!]
    
  \begin{center}
 	\begin{tabular}{ | l | l | l | l | p{2.5cm} |} 
 		\hline
 		 \multicolumn{5}{|c|}{\vphantom{\LARGE{H}} ratios along path}       
 		 \\[4pt] \hline
 		paths between singular points & 
 		$Z_{\gamma_{1}}(u)$ & $Z_{\gamma_{2}}(u)$ & $  Z_{\gamma_{1}}(u)/Z_{\gamma_{2}}(u) $ &  $Z_{\gamma_{2}}(u)/Z_{\gamma_{1}}(u)$ \\ \hline
 		$1$ & $  0 $ & $\infty$ & $0$ &  $+\infty$ \\ \hline
 		$-1$ & $\infty$ & $ 0 $  & $+\infty$ &  $0$ \\ \hline
 		$\infty$ &  $ \infty$ &  $\infty$ &  $\pm 1$ &  $\pm1$ \\ \hline
 		$1$ & $0$ & $\infty$ & $0$ &  $-\infty$ \\ \hline
 		$-1$ & $\infty$ & $ 0$  & $-\infty$ &  $0$ \\ \hline
 		$\infty$ &  $\infty$ &  $\infty$ &  $\mp 1$ &  $\mp 1$ \\ \hline
 	\end{tabular}
 \end{center}

    \caption{Central charges and their ratio at the singular points.}
    \label{tableratio}
\end{table}
\begin{align*}
    \text{\textit{ Exclusion of combinations with $ \ \frac{m}{n}>1$}}
\end{align*}
From this table (\ref{tableratio}) above it can be seen that, for $Z_{\gamma_{2}}(u)/Z_{\gamma_{1}}(u)$, the arc from $[\infty,1]$ in the $u$-plane gives a range of
$Z_{\gamma_{2}}(u)/Z_{\gamma_{1}}(u)$ from $[-1,-\infty]$. Given that this ratio is a continuous analytic function, it will take any value in this range. Hence the equation 
\begin{align} \label{eq:highercombinationscentralcharges}
m Z_{\gamma_{1}}(u)+n Z_{\gamma_{2}}(u) =0 \longrightarrow \frac{Z_{\gamma_{2}}(u)}{Z_{\gamma_{1}}(u)} = -\frac{m}{n},
\end{align}
will have a solution along the $[\infty,1]$ segment corresponding to an attractor point of vanishing central charge at a regular point in the moduli space - meaning such a $(m,n), \frac{m}{n}>1$ BPS state doesn't exist. 
\begin{align*}
    \text{\textit{ Exclusion of combinations with $ \ \frac{n}{m}>1$}}
\end{align*}
Furthermore, we can see that also for $Z_{\gamma_{2}}(u)/Z_{\gamma_{1}}(u)$, the arc from $[-1, \infty]$ in the $u$-plane gives a range of $Z_{\gamma_{1}}(u)/Z_{\gamma_{2}}(u)$ from $[-\infty, -1]$. Again, because of the continuity and analyticity of the ratio, the equation (\ref{eq:highercombinationscentralcharges})
must also have a solution in the range $[-1, \infty]$ by the same argument meaning BPS states of the form $(m,n), \frac{n}{m}>1$ also don't exist. Hence $ \forall n, m \neq 0, \frac{n}{m} = 1 $. \footnote{Note that $\frac{Z_{\gamma_{2}}(u)}{Z_{\gamma_{1}}(u)} \in [- \infty, 0 ]$ along the lower $[1, -1]$ segment. However, this doesn't exclude any linear combination of BPS states as they pass through the wall or a branch cut before flowing to a termination point on the lower segment.}
\\
\subsubsection*{Flow of higher linear combinations through the branch cuts}
\begin{table}[h!]
 \begin{center}
\begin{tabular}{ | l | l | l | l | l | l | l | p{5cm} |}
		\hline
	\multicolumn{7}{|c|}{Continuation of $Z_{\gamma_{1}}(u)$ flow} \\
	\hline
		Flow line & \begin{tikzpicture}
\text{The blue line}
    \draw [red,thick,dashed,-{Stealth}] (0,-1)--(1,-1); 
\end{tikzpicture}   & \begin{tikzpicture}
\text{The blue line}
    \draw [gray,thick,-{Stealth}] (0,-1)--(1,-1); 
\end{tikzpicture}   & \begin{tikzpicture}
\text{The blue line}
    \draw [brown,thick,-{Stealth}] (0,-1)--(1,-1); 
\end{tikzpicture} &  \begin{tikzpicture}
\text{The blue line}
    \draw [gray,thick,dashed,-{Stealth}] (0,-1)--(1,-1); 
\end{tikzpicture} & \begin{tikzpicture}
\text{The blue line}
    \draw [black,thick,dashed,-{Stealth}] (0,-1)--(1,-1); 
\end{tikzpicture} & \begin{tikzpicture}
\text{The blue line}
    \draw [darkgray,thick,dashed,-{Stealth}] (0,0)--(1,0); 
\end{tikzpicture} \\ \hline
		Charge & $\gamma_{1}$ & $\gamma_{1}$ & $\gamma_{1}-2\gamma_{2}$& $-2\gamma_{2}+5\gamma_{1}$ & $-2\gamma_{2}-3\gamma_{1}$ & $\gamma_{1}-4\gamma_{2}$ \\ \hline
		Cover number & 1 & 1 & 2 & 3 & 4 & 5 \\ \hline
	\end{tabular}

 \end{center}   
    
    \caption{Continuation of $Z_{\gamma_{1}}$ flow through branch cut.}
    \label{tableZ2continued} 
\end{table}There are many possible ways that the attractor flow of a general linear combination $n\gamma_{1}+m\gamma_{2}$ can flow through a branch cut.\begin{figure}[h!]
 	\centering
 	{\includegraphics[width=0.7\textwidth]{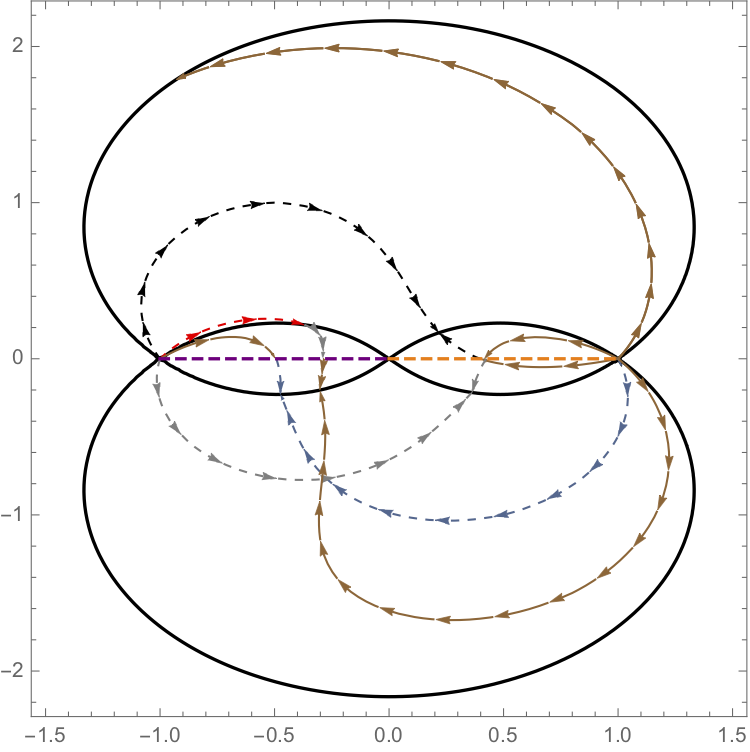}}

 	\vspace{-23ex}
	\begin{center}
	$\   \  \ \  \  \  $   C    
	\end{center}
	\vspace{31ex}
	\vspace{-78ex}
	\begin{center}
	B  $\   \  \ \  \  \  \  \ $  
	\end{center}
	\vspace{70ex}
	
	\vspace{-86ex}
	\begin{center}
	A  $\   \  \ \  \  \  \  \ \  \   \  \ \  \  \  \  \ \  \   \  \ \  \  \  \  \ \ \   \  \ \  \  \  \  \ \  \   \  \ \  \  \  \  \ \ \  \   \  \ \  \  \  \  \ \ \   \ \ \  \  \ \ \   \ \ \  $  
	\end{center}
	\vspace{64ex}	

\vspace{-49.5ex}
	\begin{center}
	D1  $\   \  \ \  \  \  \  \  \ \   \ \ \  \  \ \ \   \ \ \ \  \  \  \  \  \ \   \ \ \  \  \ \ \   \ \ \  $  
	\end{center}
	\vspace{40ex}

   \vspace{-37ex}
	\begin{center}
	D2  $\   \  \ \  \  \  \  \  \ \   \ \ \  \  \ \ \   \ \ \ \  \  \  \  \  \ \   \ \ \  \  \ \ \   \ \ \  $  
	\end{center}
	\vspace{30ex}

\vspace{-49.2ex}
	\begin{center}
	 $\   \  \ \  \  \  \  \  \ \   \ \ \  \  \ \ \   \ \ \ \  \  \  \  \  \ \   \ \ \  \  \ \ \   \ \ \  $  E1    
	\end{center}
	\vspace{40ex}

   \vspace{-36.5ex}
	\begin{center}
	 $\   \  \ \  \  \  \  \  \ \   \ \ \  \  \ \ \   \ \ \ \  \  \  \  \  \ \   \ \ \  \  \ \ \   \ \ \  $    E2    
	\end{center}
	\vspace{29ex}

 	\captionsetup{singlelinecheck=off}
 	\caption[flowline]{This diagram shows the analytic continuation of the $Z_{\gamma_{1}}(u)$, the dotted red line, in its non-existing central left region D1 on the cover on the other side of the branch cut D2. Here it becomes $Z_{\gamma_{1}}(u) - 2Z_{\gamma_{2}}(u)$, represented by the brown line. 
 	\begin{enumerate}[label = \alph*.)]
 	\item
 	This line on this cover also flows again into the branch cut from the upper half of this center left chamber D1 and becomes $-4Z_{\gamma_{2}}(u) + Z_{\gamma_{1}}(u)$ in D2 before terminating at a regular point.	The  $Z_{\gamma_{1}}(u) - 2Z_{\gamma_{2}}(u)$ line also terminates at regular points on the lower  $[-1, \infty]$ and upper $[1, -1]$ segments. 
 	
 	\item
 	It can also flow into the branch cut in the central right region E: the flow in the lower half E2 is analytically continued to  $-2Z_{\gamma_{2}}(u) - 3Z_{\gamma_{1}}(u)$ in black. Using the continuation from the upper half E1, the flow becomes $-2Z_{\gamma_{2}}(u) + 5Z_{\gamma_{1}}(u)$ below the cut in gray. In all cases the flow terminates at regular points.
   \end{enumerate}
  }
 	
 	\label{fig:5nonexflow}
 \end{figure}

It can be shown that all these flows are excluded by regular termination points after they are analytically continued through the cut, unless the states take the form $\pm \gamma_{1}, \pm \gamma_{2},\pm  \gamma_{1} \pm \gamma_{2}$. The states flow through the logarithmic branch cuts
\begin{align} 
\label{branchcuts} u^{a}\log u \  \  \  \  \  \  [-1, \infty):&  \  \  \  \begin{tikzpicture}
\text{The blue line}
  \draw [purple,thick,dashed] (0,0)--(1,0); 
\end{tikzpicture}   \  \  \  \  \  \  \      \text{and} \\ \label{branchcut2}
 (\infty, 1]:& \  \  \  \begin{tikzpicture}
\text{The blue line}
  \draw [orange,thick,dashed] (0,0)--(1,0); 
\end{tikzpicture} 
\end{align}
on Fig. \ref{fig:5nonexflow} through the 2 central chambers D and E. In one chamber, the states flow into the branch cut (\ref{branchcuts}) from both sides. These are analytically continued to states on another cover that terminate at a regular point on the segment of the wall just on the other side of the branch cut - excluding this and the initial state. In the other chamber, the flow lines flow into the branch cut (\ref{branchcuts}) in one half of the chamber where they are again analytically continued to states on a second cover that terminate at a regular point on the part of the wall bounding the other half of the chamber. \footnote{On the original cover in the lower half of this chamber the state has a flow out of the branch cut (from an analytic continuation of a state on a third cover) which then also flows to a different regular point on the wall segment bounding the lower half. This is the same attractor point that the lines in the lower large chamber flow to and are excluded by.}

\subsubsection*{Description of each flow line}

In Fig. \ref{fig:5nonexflow} above and Fig. \ref{fig:5nonexflow2} we give an example initially of $Z_{\gamma_{1}}(u)$ represented by a dashed red line
\begin{tikzpicture}
\text{The blue line}
    \draw [red,thick,dashed,-{Stealth}] (0,0)--(1,0); 
\end{tikzpicture}
 and the gray line \begin{tikzpicture}
\text{The blue line}
    \draw [gray,thick,-{Stealth}] (0,0)--(1,0); 
\end{tikzpicture} after it crosses the wall of D1:
in this case we analytically continue the flow of $Z_{\gamma_{1}}(u)$, shown in the previous diagrams, and table (\ref{tableZ2continued}), onto the cover it flows to when passing through the branch cut (\ref{branchcuts}) between D1 and D2. As mentioned before in (\ref{monodromyaction1}), when continuing through the branch cut we act with the monodromy $(M_{-1})^{-1}$ (from (\ref{AD2secondmonodromies})) in a clockwise direction and obtain 
\begin{align} \label{Z2analyticcontinuation}
(M_{-1})^{-1}: Z_{\gamma_{1}}(u) \longmapsto Z_{\gamma_{1}}(u) - 2Z_{\gamma_{2}}(u).
\end{align}
This flow then emerges from the branch cut (\ref{branchcuts}) in D2 as the brown line \begin{tikzpicture}
\text{The blue line}
    \draw [brown,thick,-{Stealth}] (0,0)--(1,0); 
\end{tikzpicture} and flows to a regular point on the segment bounding the lower half of the center left chamber D2 between $[-1, \infty]$. The flow of $Z_{\gamma_{1}}(u) - 2Z_{\gamma_{2}}(u)$ is represented by the brown line in the rest of the figure. One can see that in the large lower chamber C the state also flows to this attractor point on the lower wall of the center left chamber and is thus excluded. In the outer region A and large upper chamber B, the state flows to a regular point on the upper segment of chamber B between $[1,-1]$.
\\
\\
We now consider the various ways the state $Z_{\gamma_{1}}(u) - 2Z_{\gamma_{2}}(u)$ represented by \begin{tikzpicture}
\text{The blue line}
    \draw [brown,thick,-{Stealth}] (0,0)--(1,0); 
\end{tikzpicture} can flow through the logarithmic branch cuts between $[-1, \infty)$ and $(\infty, 1]$: 

\begin{enumerate}[label= (\roman*)]

\item 

We have already described how the flow emerges from the branch cut in the lower half of the central left chamber D2 from the
 $Z_{\gamma_{1}}(u)$ state on a second cover. However, on the same cover as the emerging flow in the lower half of the chamber, in the upper half of the chamber D1, $Z_{\gamma_{1}}(u) - 2Z_{\gamma_{2}}(u)$ from (\ref{Z2analyticcontinuation}) flows into the branch cut (\ref{branchcuts}). Here we again act with $(M_{-1})^{-1}$ in a clockwise direction, such that 
 \begin{align}
(M_{-1})^{-1}: 
Z_{\gamma_{1}}(u) - 2Z_{\gamma_{2}}(u) \longmapsto (Z_{\gamma_{1}}(u)-2Z_{\gamma_{2}}(u)) - 2Z_{\gamma_{2}}(u) \longmapsto Z_{\gamma_{1}}(u) - 4Z_{\gamma_{2}}(u).
\end{align}
 This state, represented by the dotted dark gray line \begin{tikzpicture}
\text{The blue line}
    \draw [darkgray,thick,dashed,-{Stealth}] (0,0)--(1,0); 
\end{tikzpicture}, then emerges from the cut in the lower half of the chamber, D2 again but on a new cover, on which it also terminates at a regular point (on the left of the previous attractor point) on the lower segment of the wall bounding the half chamber D2, and is hereby excluded.

 \item
 
 The state  $Z_{\gamma_{1}}(u) - 2Z_{\gamma_{2}}(u)$ from (\ref{Z2analyticcontinuation}) also flows into the logarithmic branch cut in the center right chamber E from $(\infty, 1]$ from both above and below on the same cover. When it flows up into the branch cut (\ref{branchcut2}) from the lower half of the chamber E2 one acts with $(M_{+1})^{-1}$ and 
 \begin{align} \label{furtheranalyticcontinuation}
(M_{+1})^{-1}: Z_{\gamma_{1}}(u) - 2Z_{\gamma_{2}}(u)\longmapsto Z_{\gamma_{1}}(u) - 2( Z_{\gamma_{2}}(u)+2Z_{\gamma_{1}}(u)) \longmapsto - 2Z_{\gamma_{2}}(u)-3Z_{\gamma_{1}}(u),
\end{align}
 which emerges in E1 as the dashed black line \begin{tikzpicture}
\text{The blue line}
    \draw [black,thick,dashed,-{Stealth}] (0,0)--(1,0); 
\end{tikzpicture} on a new cover in the Figures \ref{fig:5nonexflow}, \ref{fig:5nonexflow2} below. This flow then terminates on a regular point on the upper segment bounding the upper half of E1, and is excluded. 

\item 

In the upper half of the center right chamber E1 on the initial cover,  $Z_{\gamma_{1}}(u) - 2Z_{\gamma_{2}}(u)$ flows downwards into the branch cut (\ref{branchcut2}). This time we act with $M_{+1}$ in a counterclockwise direction and the state becomes:
\begin{align}
M_{+1}: Z_{\gamma_{1}}(u) - 2(Z_{\gamma_{2}}(u)-2Z_{\gamma_{1}}(u)) \longmapsto -2Z_{\gamma_{2}}(u) + 5Z_{\gamma_{1}}(u).
\end{align}
This flow, represented by a dashed grey line \begin{tikzpicture}
\text{The blue line}
    \draw [gray,thick,dashed,-{Stealth}] (0,0)--(1,0); 
\end{tikzpicture}, then emerges from the branch cut on another cover in E2. Again, the line terminates at a regular point - this time on the lower segment of the wall bounding the lower half of E2, and the state is once again excluded.

\end{enumerate}

 \subsubsection*{Summary}
 
\begin{figure}[h!]
 	\centering
 	{\includegraphics[width=0.7\textwidth]{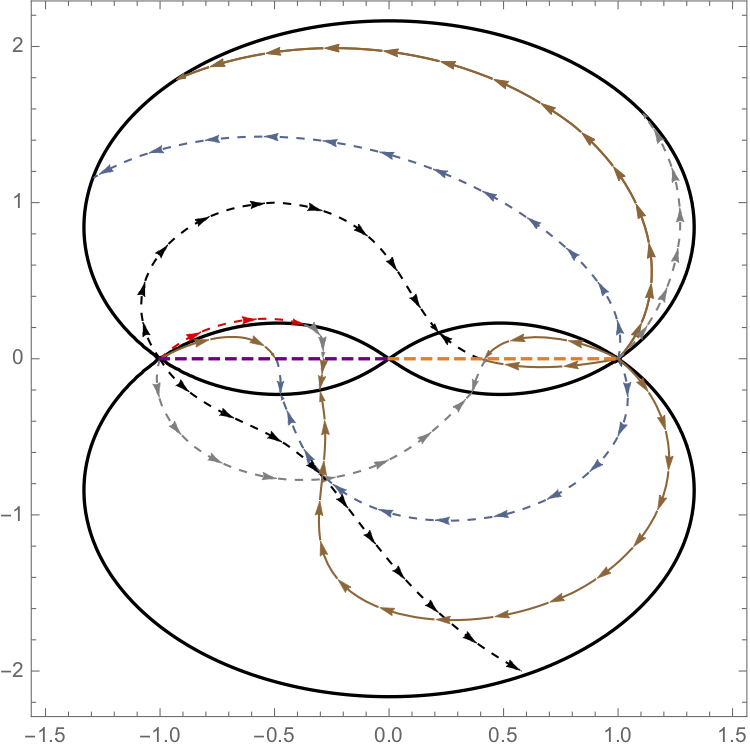}}

 \vspace{-23ex}
	\begin{center}
	$\   \  \ \  \  \  \   \  \ \  \  \  \  \ \  \   \  \ \  \  \  \  \ \  \   \  \ \  \  \    $   C    
	\end{center}
	\vspace{31ex}
	\vspace{-78ex}
	\begin{center}
	 $\   \  \ \  \  \  \  \  \   \  \ \  \  \  \  \ \  \   $   B   
	\end{center}
	\vspace{70ex}
	
	\vspace{-86ex}
	\begin{center}
	A  $\   \  \ \  \  \  \  \ \  \   \  \ \  \  \  \  \ \  \   \  \ \  \  \  \  \ \ \   \  \ \  \  \  \  \ \  \   \  \ \  \  \  \  \ \ \  \   \  \ \  \  \  \  \ \ \   \ \ \  \  \ \ \   \ \ \  $  
	\end{center}
	\vspace{64ex}	

\vspace{-49.5ex}
	\begin{center}
	D1  $\   \  \ \  \  \  \  \  \ \   \ \ \  \  \ \ \   \ \ \ \  \  \  \  \  \ \   \ \ \  \  \ \ \   \ \ \  $  
	\end{center}
	\vspace{40ex}

   \vspace{-35ex}
	\begin{center}
	D2  $\   \  \ \  \  \  \  \  \ \   \ \ \  \  \ \ \   \ \ \ \  \  \  \  \  \ \   \ \ \  \  \ \ \   \ \ \  $  
	\end{center}
	\vspace{30ex}

\vspace{-51.2ex}
	\begin{center}
	 $\   \  \ \  \  \  \  \  \ \   \ \ \  \  \ \ \   \ \ \ \  \  \  \  \  \ \   \ \ \  \  \ \ \   \ \ \  $  E1    
	\end{center}
	\vspace{40ex}

   \vspace{-36.5ex}
	\begin{center}
	 $\   \  \ \  \  \  \  \  \ \   \ \ \  \  \ \ \   \ \ \ \  \  \  \  \  \ \   \ \ \  \  \ \ \   \ \ \  $    E2    
	\end{center}
	\vspace{29ex}

 	\caption{This diagram shows the continuation of the higher combinations on the other side of the branch cuts to their second termination points on that cover: $-4Z_{\gamma_{1}}(u) + Z_{\gamma_{2}}(u)$ is the light blue line,  $-2Z_{\gamma_{1}}(u) - 3Z_{\gamma_{2}}(u)$ in black, and $-2Z_{\gamma_{1}}(u) + 5Z_{\gamma_{2}}(u)$ in grey. In all cases the flow terminates at a regular point on a segment bounding one of the inner chambers D,E as well as one on the segment bounding the large outer chamber A opposite to the first regular point - hence such states are excluded. }
 	
 	\label{fig:5nonexflow2}
 \end{figure}

 Hence we have determined that the state $Z_{\gamma_{1}}(u) - 2Z_{\gamma_{2}}(u)$ is always excluded because all possible flows in all possible regions of the moduli space end at a regular point, including all possible flows through branch cuts. This means this state can never exist as part of the BPS spectrum. As shown in the diagram below (Fig. \ref{fig:5nonexflow2}) this can be successively continued to other combinations, such as $ - 2Z_{\gamma_{2}}(u)-3Z_{\gamma_{1}}(u) $ from (\ref{furtheranalyticcontinuation}), that are also excluded as BPS states (Fig. \ref{fig:5nonexflow2} below shows that, like $Z_{\gamma_{1}}(u) - 2Z_{\gamma_{2}}(u)$, this state has a second regular termination point on the outer arc). If one also considers analytic continuations of the non-existing $Z_{\gamma_{1}}(u)+Z_{\gamma_{2}}(u)$ flow, this process of flowing through the cuts can continue until all linear combinations except for the charges  $\pm\gamma_{1}, \pm\gamma_{2},\pm  \gamma_{1} \pm \gamma_{2}$ are excluded.

\subsubsection*{Final existing states in each chamber}
 
Therefore we now know the combination of states that exist in each region of the moduli space. We find 3 BPS states existing in the 2 chambers below the outer arc and 2 BPS states existing in the remaining 3 chambers. This is as we expect from the literature e.g. in Shapere and Vafa \cite{Shapere:1999xr}. Each cycle $\gamma_{1}, \gamma_{2}, \gamma_{3}$ exists in 4 out of 5 regions in the moduli space. The exact description of the BPS existence in the moduli space is given in the table (\ref{existingchargestable}) below.\begin{table}[h!]
    
 \begin{center} 
	\begin{tabular}{ | l | l | l | l | p{5cm} |}
		\hline
		Chamber & Existing charges   & Count \\ \hline
		D: Central left & $\gamma_{2} , \gamma_{3}$ & 2 \\ \hline
		E: Central right& $\gamma_{1}, \gamma_{3}$ &2 \\ \hline
		B: Upper arc &  $\gamma_{1}, \gamma_{2}, \gamma_{3}$ & 3\\
		\hline C: Lower arc &  $\gamma_{1}, \gamma_{2}, \gamma_{3}$ &3 \\
		\hline
		A: Outside wall &  $\gamma_{1}, \gamma_{2}$ & 2\\
		\hline
	\end{tabular}
\end{center}   
\caption{Existing BPS states in each chamber labelled by $\gamma_{1},\gamma_{2}$ and $\gamma_{3}$.}
    \label{existingchargestable}
\end{table} \\ Now we take into account the branch cuts on the diagram and remember that in the region on the diagram below the 2 branch cuts (\ref{branchcuts}, \ref{branchcut2}) but still above the outer lower arc \footnote{This contains the lower part of the central 2 chambers D2, E2 and the full lower chamber C above the lower arc.} the central charge of the sum $Z_{\gamma_{1}}(u)+Z_{\gamma_{2}}(u)$ becomes either 
$Z_{\gamma_{1}}(u)-Z_{\gamma_{2}}(u)$ or $-Z_{\gamma_{1}}(u)+Z_{\gamma_{2}}(u)$ depending on which branch cut the analytic continuation is done through (\ref{sumanalyticcontinuation}) and hence which of the two possible covers one considers for the central charge. The complete tabulation is shown in the table (\ref{completetabulationfirst}).

\begin{table}[h!]
    
 \begin{center} 
	\begin{tabular}{ | l | l | l | l | p{5cm} |}
		\hline
		Chamber & Existing charges cover 1  & Existing charges cover 2    & Count \\ \hline
		D1: Central left upper half & $\gamma_{2} , \ \ \gamma_{1}+\gamma_{2}$ & $\gamma_{2} , \ \ \gamma_{1}+\gamma_{2}$ & 2 \\ \hline
		D2: Central left lower half& $\gamma_{2} , \ \ -\gamma_{1}+\gamma_{2} $ & $\gamma_{2} , \ \  \gamma_{1}-\gamma_{2}  $  & 2 \\ \hline
		E1: Central right upper half & $\gamma_{1},  \ \ \gamma_{1}+\gamma_{2}$ & $\gamma_{1},  \ \ \gamma_{1}+\gamma_{2}$ &2 \\ \hline
		E2: Central right lower half& $\gamma_{1},\ \ \gamma_{2}-\gamma_{1}$ & $\gamma_{1}, \ \ \gamma_{1}-\gamma_{2}$ &2 \\ \hline
		B: Upper arc &  $\ \ \gamma_{1}, \ \  \gamma_{2}, \ \ \gamma_{1}+\gamma_{2}$ &$\ \ \gamma_{1}, \ \  \gamma_{2}, \ \ \gamma_{1}+\gamma_{2}$& 3\\
		\hline C: Lower arc &  $ \ \ \gamma_{1}, \ \ \gamma_{2}, \ \ \gamma_{2}-\gamma_{1} $ & $\ \ \gamma_{1}, \ \ \gamma_{2}, \ \ \gamma_{1}-\gamma_{2}$  &3 \\
		\hline
		A: Outside wall &  $\ \ \gamma_{1}, \ \ \gamma_{2}$ &$\ \ \gamma_{1}, \ \ \gamma_{2}$& 2\\
		\hline
	\end{tabular}
\end{center}   
    
    \caption{Existing BPS states, this time distinguishing $\gamma_{3}=\gamma_{1}+\gamma_{2}$ from $\gamma_{3}=\gamma_{2}-\gamma_{1}$. }
    \label{completetabulationfirst}
\end{table}

\newpage

\begin{align*}
\\
\end{align*}

\subsection{Attractor flow in Seiberg-Witten SU(2)}
  
We also carry out the analysis of deriving the Picard-Fuchs equation, the solutions and the BPS central charges for Seiberg-Witten $SU(2)$ \cite{SW}.  As for the Argyres--Douglas theory, we determine the attractor flow and use it to reproduce the spectrum of BPS states in each chamber. We hereby reproduce the BPS spectrum of the quiver theory \cite{Alim:2011kw} with just 2 basis states $\gamma_{1}, \gamma_{2}$ in one chamber, and infinitely many in the other chamber, with charges of the form $n\gamma_{1}+(n+1)\gamma_{2}$. We use a similar set of steps as for the $A_{2}$ theory.\\ 
\\
The central charges (\ref{swsZ1}-\ref{swsZ2}) chosen from the solutions of the Picard-Fuchs in this section are: \footnote{This is modified up to normalisation from the central charges in (\ref{swcharge1}-\ref{swcharge2}).}
 \begin{align} 
 Z_{\gamma_{1}}(u)= \label{swsZ1}
 & -\frac{-i \sqrt{\pi} \Gamma(-\frac{1}{4})}{8 \Gamma(\frac{5}{4})}  F_{2}^{1}(-\frac{1}{4},-\frac{1}{4},\frac{1}{2}, u^{2})\\ \nonumber & - \frac{i \sqrt{\pi} \Gamma(\frac{5}{4})}{\Gamma(\frac{3}{4})} u F_{2}^{1}(\frac{1}{4},\frac{1}{4},\frac{3}{2}, u^{2}), \  \  \  \  \  \ u \in \mathbbm{P}^1\setminus  [-1, \infty),  \\ \nonumber
 \\
 Z_{\gamma_{2}}(u) = \label{swsZ2}
&  -\frac{2^{\frac{1}{2}}}{ \pi^{\frac{1}{2}}} \Gamma(\frac{3}{4})^{2} F_{2}^{1}(-\frac{1}{4},-\frac{1}{4},\frac{1}{2}, u^{2})\\ \nonumber
 & -u \frac{\Gamma(\frac{1}{4})^{2}}{4 \sqrt{2\pi}}    F_{2}^{1}(\frac{1}{4},\frac{1}{4},\frac{3}{2}, u^{2}), \  \  \  \  \  \  \  \  \  \   u \in \mathbbm{P}^1\setminus  [1, \infty),  \\ \nonumber \end{align} 
on a particular cover. There are $u^{c} \log u$ branch cuts at $[-1, \infty)$ and $[1, \infty)$ that are represented by 
\begin{tikzpicture}
\text{The blue line}
    \draw [purple,thick,dashed] (0,0)--(1,0); 
\end{tikzpicture}
and \begin{tikzpicture}
\text{The blue line}
    \draw [orange,thick,dashed] (0,0)--(1,0); 
\end{tikzpicture}
on Fig. \ref{SWSU2sample}.

\subsubsection{Spectrum from attractor flow}

This behaves very similarly to the $A_{2}$ case in the new parameterisation, looking at the plot below (Fig. \ref{SWSU2sample}), the behavior is almost identical. However, (e.g. from the quiver theory) we expect the spectrum to contain infinite BPS states at the chamber B around infinity. To verify this, we first  consider the monodromies from (\ref{SWmonodromies}) and transformations of the central charges around the singular points and branch cuts ending there. \\
\\
We recall from (\ref{SWmonodromies}) that at $-1$ the transformations are: \footnote{The sign on the $\pm 2$ is determined by the direction of the loop taken around the singular point.}  
\begin{align} \label{SWmonodromies2}
& \text{\underline{Monodromy transformations}} \\ \nonumber \\ \nonumber
  & 
\begin{pmatrix}
1 & \pm 2\\
0 &  1
\end{pmatrix}
\begin{pmatrix}
Z_{\gamma_{1}}(u) \\ Z_{\gamma_{2}}(u)
\end{pmatrix},   
& \\ \nonumber 
 & \text{such that $Z_{\gamma_{2}}(u) \longmapsto Z_{\gamma_{2}}(u)$ and $Z_{\gamma_{1}}(u) \longmapsto Z_{\gamma_{1}}(u)+2Z_{\gamma_{2}}(u)$.} 
 &\\ \nonumber
 &\\ \nonumber 
 &\text{At $+1$ we have:} 
 &\\   
 & 
\begin{pmatrix}
1 &  0 \\
\mp 2 &  1
\end{pmatrix}
\begin{pmatrix}
Z_{\gamma_{1}}(u) \\ Z_{\gamma_{2}}(u)
\end{pmatrix},
& \\ \nonumber
& \text{such that the transformations can be written as:} 
&\\ \nonumber
& Z_{\gamma_{1}}(u)\longmapsto Z_{\gamma_{1}}(u), \ \ \text{and} \ \  Z_{\gamma_{2}}(u) \longmapsto Z_{\gamma_{2}}(u)-2Z_{\gamma_{1}}(u). 
\end{align}
Now consider the attractor flow: combinations of the form $\pm m Z_{\gamma_{1}}(u)\pm nZ_{\gamma_{2}}(u)$ always flow to a point on the wall for $n,m \geq 1$. For $\frac{n}{m} \geq 0$ the flow terminates on the lower arc and for  $\frac{n}{m} \leq 0$ on the upper arc. 
To avoid all states being excluded from existence by flowing to the regular point, we consider the flow entering the branch cuts $[-1, \infty)$  and $(\infty, 1]$ between B1 and B2. At the branch cuts we can combine the flows by acting with the transformations (\ref{SWmonodromies}).
These must have flows continuous with that on the other side of the branch cut and cannot terminate at a regular point if the original state exists. This can only happen if the ratio changes sign from $\frac{n}{m} \geq 0$ to  $\frac{n}{m} \leq 0$ or vice versa.

\subsubsection*{Infinite tower of existing BPS states}

Hence we can use the monodromies in (\ref{SWmonodromies}) to generate the set of all existing states not excluded by the regular points by acting with the monodromy transformations that reverse this ratio. Initially we act in a similar way to the $A_{2}$ case when considering a rotation (starting in the upper half plane) of the form $u= -1 + \epsilon e^{-i \theta}, \ \theta: 0 \rightarrow +2\pi$ around $-1$ and $u= +1 + \epsilon e^{-i \theta}, \ \theta: -\pi \rightarrow +\pi$ around $+1$, where $\epsilon \in \mathbb{R}^{+}$.\\
\\
Therefore the transformations become $+1$: $Z_{\gamma_{2}}(u) \longmapsto Z_{\gamma_{2}}(u)-2Z_{\gamma_{1}}(u) $ and $-1$: $Z_{\gamma_{1}}(u) \longmapsto Z_{\gamma_{1}}(u)+2Z_{\gamma_{2}}(u) $, a change of sign happening because we are considering a rotation from the lower half plane. Knowing that the basis states $Z_{\gamma_{1}}(u), Z_{\gamma_{2}}(u)$ exist in the chamber B around infinity, we can consider the transformations that generate the full spectrum by acting with these monodromies:

\begin{align}
& \text{\underline{First examples from first basis state}} \\ \nonumber \\ \nonumber
+1:& \ Z_{\gamma_{2}}(u) \longmapsto Z_{\gamma_{2}}(u)-2Z_{\gamma_{1}}(u), \\ \nonumber
-1:& \ Z_{\gamma_{2}}(u)-2Z_{\gamma_{1}}(u) \longmapsto Z_{\gamma_{2}}(u)-2(Z_{\gamma_{1}}(u)+2Z_{\gamma_{2}}(u))  \longmapsto -3Z_{\gamma_{2}}(u)-2Z_{\gamma_{1}}(u), \\ \nonumber
+1:& \ -3Z_{\gamma_{2}}(u)-2 Z_{\gamma_{1}}(u) \longmapsto -3(Z_{\gamma_{2}}(u)-2Z_{\gamma_{1}}(u))-2Z_{\gamma_{1}}(u)    \longmapsto -3Z_{\gamma_{2}}(u)+4Z_{\gamma_{1}}(u),\\ \nonumber
-1:& \ -3Z_{\gamma_{2}}(u)+4Z_{\gamma_{1}}(u)   \longmapsto -3Z_{\gamma_{2}}(u)+4(Z_{\gamma_{1}}(u)+2 Z_{\gamma_{2}}(u))\longmapsto 5Z_{\gamma_{2}}(u)+4Z_{\gamma_{1}}(u), \\ \nonumber
+1:& \ 5Z_{\gamma_{2}}(u)+4Z_{\gamma_{1}}(u)  \longmapsto 5(Z_{\gamma_{2}}(u)-2Z_{\gamma_{1}}(u))+4Z_{\gamma_{1}}(u) \longmapsto 5Z_{\gamma_{2}}(u)-6Z_{\gamma_{1}}(u), \ \ \ ...   \\ \nonumber \\ \nonumber   & \text{\underline{In general}} \\ \nonumber \\ \nonumber &
(n+1)Z_{\gamma_{2}}(u)+n Z_{\gamma_{1}}(u)  \longmapsto (n+1)(Z_{\gamma_{2}}(u)-2Z_{\gamma_{1}}(u))+nZ_{\gamma_{1}}(u)   \longmapsto           \\ \nonumber  & (n+1)Z_{\gamma_{2}}(u)-(n+2)Z_{\gamma_{1}}(u) \ \ m=n+1     \longmapsto    mZ_{\gamma_{2}}(u)-(m+1)Z_{\gamma_{1}}(u),  \ \ \ ... 
\end{align}

\begin{align}
& \text{\underline{First examples from second basis state}}
\\ \nonumber
\\ \nonumber
-1:& \ Z_{\gamma_{1}}(u) \longmapsto Z_{\gamma_{1}}(u)-2Z_{\gamma_{2}}(u), \\ \nonumber
+1:& \ Z_{\gamma_{1}}(u)-2Z_{\gamma_{2}}(u) \longmapsto Z_{\gamma_{1}}(u)-2(Z_{\gamma_{2}}(u)+2Z_{\gamma_{1}}(u))  \longmapsto -3Z_{\gamma_{1}}(u)-2Z_{\gamma_{2}}(u), \\ \nonumber
-1:& \ -3Z_{\gamma_{1}}(u)-2Z_{\gamma_{2}}(u)  \longmapsto -3(Z_{\gamma_{1}}(u)-2Z_{\gamma_{2}}(u))-2Z_{\gamma_{2}}(u) \longmapsto -3Z_{\gamma_{1}}(u)+4Z_{\gamma_{2}}(u), \\ \nonumber
+1:& \ -3Z_{\gamma_{1}}(u)+4Z_{\gamma_{2}}(u)   \longmapsto -3Z_{\gamma_{1}}(u)+4(Z_{\gamma_{2}}(u)+2Z_{\gamma_{1}}(u)) \longmapsto 5Z_{\gamma_{1}}(u)+4Z_{\gamma_{2}}(u), \\ \nonumber
-1:& \ 5Z_{\gamma_{1}}(u)+4Z_{\gamma_{2}}(u)  \longmapsto 5(Z_{\gamma_{1}}(u)-2Z_{\gamma_{2}}(u))+4Z_{\gamma_{2}}(u) \longmapsto 5Z_{\gamma_{1}}(u)-6Z_{\gamma_{2}}(u), \ \ \ ... \\ \nonumber
\\ \nonumber  &  \text{\underline{In general}} \\ \nonumber \\ \nonumber
&(n+1)Z_{\gamma_{1}}(u)+nZ_{\gamma_{2}}(u)   \longmapsto (n+1)(Z_{\gamma_{1}}(u)-2Z_{\gamma_{2}}(u))+nZ_{\gamma_{2}}(u)  \longmapsto \\ \nonumber &    (n+1)Z_{\gamma_{1}}(u)-(n+2)Z_{\gamma_{2}}(u) ,\ \ m=n+1 \longmapsto mZ_{\gamma_{1}}(u)-(m+1)Z_{\gamma_{2}}(u)   \ \ \ ... 
\end{align}
These are the existing states in the model that are not excluded by termination at a regular point. One obtains the same pattern starting with $Z_{\gamma_{2}}(u)+2Z_{\gamma_{1}}(u)$. Hence all the combinations are of the form: $nZ_{\gamma_{1}}(u) \pm (n+1)Z_{\gamma_{2}}(u)$ and $nZ_{\gamma_{2}}(u) \pm (n+1)Z_{\gamma_{1}}(u)$, which was previously expected. We also expect another state to exist in chamber B around infinity. This corresponds to a W-boson. In our basis its central charge is the sum of the central charges $Z_{\gamma_{1}}(u)+Z_{\gamma_{2}}(u)$. Some examples of sample attractor flow lines for some central charges are shown in the diagrams (Figs. \ref{SWSU2sample}, \ref{SWSU2zoom}) below:

\begin{table}[h!]

 \begin{center}
	\begin{tabular}{ | l | l | l | l | l | l | l | l | l |  p{5cm} |}
		\hline
	\multicolumn{9}{|c|}{Split flow lines for example existing BPS states}   \\
	\hline
		Flow line & \multicolumn{2}{|c|}{\begin{tikzpicture}
\text{The blue line}
    \draw [green,thick,-{Stealth}] (0,-1)--(1,-1); 
\end{tikzpicture} }  & \multicolumn{2}{|c|}{\begin{tikzpicture}
\text{The blue line}
    \draw [blue,thick,dashed,-{Stealth}] (0,-1)--(1,-1); 
\end{tikzpicture}} & \multicolumn{2}{|c|}{\begin{tikzpicture}
\text{The blue line}
    \draw [red,thick,dashed,-{Stealth}] (0,-1)--(1,-1); 
\end{tikzpicture}} & \multicolumn{2}{|c|}{\begin{tikzpicture}
\text{The blue line}
    \draw [green,thick,dashed,-{Stealth}] (0,-1)--(1,-1); 
\end{tikzpicture} }  \\ \hline
		Charges & \multicolumn{2}{|c|}{$\gamma_{1}+\gamma_{2}$} &    \multicolumn{2}{|c|}{$\gamma_{2}-2\gamma_{1}$} & \multicolumn{2}{|c|}{$\gamma_{1}-2\gamma_{2}$} & \multicolumn{2}{|c|}{$\gamma_{2}-\gamma_{1}$} \\ \hline
		Split flow lines & \begin{tikzpicture}
\text{The blue line}
    \draw [red,thick,-{Stealth}] (0,-1)--(1,-1); 
\end{tikzpicture}  & \begin{tikzpicture}
\text{The blue line}
    \draw [blue,thick,-{Stealth}] (0,-1)--(1,-1); 
\end{tikzpicture} & \begin{tikzpicture}
\text{The blue line}
    \draw [red,thick,-{Stealth}] (0,-1)--(1,-1); 
\end{tikzpicture}  &     \begin{tikzpicture}
\text{The blue line}
    \draw [blue,thick,-{Stealth}] (0,-1)--(1,-1); 
\end{tikzpicture}                      & \begin{tikzpicture}
\text{The blue line}
    \draw [red,thick,-{Stealth}] (0,-1)--(1,-1); 
\end{tikzpicture} &   \begin{tikzpicture}
\text{The blue line}
    \draw [blue,thick,-{Stealth}] (0,-1)--(1,-1); 
\end{tikzpicture}    & \begin{tikzpicture}
\text{The blue line}
    \draw [red,thick,-{Stealth}] (0,-1)--(1,-1); 
\end{tikzpicture} & \begin{tikzpicture}
\text{The blue line}
    \draw [blue,thick,-{Stealth}] (0,-1)--(1,-1); 
\end{tikzpicture}  \\ \hline
		Charges & $\gamma_{1}$ & $\gamma_{2}$ &  $\gamma_{1}$ &     $\gamma_{2}$              & $\gamma_{1}$ &   $\gamma_{2}$       & $\gamma_{1}$ & $\gamma_{2}$ \\ \hline
		\multicolumn{9}{|c|}{Non-existing BPS states} \\ \hline
		Flow line &  \multicolumn{4}{|c|}{\begin{tikzpicture}
\text{The blue line}
    \draw [brown,thick,-{Stealth}] (0,-1)--(1,-1); 
\end{tikzpicture}}
		&  \multicolumn{4}{|c|}{\begin{tikzpicture}
\text{The blue line}
    \draw [brown,thick,dashed,-{Stealth}] (0,-1)--(1,-1); 
\end{tikzpicture}} \\ \hline
 Charges  & \multicolumn{4}{|c|}{$3\gamma_{2}+\gamma_{1}$} & \multicolumn{4}{|c|}{$\gamma_{1}+\gamma_{2}$}                  \\ \cline{1-5}
 
 Forked flow lines  & \multicolumn{2}{|c|}{\begin{tikzpicture}
\text{The blue line}
    \draw [red,thick,-{Stealth}] (0,-1)--(1,-1); 
\end{tikzpicture}} & \multicolumn{2}{|c|}{\begin{tikzpicture}
\text{The blue line}
    \draw [blue,thick,-{Stealth}] (0,-1)--(1,-1); 
\end{tikzpicture}} & \multicolumn{4}{|c|}{}                  \\ \cline{1-5}

Charges  & \multicolumn{2}{|c|}{$\gamma_{1}$} & \multicolumn{2}{|c|}{$\gamma_{2}$} & \multicolumn{4}{|c|}{}                  \\ \hline
 
	\end{tabular}
\end{center}

    \caption{Split flow lines of BPS states on Fig. \ref{SWSU2sample}.}
    
\end{table}

\begin{figure}
	\centering
	{\includegraphics[width=0.7\textwidth]{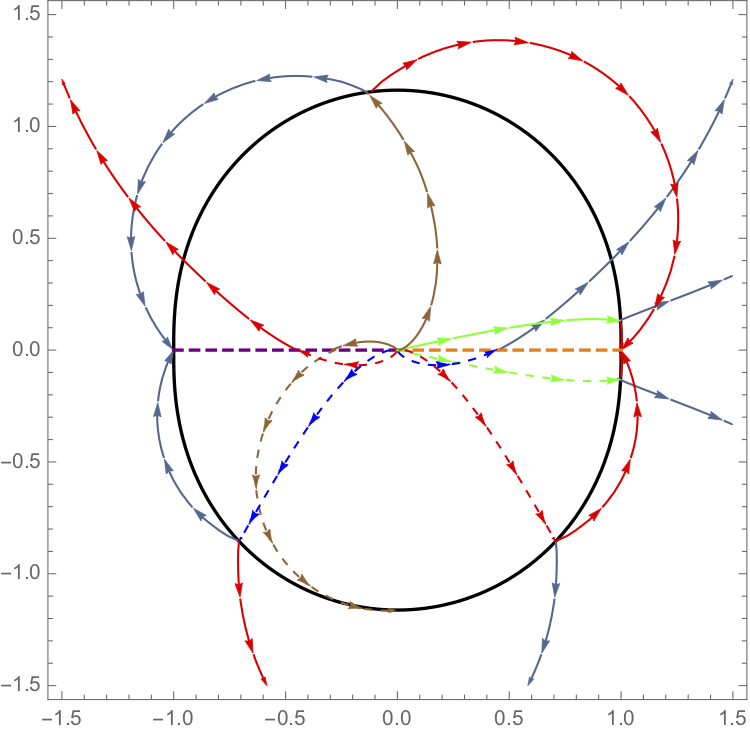}}

	\vspace{-52.5ex}
	\begin{center}
	B1  $\   \  \ \  \  \  \  \   \  \  \ $  
	\end{center}
	\vspace{47ex}
	
	\vspace{-30ex}
	\begin{center}
	 $\   \  \ \  \ $    B2    
	\end{center}
	\vspace{23ex}

	\vspace{-21ex}
	\begin{center}
	A  $\   \  \ \  \  \  \  \ \  \   \  \ \  \  \  \  \ \  \   \  \ \  \  \  \  \ \ \   \  \ \  \  \  \  \ \  \   \  \ \  \  \  \  \ \  \  \  \  \  \ \ \  \   \  \ \ $  
	\end{center}
	\vspace{10ex}

	\captionsetup{singlelinecheck=off}
	\caption[flowline3]{Sample attractor flow lines at infinity for Seiberg-Witten $SU(2)$. The wall of marginal stability is in black.
	
	\begin{enumerate}[label = \alph*.)]
	
	\item
		The solid red and blue lines represent the flows of $Z_{\gamma_{1}}(u)$ and $Z_{\gamma_{2}}(u)$ respectively flowing to $\pm 1$. 
		
	\item	
		The dashed blue and red lines correspond to $Z_{\gamma_{2}}(u)-2Z_{\gamma_{1}}(u)$ and $-2Z_{\gamma_{2}}(u)+Z_{\gamma_{1}}(u)$ - they flow through the branch cut and become the respective basis charges. 
	\item	
		The green line corresponds to $Z_{\gamma_{1}}(u)+Z_{\gamma_{2}}(u)$ this flows parallel to the branch cut and its analytic continuation $Z_{\gamma_{2}}(u)-Z_{\gamma_{1}}(u)$ is shown by the dashed green line. 
	\item	
		The brown line corresponds to $3Z_{\gamma_{2}}(u)+Z_{\gamma_{1}}(u)$ and flows through the branch cut to become  $Z_{\gamma_{1}}(u)+Z_{\gamma_{2}}(u)$ on a new cover, where it flows to a regular point on the lower wall. 
	
	\end{enumerate}	
		
		All higher linear combinations split at the wall into the basis flows.  }
	
	\label{SWSU2sample}
\end{figure}

\subsubsection*{Analytic continuation for monopole and dyon}

The diagram Fig. \ref{SWSU2sample} shows the first example of the existing states generated by analytically continuing the central charges of the basis states (the monopole and dyon) through the branch cuts - the diagram shows  
\begin{align}
 & (M_{+1})^{-1}: \ \ \ Z_{\gamma_{2}}(u)-2Z_{\gamma_{1}}(u) \longmapsto Z_{\gamma_{2}}(u) \ \ \ \ \ \text{and}
 \\ \nonumber
& M_{-1}: \ \ \ -2Z_{\gamma_{2}}(u)+Z_{\gamma_{1}}(u)  \longmapsto Z_{\gamma_{1}}(u) 
\end{align}
as dashed blue \begin{tikzpicture}
\text{The blue line}
    \draw [blue,thick, dashed,-{Stealth}] (0,-1)--(1,-1); 
\end{tikzpicture} and red \begin{tikzpicture}
\text{The blue line}
    \draw [red,thick, dashed,-{Stealth}] (0,-1)--(1,-1); 
\end{tikzpicture} lines in B2, becoming the basis charges in B1 shown by \begin{tikzpicture}
\text{The blue line}
    \draw [blue,thick,-{Stealth}] (0,-1)--(1,-1); 
\end{tikzpicture} and \begin{tikzpicture}
\text{The blue line}
    \draw [red,thick,-{Stealth}] (0,-1)--(1,-1); 
\end{tikzpicture} that flow to singular points and exist everywhere. This process can be continued indefinitely, generating the $nZ_{\gamma_{1}}(u) \pm (n+1)Z_{\gamma_{2}}(u)$ tower. However, these higher combinations only exist in the central chamber B - all such states split at the wall into their composite basis states and flow to $\pm 1$. If the linear combination were to be continued in the outer chamber A around $u=0$ it would flow to a regular point on the wall and be excluded.

\subsubsection*{Flow for W-boson}

The combination $ Z_{\gamma_{1}}(u)+Z_{\gamma_{2}}(u) $ 
(the green line \begin{tikzpicture}
\text{The blue line}
    \draw [green,thick,-{Stealth}] (0,-1)--(1,-1); 
\end{tikzpicture}) exists in chamber B1 and is interesting because it flows in parallel to the branch cut rather than flowing through it. When analytically continued through the cut to $Z_{\gamma_{2}}(u)-Z_{\gamma_{1}}(u) $ in B2 (represented by the dashed green line  \begin{tikzpicture}
\text{The blue line}
    \draw [green,thick,dashed,-{Stealth}] (0,-1)--(1,-1); 
\end{tikzpicture}). The flow is symmetric with that above the cut. Therefore this state exists within the central chamber B on 2 covers, but again splits at the wall into the basis states for the same reason as the other higher combinations. It therefore doesn't exist in the outer region A. Physically this should correspond to the W-boson in the spectrum.

\subsubsection*{Example of flow for non-existing state}

\begin{figure}[h!]
	\centering
	{\includegraphics[width=0.7\textwidth]{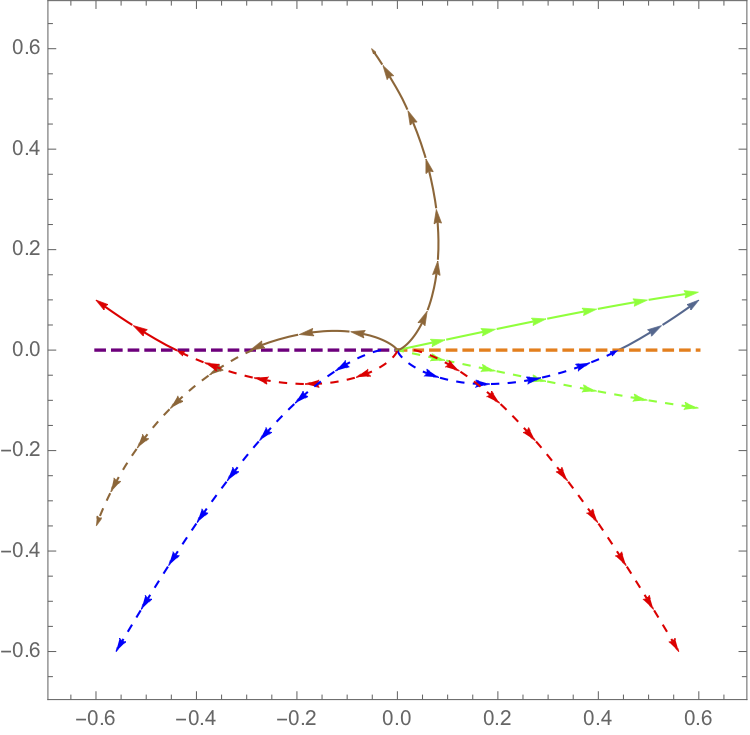}}

	\vspace{-52.5ex}
	\begin{center}
	B1  $\   \  \ \  \  \  \  \   \  \  \ $  
	\end{center}
	\vspace{47ex}
	
	\vspace{-30ex}
	\begin{center}
	 $\   \  \ \  \ $    B2    
	\end{center}
	\vspace{23ex}

	\caption{Zoom in around infinity of the attractor flow on the previous diagram: 
	This shows the flow of  $ Z_{\gamma_{2}}(u)-2Z_{\gamma_{1}}(u)  \rightarrow Z_{\gamma_{2}}(u)$,   $ -2Z_{\gamma_{2}}(u)+Z_{\gamma_{1}}(u)  \rightarrow Z_{\gamma_{1}}(u)$,  $ Z_{\gamma_{1}}(u)+Z_{\gamma_{2}}(u) \rightarrow Z_{\gamma_{2}}(u)-Z_{\gamma_{1}}(u)$  and $3Z_{\gamma_{2}}(u)+Z_{\gamma_{1}}(u) \rightarrow Z_{\gamma_{1}}(u)+Z_{\gamma_{2}}(u)$ in blue, red, green and brown respectively.}

	\label{SWSU2zoom}
\end{figure}

Finally the state $3Z_{\gamma_{2}}(u)+Z_{\gamma_{1}}(u) $ (represented by the brown line \begin{tikzpicture}
\text{The blue line}
    \draw [brown,thick,-{Stealth}] (0,-1)--(1,-1); 
\end{tikzpicture} in B1) flows into the branch cut onto a new cover where it becomes $ Z_{\gamma_{1}}(u)+Z_{\gamma_{2}}(u) $  \begin{tikzpicture}
\text{The blue line}
    \draw [brown,thick,dashed,-{Stealth}] (0,-1)--(1,-1); 
\end{tikzpicture}    in B2 and terminates at a regular point on the lower part of the wall. This means it is one of the states excluded by the existence conditions and is not in the spectrum of BPS states. Other non existing higher combinations follow a similar flow pattern.\\
\\
Below we show a diagram, Fig. \ref{SWSU2zoom}, showing more closely the central region B with the flow lines passing through the branch cuts. This table describes the flow lines on this diagram:

\begin{table}[h!]

 \begin{center}
	\begin{tabular}{| l | l | l | l | l | l | l | l | l | l | p{5cm} |}
		\hline
	\multicolumn{8}{|c|}{Existing BPS states} & \multicolumn{2}{|c|}{Non-existing BPS states} \\
	\hline
Cover 1		& Flow line & \multicolumn{2}{|c|}{\begin{tikzpicture}
\text{The blue line}
    \draw [green,thick,dashed,-{Stealth}] (0,-1)--(1,-1); 
\end{tikzpicture} }  & \multicolumn{2}{|c|}{\begin{tikzpicture}
\text{The blue line}
    \draw [blue,thick,dashed,-{Stealth}] (0,-1)--(1,-1); 
\end{tikzpicture}} & \multicolumn{2}{|c|}{\begin{tikzpicture}
\text{The blue line}
    \draw [red,thick,dashed,-{Stealth}] (0,-1)--(1,-1); 
\end{tikzpicture}} & \multicolumn{2}{|c|}{\begin{tikzpicture}
\text{The blue line}
    \draw [brown,thick,dashed,-{Stealth}] (0,-1)--(1,-1); 
\end{tikzpicture} } \\ \cline{2-10}
	&	Charges & \multicolumn{2}{|c|}{$\gamma_{2}-\gamma_{1}$} &    \multicolumn{2}{|c|}{$\gamma_{2}-2\gamma_{1}$} & \multicolumn{2}{|c|}{$\gamma_{1}-2\gamma_{2}$} & \multicolumn{2}{|c|}{$\gamma_{1}+\gamma_{2}$} \\ \hline
Cover 2	&	Split flow lines &    \multicolumn{2}{|c|}{\begin{tikzpicture}
\text{The blue line}
    \draw [green,thick,-{Stealth}] (0,-1)--(1,-1); 
\end{tikzpicture} }  &   \multicolumn{2}{|c|}{\begin{tikzpicture}
\text{The blue line}
    \draw [red,thick,-{Stealth}] (0,-1)--(1,-1); 
\end{tikzpicture} }                        &  \multicolumn{2}{|c|}{\begin{tikzpicture}
\text{The blue line}
    \draw [blue,thick,-{Stealth}] (0,-1)--(1,-1); 
\end{tikzpicture} }     &  \multicolumn{2}{|c|}{\begin{tikzpicture}
\text{The blue line}
    \draw [brown,thick,-{Stealth}] (0,-1)--(1,-1); 
\end{tikzpicture} }  \\ \cline{2-10}
	&	Charges & \multicolumn{2}{|c|}{$\gamma_{1}+\gamma_{2}$} &  \multicolumn{2}{|c|}{$\gamma_{1}$}              & \multicolumn{2}{|c|}{$\gamma_{2}$}        & \multicolumn{2}{|c|}{$3\gamma_{2}+\gamma_{1}$} \\ \hline
	\end{tabular}
\end{center}

    \caption{Flow lines through branch cuts on Fig. \ref{SWSU2zoom}.}
    \label{tab:higherstates2}
\end{table}

\subsubsection*{Final existing states in each chamber}

We now have all the required information to write down the spectrum of existing BPS states in each chamber, which we present in table (\ref{SWfinaltableexistence}) below.

\begin{table}[h!]

\begin{center} 
	\begin{tabular}{ |  p{5cm}  | p{3.8cm}| p{3.8cm} | p{2cm}| p{2cm} |}
		\hline
	\multicolumn{4}{|c|}{\vphantom{\Huge{H}} All existing states in all chambers} 
	    \\[7pt] \hline
	    
		Chamber & Existing charges $\ \ $ cover 1  & Existing charges $\ \ $ cover 2    & Count \\ \hline
		B1: Central region upper half & $  -\gamma_{1} ,  -\gamma_{2}, \   -\gamma_{1}-\gamma_{2} \ \ \ \ \ $  \   \    $-(n+2)\gamma_{1}-(n+3)\gamma_{2} $ $-(n+2)\gamma_{2}-(n+3)\gamma_{1} $ & $ \  \ \gamma_{1} , \ \gamma_{2}, \ \ \gamma_{1}+\gamma_{2} \   \  \ \ \  $      $ \  \ n\gamma_{1}+(n+1)\gamma_{2} \  \   \  \   \  \ $  $\   \    \   n\gamma_{2}+(n+1)\gamma_{1} $& Infinite \\ \hline
		B2: Central region lower half & $ \ \ \ -\gamma_{1} , \   -\gamma_{2}, \  \gamma_{2}-\gamma_{1} \  \  \ $   \  \  \   $ (n+4)\gamma_{1}-(n+3)\gamma_{2} \   \   \   \ $  $\  \  \ (n+4)\gamma_{2}-(n+3)\gamma_{1} $&  $ \ \ \gamma_{1} , \  \gamma_{2}, \  \gamma_{1}-\gamma_{2}$ \newline  \   \   $-(n+2)\gamma_{1}+(n+1)\gamma_{2} $   $-(n+2)\gamma_{2}+(n+1)\gamma_{1} $  & Infinite \\ \hline
		A: Outer region & $ \  \ \pm\gamma_{1},  \ \ \pm\gamma_{2}$ & $ \ \ \pm\gamma_{1},  \ \ \pm\gamma_{2}$ &2 \\ \hline
		
	\end{tabular}
\end{center}

    \caption{Existing states in Seiberg-Witten theory on 2 covers.}
    \label{SWfinaltableexistence}
\end{table}

\subsection{Summary and discussion of results on attractor flow}

We have used the attractor flow equations of \cite{Ferrara_1995,Ferrara_1996} derived from the type IIB supergravity limit to determine and reproduce the spectra of BPS states in $\mathcal{N}=2$ theories such as Argyres-Douglas theories and Seiberg-Witten theory. We followed the methods developed in the literature \cite{Moore:1998pn,Denef:1998sv,Denef:2000nb,Denef:2001xn,Denef:2007vg} to count BPS states from existence conditions on the endpoint of the flow. This was done by approximating the flow as that described by the spherically symmetric supergravity equations. Given that we are working with theories with complex 1-dimensional moduli spaces $\mathcal{B}$ we could reduce the attractor flow equations to gradient flow lines on this moduli space. The gradient flow was found in $|Z_{\gamma_{i}}(u)|$, where the central charges were derived by solving the Picard-Fuchs equations for the periods of the elliptic curve $\Sigma$ describing the theory. The linear combinations of periods that  
correspond to a basis of BPS central charges were determined by identifying the vanishing cycles at the singular points.

We found that these gradient flow lines encode the spectrum of BPS states in each region of the moduli space with the wall crossing being described by split flow lines (as described for example in \cite{Denef:2000nb} for Seiberg-Witten theory) at the wall of marginal stability. If a flow line entered a branch cut we acted with a monodromy to analytically continue the central charge through the cut. This then allowed us the determine the spectrum of BPS states on a new cover. This method reproduced the known spectrum of BPS states at strong and weak coupling. For Seiberg-Witten theory \cite{SW} at strong coupling there is just a monopole and dyon whereas at weak coupling there are infinitely many additional dyons and a W-boson. For the Argyres-Douglas $A_{2}$ model \cite{Argyres:1995jj} (see also \cite{Shapere:1999xr}) there is a monopole and dyon at strong coupling that combine into a dyon on the other side of the wall of marginal stability.

These wall crossing phenomena have also been described previously in the literature on quivers \cite{Douglas:2000qw,Denef:2007vg,Dimofte:2009tm,Alim:2011kw}, via quiver representations as well as the mutation method. A further interpretation in terms of scattering diagrams into which the split attractor flow lines can be embedded was developed by Bridgeland in \cite{Bridgelandscattering} for the Argyres-Douglas and Seiberg-Witten examples.  These scattering diagrams are given again in the context of attractor flow in \cite{Bousseau:2022snm}. 

The Argyres-Douglas (AD) $A_{2}$ models we are looking at are deformations of $A_{2}$ singularities at AD points in one complex variable $u \in \mathbb{C}$. In general the Argyres-Douglas points exist within the moduli space of SU(3) gauge theory \cite{Argyres:1995jj}. This has a complex 2d moduli space which we can call $\tilde{u}, \tilde{v} \in \mathbb{C}$. One can recover the Argyres-Douglas model by choosing slices in this moduli space that pass through the AD points. These exist at $(\tilde{u}, \tilde{v}) = (0,\pm 1)$. This extended moduli space has been studied extensively in \cite{DDN}. In this case as in our work the periods were computed, initially in \cite{Klemm:1995wp}, using solutions to Picard-Fuchs equations. However, because in this case there are 2 complex variables the hypergeormetric functions are generalised to Appell $F_{4}$ functions.

In this case as in our example the periods are a linear combination of these functions and the central charges of the BPS states are  now given by charge multiples of these periods. In the full $SU(3)$ theory there are 6 BPS states in the strong coupling region. As with the BPS states in our example these BPS states can become massless at Argyres-Douglas points. In fact 3 of these states become massless at the point $(\tilde{u}, \tilde{v}) = (0,+1)$  and the other 3 at $(\tilde{u}, \tilde{v}) = (0,-1)$. As with our examples the walls of marginal stability occur when the central charges align. However, there are now 2 complex variables so one must take complex one dimensional slices on which one can plot the walls.

In \cite{DDN} the walls are plotted on the $\tilde{u} = 0$ and the $\tilde{v} = 0$ planes. For the $\tilde{u} = 0$ slice the walls look like those in our examples and pass through both AD points where the central charges of the 2 particles vanish. However, in this case the walls are not symmetric about $\text{Im}[\tilde{v}] = 0$. This symmetry is restored when one overlays the walls (Fig. 8 of \cite{DDN}) or if one plots the locus of vanishing Kähler potential. For the $\tilde{v} = 0$ slice the situation simplifies as all central charges can be written in terms of one period and its dual. Therefore there is only one wall on this slice given by the real ratio of the period with its dual.  In the full $SU(3)$ theory there are additional 3 points called multi-monopole points where 3 pairs of BPS states can become simultaneously massless. These are found on this wall in the $\tilde{v}=0$ slice. However, these are not present in our examples which are outside this slice. 

It would be interesting to apply the attractor flow method to such theories with higher dimensional moduli spaces. This should also be generalisable to other ADE type Argyres-Douglas theories if the moduli dependent central charges can be found.

\section{Conclusion and discussion}\label{seccon}
In this paper we investigated how the stability and existence data of a given BPS structure is encoded in its moduli space. We introduced a notion of BPS variation of Hodge structure guided by the variation of Hodge structure governing the periods of CY threefolds when the BPS structures are geometrically realized by type IIB string theory on CY threefolds. Our goal is to give the data of the BPS-VHS as an abstract part of the BPS structure setting which is independent of the specific geometric realization considered. We expect this to shed more light on the additional structure required on Bridgeland's spaces of stability conditions \cite{Bridgelandspaces} which are generically much larger than the ones considered in the physical context since the notion of special geometry obeyed by the central charges is missing. We hope that our setup and the concrete examples will guide the way towards developing this further. It would be interesting in particular to investigate whether the admissibility of a BPS-VHS puts constraints on the allowed combinations of BPS structure, lattices and pairings considered on these lattices. A similar spirit of classification of allowed physical theories given topological data was undertaken by Cecotti and Vafa in \cite{Cecotti:1992rm} and was also a motivation of distinguishing BPS quivers corresponding to BPS structures which can be physically realized from the ones which cannot in \cite{Cecotti:2011rv,Alim:2011ae,Alim:2011kw}.

Using the BPS VHS we derived Picard-Fuchs equations for two geometric realizations of $A_2$ Argyres-Douglas theory which allowed us to explore the exact duality groups as well as the structure in terms of quasi-modular forms of the central charges of these theories. These results can be used as the input data of study of both topological string theory at higher genus on the corresponding non-compact threefold geometry along the lines of \cite{Huang:2006si,Aganagic:2006wq,ASYZ} as well as topological string theory at higher orders of an $\hbar$ deformation corresponding to the quantum mechanical context given by the NS limit as has been addressed for instance in \cite{Codesido:2017dns}. An interesting outcome of our analysis are the walls of marginal stability which were known in the first realization of AD theory which we considered as well as in the SW case. The second realization of AD theory led to an interesting picture dividing the moduli space into 5 chambers. Using the attractor flow equations we subsequently studied in details the attractor flow in our examples showing in particular that in these cases the analysis of the attractor flow gives access to the complete information of the spectrum as well as the wall-crossing data. In more general non-compact threefold geometry is a much more formidable task and has been recently revisited for instance in \cite{Bousseau:2022snm}.

\subsection*{Acknowledgements}
The work of the Murad Alim, Florian Beck and Daniel Bryan is supported through the DFG Emmy Noether grant AL 1407/2-1. Daniel Bryan was furthermore supported by the Friedrich-Naumann-Foundation (FNF) scholarship. We would like to thank Arpan Saha and Ivan Tulli for discussions.


\providecommand{\href}[2]{#2}\begingroup\raggedright\endgroup

\end{document}